\documentclass[aps,prd,twocolumn,showpacs,floats,nofootinbib]{revtex4-1}

\usepackage{graphicx}
\usepackage{amsmath,amsfonts}
\usepackage{dcolumn}
\usepackage{hyperref}

\mathchardef\minus = "002D

\newcommand{\swS}[5][]{{}_{{}_{#2}}S^{#1}_{#3}(#4;#5)}
\newcommand{\scA}[4][]{{}_{{}_{#2}}A^{#1}_{#3}(#4)}

\newtheorem{theorem}{Theorem}

\newcolumntype{f}[1]{D{.}{.}{#1}}
\newcommand{\Chead}[1]{\multicolumn{1}{c}{#1}}

\begin{document}

\title{Modes of the {K}err geometry with purely imaginary frequencies}

\author{Gregory B. Cook}\email{cookgb@wfu.edu}
\affiliation{Department of Physics, Wake Forest University,
		 Winston-Salem, North Carolina 27109}
\author{Maxim Zalutskiy}\email{zalump8@wfu.edu}
\affiliation{Department of Physics, Wake Forest University,
		 Winston-Salem, North Carolina 27109}

\date{\today}

\begin{abstract}
In this paper, we examine the behavior of modes of the Kerr geometry
when the mode's frequency is purely imaginary.  We demonstrate that
quasinormal modes must be polynomial in nature if their frequency is
purely imaginary, and present a method for computing such modes.  The
nature of these modes, however, is not always easy to determine.  Some
of the polynomial modes we compute are quasinormal modes.  However,
some are simultaneously quasinormal modes and total transmission
modes, while others fail to satisfy the requisite boundary conditions
for either.  This analysis is, in part, an extension of the results
known for Schwarzschild black holes, but clarifies misconceptions for
the behavior of modes when the black hole has angular momentum.  We
also show that the algebraically special modes of Kerr with $m=0$ have
an additional branch of solutions not seen before in the literature.
All of these results are in precise agreement with new numerical
solutions for sequences of gravitational quasinormal modes of Kerr.
However, we show that some prior numerical and analytic results
concerning the existence of quasinormal modes of Kerr with purely
imaginary frequencies were incorrect.
\end{abstract}

\pacs{04.20.-q,04.70.Bw,04.20.Cv,04.30.Nk}

\maketitle

\section{Introduction}
\label{sec:introduction}

The Kerr geometry, representing an isolated black hole with angular
momentum, was introduced more than 50 years ago\cite{kerr-1963}.  It
is arguably the most astrophysically important solution of Einstein's
equations, and has been extensively studied (see
Ref.\cite{teukolsky_kerr-2015} for a recent review).  In this paper,
we are primarily interested in the quasinormal modes (QNMs) of the
Kerr spacetime.  They represent the natural resonant vibrations of
the black hole and are linear perturbations with boundary conditions
that demand no wave travel in from infinity or travel out of the black
hole.  QNMs of black holes in general have been extensively studied,
and Refs.\cite{berti-QNM-2009,nollert-qnm-1999} offer excellent
reviews of the subject.  The QNMs of Kerr are of great importance to
the new field of gravitational-wave astronomy as they describe the
late-time ring-down of a remnant black hole following some violent
astrophysical event such as the collision of two black
holes\cite{GW150914-2016,GW151226-2016}.  They may also offer clues to
the transition between classical and quantum gravity\cite{Hod-1998}.

Our understanding of the nature of a particular set of gravitational
modes of the Kerr geometry has remained confused for some time.
Leaver's\cite{leaver-1985} numerical results based on the
Regge-Wheeler equation\cite{regge-wheeler-1957} first suggested that
certain QNMs of the Schwarzschild geometry might have frequencies that
exist on the negative imaginary axis (NIA).  These were found with
frequencies, $\omega$ which were consistent with the algebraically
special modes of Kerr in the Schwarzschild limit\cite{chandra-1984}.
The frequencies of the algebraically special modes of Schwarzschild
will be denoted $\omega=\Omega_\ell$ (see
Eq.~(\ref{eq:alg-spec-sch})).  The algebraically special modes of Kerr
are, in general, total transmission modes (TTMs).  There are two types
of TTMs, distinguished by their behavior at the boundaries.  If the
QNM boundary conditions are changed at the black-hole boundary to
demand no waves travel into the black hole, then we have ``left''-TTMs
(TTM${}_L$s).  If, instead, the QNM boundary conditions are changed at
infinity to demand no waves travel out at infinity, then we have
``right''-TTMs (TTM${}_R$s).  While the behavior of the algebraically
special solutions at infinity was clear, their behavior at the event
horizon was not explicitly considered in Ref.~\cite{chandra-1984}.  On
the other hand, in Ref.\cite{andersson-1994} the author examined the
first few algebraically special modes of the Regge-Wheeler equation
and found them to be TTM${}_R$s.  Then, in Ref.\cite{onozawa-1997},
based on more refined numerical studies, the author found that the
frequencies of certain QNMs of Kerr do, in the Schwarzschild limit,
approach the $\Omega_\ell$.  However, the author also suggested that
the QNMs should disappear at the $\Omega_\ell$ where they are replaced
by the left and right TTMs.  At this point, the literature suggested
that both TTMs exist at the $\Omega_\ell$, and there was no conclusive
evidence for QNMs at the $\Omega_\ell$.  This confusion was set to
rest by Maassen van den Brink\cite{van_den_brink-2000} who rigorously
proved that at the $\Omega_\ell$, Schwarzschild black hole modes are
simultaneously QNMs and TTM${}_L$s, but are not TTM${}_R$s.  More
precisely, using the supersymmetric relationship between the
Zerilli\cite{zerilli-1970} and Regge-Wheeler\cite{regge-wheeler-1957}
equations, he showed that the even-parity $\Omega_\ell$ modes of the
Zerilli equation are simultaneously QNMs and TTM${}_L$s, while the
odd-parity $\Omega_\ell$ modes of the Regge-Wheeler equation are
neither QNMs nor TTMs.  This fully clarified our understanding of the
modes of Schwarzschild with frequencies on the NIA, but how the modes
of Kerr, when the angular momentum is non-zero, behave when their
frequencies are on the NIA was still not understood.  Onozawa and
collaborators\cite{onozawa-1997,berticardoso-2003} found the first
clear numerical evidence that QNMs of Kerr can have frequencies that
approach the NIA at locations that, while close to the $\Omega_\ell$,
are not at the $\Omega_\ell$.  Their results were refined in
Ref.~\cite{cook-zalutskiy-2014}, where we found solutions with
frequencies much closer to the NIA.  These results are reproduced as
part of our Fig.~\ref{fig:malll2n8}.

Continuing our exploration of the Kerr QNMs first reported in
Ref.~\cite{cook-zalutskiy-2014}, we have found numerous new examples
where the frequencies of sequences of QNMs (parameterized by the
angular momentum of the black hole) get arbitrarily close to the NIA.
We will examine these numerical results in
Sec.~\ref{sec:numerical_results}.  As we were beginning our numerical
investigations of Kerr QNMs, Yang {\em et
  al}\cite{Yang-et-al-2013a,Yang-et-al-2013b} and then
Hod\cite{Hod-2013} reported finding a continuum of Kerr QNMs with
frequencies on the NIA in the limit of small frequency $|\omega|\ll1$
and for angular momenta near the extreme limit.  However, we have
observed no numerical evidence of this family of solutions.

Our goal in this paper is to develop a clear understanding of the
behavior of the set of modes of the Kerr geometry which have
frequencies that lie on the NIA.  A brief description of our main
results can be found in a shorter paper\cite{cook-zalutskiy-2016a}.
We have been primarily interested in the QNMs, but a clear
understanding of these modes requires that we consider QNMs {\em and}
TTMs.  Using the framework of the confluent Heun equation, and in
particular the theory of confluent Heun polynomial solutions outlined
in Ref.\cite{cook-zalutskiy-2014}, we will show that any modes with
frequencies on the NIA must be polynomial.  We will show that
potential QNMs on the NIA fall into two categories, both of which
consist of countably infinite sets of solutions at discrete values of
the black hole's angular momentum.  One category yields solutions that
are QNMs.  This category of solutions has never before been
recognized.  A second category of solutions is more complicated,
itself split into two different behaviors.  One subset of polynomial
solutions are neither QNMs nor TTMs, but are an inseparable 
combination of QNM and TTM${}_L$ behaviors.  The other subset of
polynomial solutions are {\em simultaneously} QNMs and TTM${}_L$s.
This subset can be considered a direct extension, to non-vanishing
angular momenta, of the modes of Schwarzschild with frequencies
$\Omega_\ell$.  These special modes occur at discrete frequencies
along the $m=0$ sequences of the algebraically special modes of Kerr.
Maassen van den Brink\cite{van_den_brink-2000} argued that this would
not happen except in the Schwarzschild case.  We have shown this not
to be true and, further, have found an additional branch of the $m=0$
algebraically special modes of Kerr that, to our knowledge, has never
before been recognized.

This paper is organized as follows.  In
Sec.~\ref{sec:teukolsky_equations}, we provide a very brief overview
of the Teukolsky equations governing perturbations in the Kerr
geometry.  We also review the definitions of QNMs and TTMs.  In
Sec.~\ref{sec:numerical_results}, we provide an overview of the new
numerical results that strongly suggested to us that many QNMs with
frequencies on the NIA might exist.  In Sec.~\ref{sec:modes_on_NIA},
we present the theory behind, and main results of the paper.  The
general formalism for the confluent Heun equation is reviewed in
Sec.~\ref{sec:sol_of_CHE}.  This is applied to the Teukolsky radial
equation in Sec.~\ref{sec:radial-teukolsky-Leaver}, where we also
demonstrate that QNMs on the NIA must be polynomial.  We discuss the
details of how we find potential polynomial QNMs in
Sec.~\ref{sec:Heun_polynomials}.  Then we discuss how we characterize
these modes in Sec.~\ref{sec:Generic_Anomalous_Miraculous} by using
methods outlined by Maassen van den
Brink\cite{van_den_brink-2000}. Finally, we summarize and discuss our
results in Sec.~\ref{sec:summary}.  In particular, we discuss how our
results reveal a new branch of the $m=0$ algebraically special modes
of Kerr in Sec.~\ref{sec:summary_anom}.  In
Sec.~\ref{sec:summary_mirac}, we provide some insights into the
incorrect results of
Refs.\cite{Yang-et-al-2013a,Yang-et-al-2013b,Hod-2013} claiming
to find a continuum of QNMs with frequencies on the NIA.

\section{The Teukolsky Equations}
\label{sec:teukolsky_equations}
Perturbations of the Kerr geometry obey the Teukolsky master equation
which governs a complex function ${}_s\psi$ of spin-weight
$s$\cite{teukolsky-1973}.  Assuming the vacuum case, the master
equation separates using
\begin{equation}\label{eq:Teukolsky_separation_form}
  {}_s\psi(t,r,\theta,\phi) = e^{-i\omega{t}} e^{im\phi}S(\theta)R(r).
\end{equation}
The radial function $R(r)$ then satisfies the radial Teukolsky
equation
\begin{subequations}
\begin{align}\label{eqn:radialR:Diff_Eqn}
\Delta^{-s}\frac{d}{dr}&\left[\Delta^{s+1}\frac{dR(r)}{dr}\right]
 \\
&+ \left[\frac{K^2 -2is(r-M)K}{\Delta} + 4is\omega{r} - \lambdabar\right]R(r)=0,
\nonumber
\end{align}
where
\begin{align}
  \Delta &\equiv r^2-2Mr+a^2, \\
  K &\equiv (r^2+a^2)\omega - am, \\
  \lambdabar &\equiv \scA{s}{\ell{m}}{a\omega} + a^2\omega^2 - 2am\omega.
\end{align}
\end{subequations}
Here, Boyer-Lindquist coordinates are used.  $M$ is the mass of the
black hole and $a=J/M$ is the angular momentum parameter.  Finally,
$\scA{s}{\ell{m}}{a\omega}$ is the angular separation constant
associated with the angular Teukolsky equation governing $S(\theta)$.
With $x=\cos\theta$, the function
$S(\theta)=\swS{s}{\ell{m}}{x}{a\omega}$ is the spin-weighted
spheroidal function satisfying
\begin{align}\label{eqn:swSF_DiffEqn}
\partial_x \Big[ (1-x^2)\partial_x [\swS{s}{\ell{m}}{x}{c}]\Big] %\hspace{0.6in}
& \nonumber \\ 
    + \bigg[(cx)^2 - 2 csx + s& + \scA{s}{\ell m}{c} %\hspace{0.2in}
 \\ 
      & - \frac{(m+sx)^2}{1-x^2}\bigg]\swS{s}{\ell{m}}{x}{c} = 0,
\nonumber
\end{align}
where $c\ (=a\omega)$ is the oblateness parameter and $m$ the azimuthal
separation constant.

With appropriate boundary conditions, the Teukolsky equations can be
solved to determine various modes of the Kerr geometry.  For example,
if we demand that no waves travel into the domain from infinity and
that no waves travel out from the black hole horizon, then the
solutions of the Teukolsky equation will be QNMs.  These are the
natural resonance frequencies of a black hole.  If we reverse one of
these boundary conditions, our solutions will represent TTMs.  If we
demand that no waves travel into the black hole horizon, then the
solutions are referred to as ``left''-TTMs (TTM${}_{\rm L}$).  Loosely
speaking, such a wave travels out from the vicinity of the black hole
and to infinity with no net reflection.  If, instead, we demand that
no waves travel out of the domain at infinity, then the solutions are
referred to as ``right''-TTMs (TTM${}_{\rm R}$).  Loosely speaking,
such a wave travels in from infinity and then into the black hole with
no net reflection.  If we reverse both QNM boundary conditions, a
solution would represent a bound state which is not possible in the
Kerr geometry.

QNM and TTM${}_{\rm L}$ modes can be obtained when we choose $s\le0$,
while TTM${}_{\rm R}$ modes require $s\ge0$\cite{teukolsky-1973}.  With
$|s|=2$, solutions represent gravitational perturbations. $|s|=1$
gives electromagnetic perturbations, and $s=0$ gives scalar
perturbations.  Half-integer values of $s$ are also allowed.

The details of our method for computing QNMs are given in
Ref.~\cite{cook-zalutskiy-2014}, where we also describe a
high-accuracy study of the gravitational ($s=-2$) QNMs.  We will
repeat those details here only where they are directly relevant to our
current investigations.

\section{Numerical results.}
\label{sec:numerical_results}
We next show several overview plots for the complex frequencies of
gravitational ($s=-2$) QNMs of the Kerr geometry.  In
Ref.~\cite{cook-zalutskiy-2014}, we showed general results for the
first 8 overtones, $0\le n\le7$, but for all modes with
$2\le\ell\le16$.  We also considered selected sequences with $n=8$.
In this paper, we will restrict ourselves to modes with $\ell=2,3,4$,
but will consider overtones up to $n=31$.  An individual solid line in
these plots represents a sequence of mode frequencies parameterized by
the dimensionless angular momentum of the black hole $\bar{a}=a/M$ in
the range $0\le\bar{a}<1$.  In these figures, a dimensionless version
of the mode frequency $\bar\omega\equiv M\omega$ is used.

Only two significant changes were made to the numerical
methods\cite{cook-zalutskiy-2014} used in computing these results.
First, the method for choosing the step size in $\bar{a}$ when
computing sequences of modes was modified to be more efficient and
effective\footnote{A maximum step size of $\Delta\bar{a}=0.001$ is
  maintained, and refinement by bisection is controlled by comparing
  the change in the mode frequency, $\Delta\bar\omega$, between steps
  to the local radius of curvature of the sequence}.  The second
change was to the method for choosing the truncation depth of the
continued fraction used in locating the QNM frequencies\footnote{We
  now estimate the error as a function of the truncation depth and
  choose the depth to maintain a desired accuracy.  A maximum absolute
  error in $\bar\omega$ of $10^{-8}$ is maintained along the sequence,
  but this error is decreased as necessary as adaptive refinement
  causes $\Delta\bar\omega$ between adjacent solutions to decrease.}.

\begin{figure}
\includegraphics[width=\linewidth,clip]{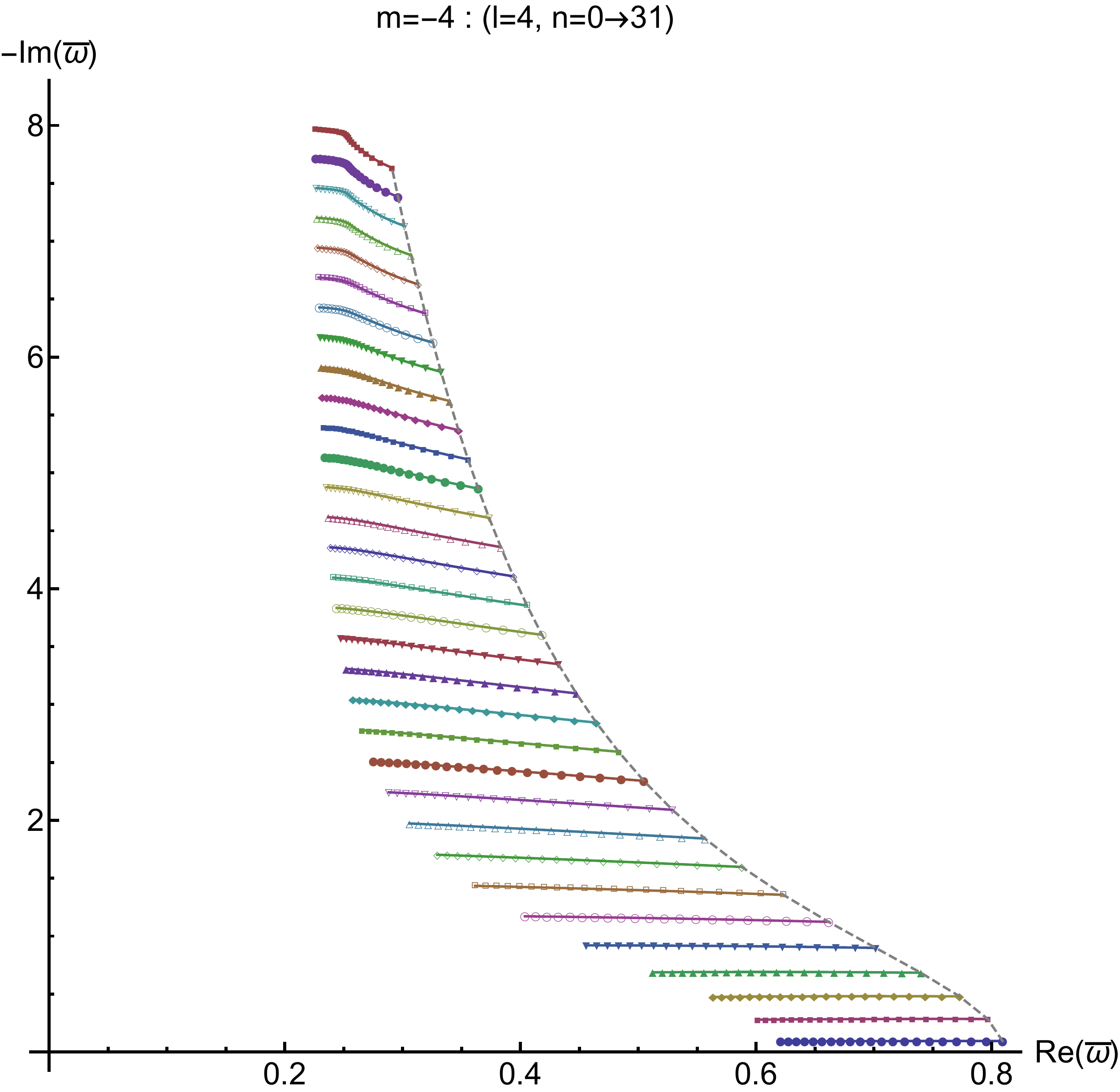}
\caption{\label{fig:mn4l4n00-31} Kerr QNM mode sequences for $m=-4$.
  The complex frequency $\bar\omega$ is plotted for the cases $\ell=4$
  and $0\le n\le31$.  Note that the imaginary axis is inverted.  Each
  sequence covers the range $0\le\bar{a}<1$, with markers on each
  sequence denoting a change in $\bar{a}$ of $0.05$.  The $\bar{a}=0$
  element of overtone are connected by a dashed line.  The overtone
  index $n$ increases monotonically as we move up the dashed line.}
\end{figure}

\begin{figure}
\includegraphics[width=\linewidth,clip]{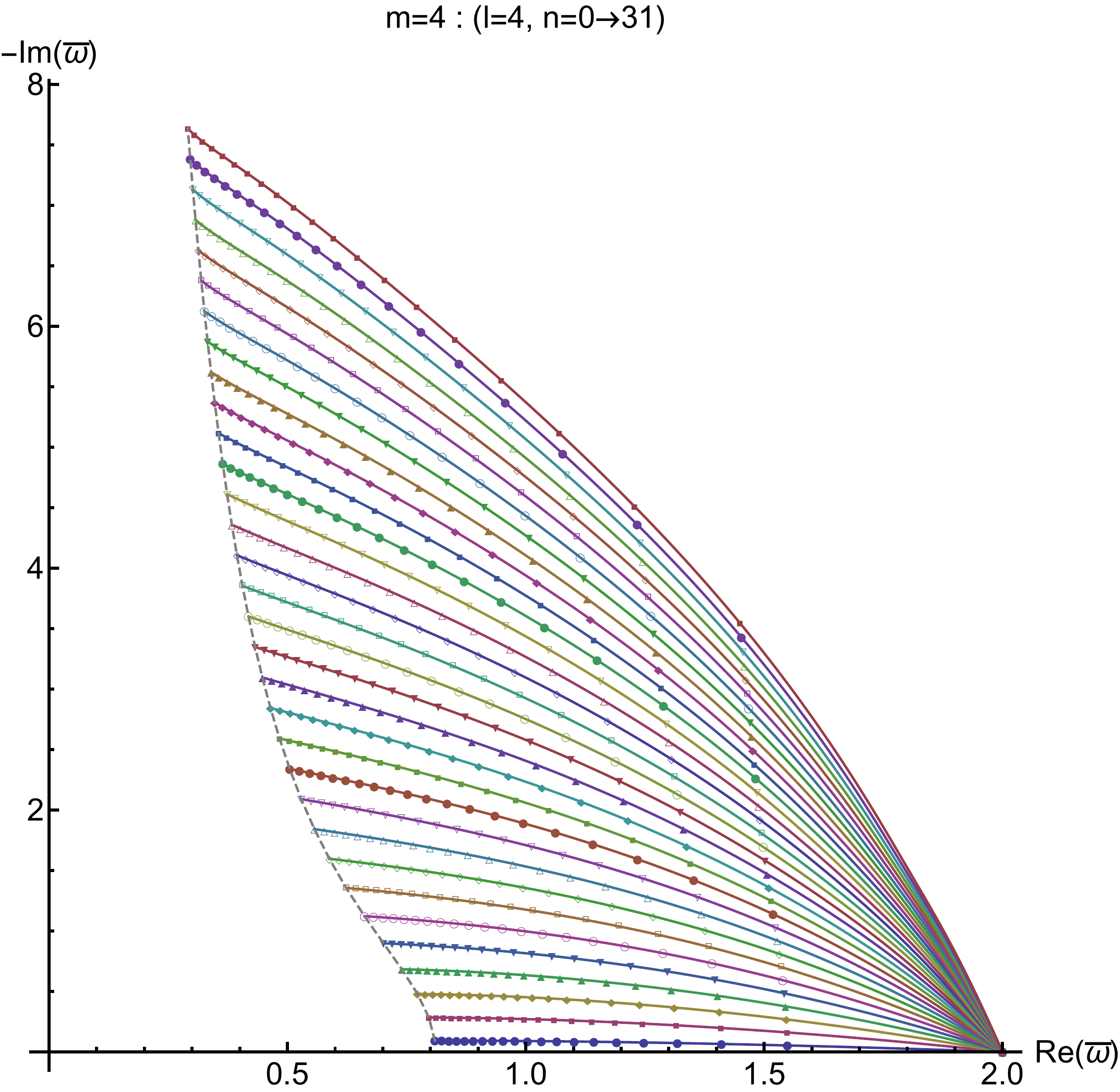}
\caption{\label{fig:mp4l4n00-31} Kerr QNM mode sequences for $m=4$.
  See Fig.~\ref{fig:mn4l4n00-31} for a full description.  Note that as
  $\bar{a}\rightarrow1$ the sequences approach an accumulation point
  at $\bar\omega=m/2$.}
\end{figure}

Figures~\ref{fig:mn4l4n00-31} and \ref{fig:mp4l4n00-31} display
sequences for $\{\ell,m,n\}=\{4,-4,0\!\!\!\rightarrow\!\!\!31\}$ and
$\{4,4,0\!\!\!\rightarrow\!\!\!31\}$ respectively.  The Schwarzschild
limit, $\bar{a}=0$, for each overtone of a given $\ell$ and $m$ is
connected by a dashed gray line.  In the Schwarzschild limit, the
overtone number increases with increasing $-{\rm Im}(\bar\omega)$.
Finally, for sequences with $m>0$ (and ${\rm Re}(\bar\omega)>0$), we
find that many of the sequences approach an accumulation point at
$\bar\omega=m/2$ as clearly seen in Fig.~\ref{fig:mp4l4n00-31}.

\begin{figure}
\includegraphics[width=\linewidth,clip]{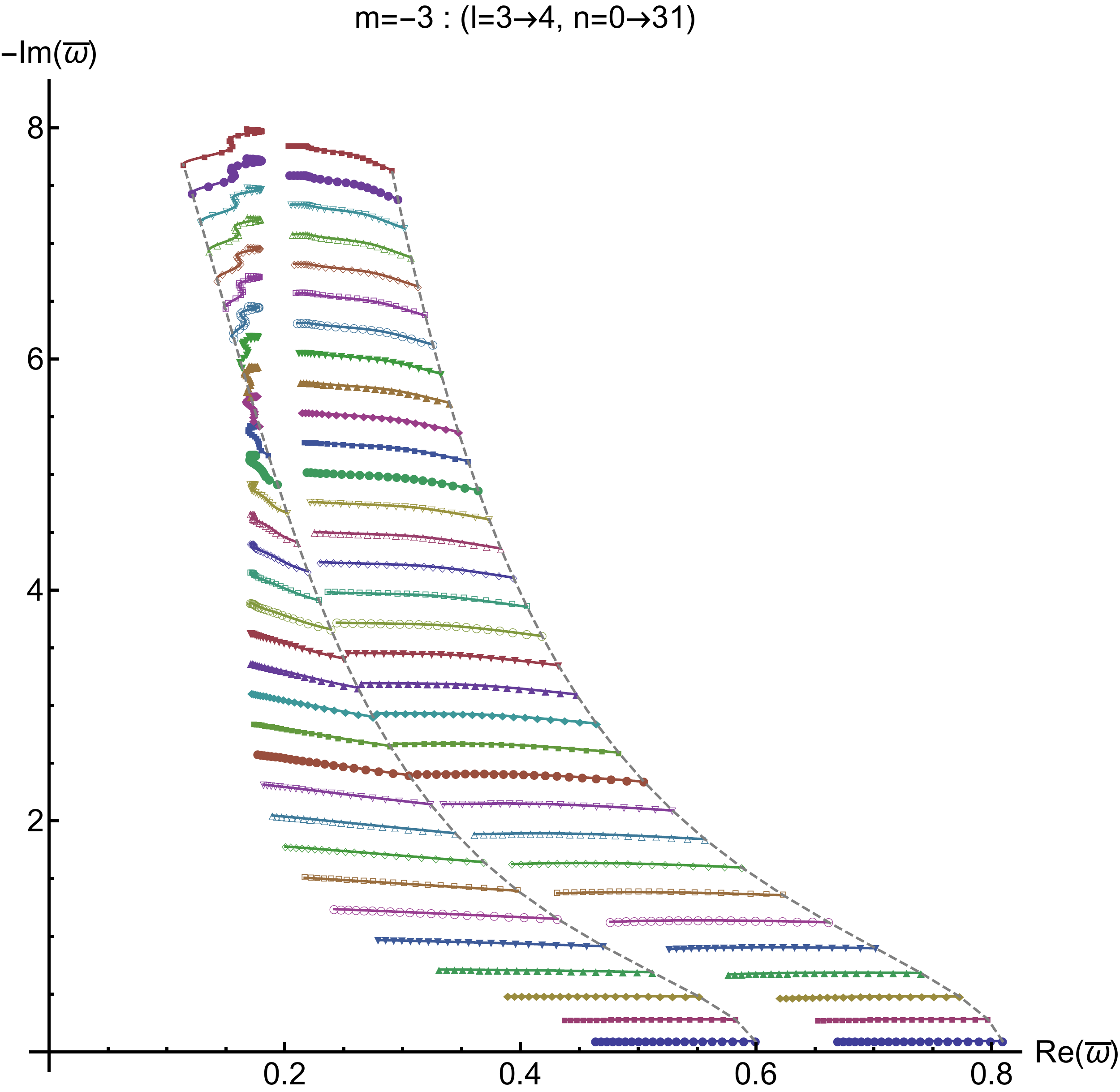}
\caption{\label{fig:mn3l3-4n00-31} Kerr QNM mode sequences for $m=-3$.
  See Fig.~\ref{fig:mn4l4n00-31} for a full description.  In this
  case, we plot the $\ell=3$ and $\ell=4$ sequences.  Sequences with
  lower $\ell$ are generally to the left of sequences with higher
  $\ell$.}
\end{figure}

\begin{figure}
\includegraphics[width=\linewidth,clip]{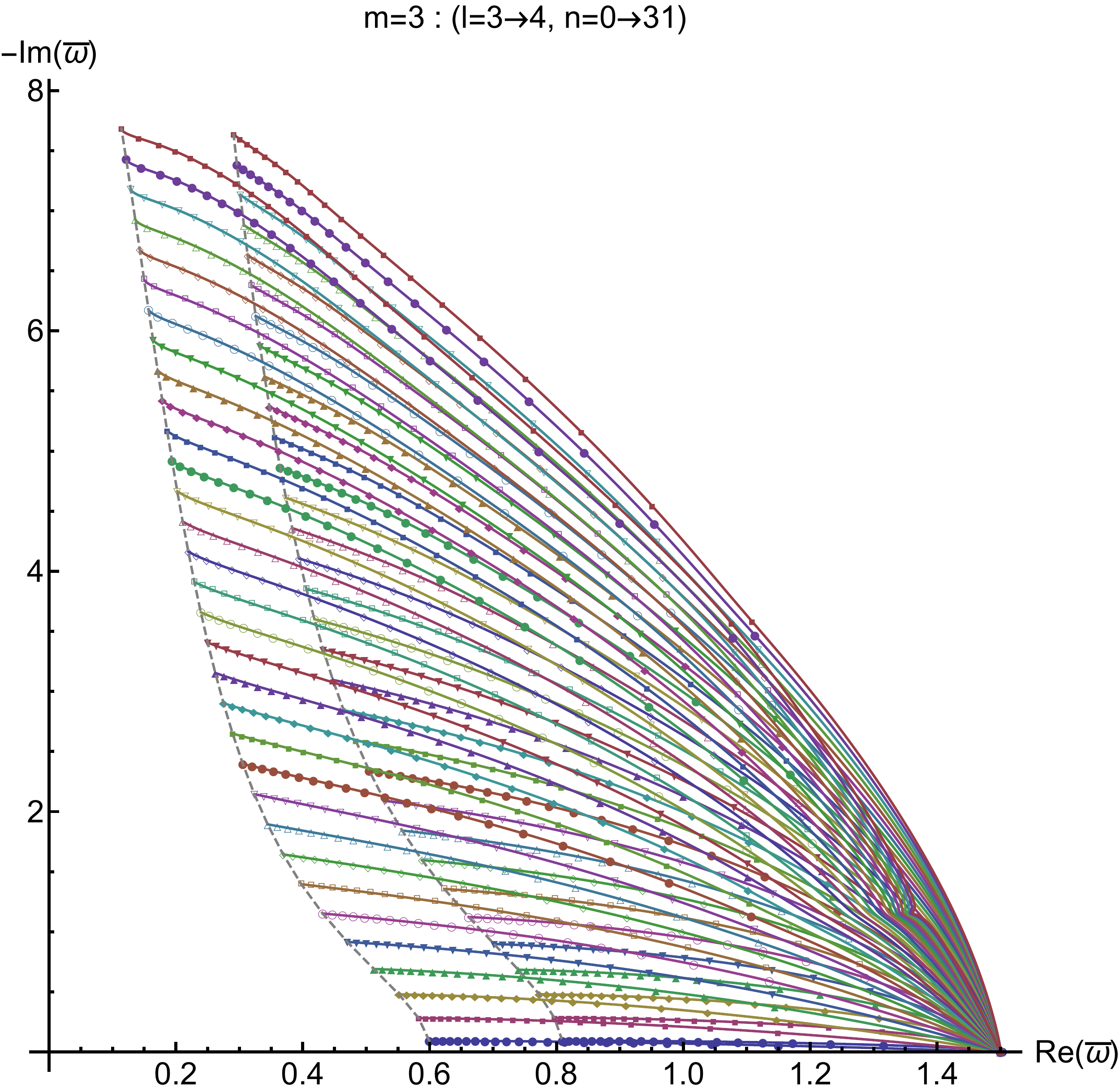}
\caption{\label{fig:mp3l3-4n00-31} Kerr QNM mode sequences for $m=3$.
  See Fig.~\ref{fig:mn4l4n00-31} for a full description.  In this
  case, we plot the $\ell=3$ and $\ell=4$ sequences.  Sequences with
  lower $\ell$ are generally to the left of sequences with higher
  $\ell$.  Note that as $\bar{a}\rightarrow1$ the sequences approach
  an accumulation point at $\bar\omega=m/2$.}
\end{figure}

Figures~\ref{fig:mn3l3-4n00-31} and \ref{fig:mp3l3-4n00-31} display
sequences for $m=-3$ and $m=3$ respectively.  Each figure shows both
$\ell=3$ and $4$, with overtones at the Schwarzschild limit for 
sequence at each $\ell$ connected by a dashed gray line.  At the
Schwarzschild limit, modes with larger values of $\ell$ generally
have larger values of ${\rm Re}(\bar\omega)$.

\begin{figure}
\includegraphics[width=\linewidth,clip]{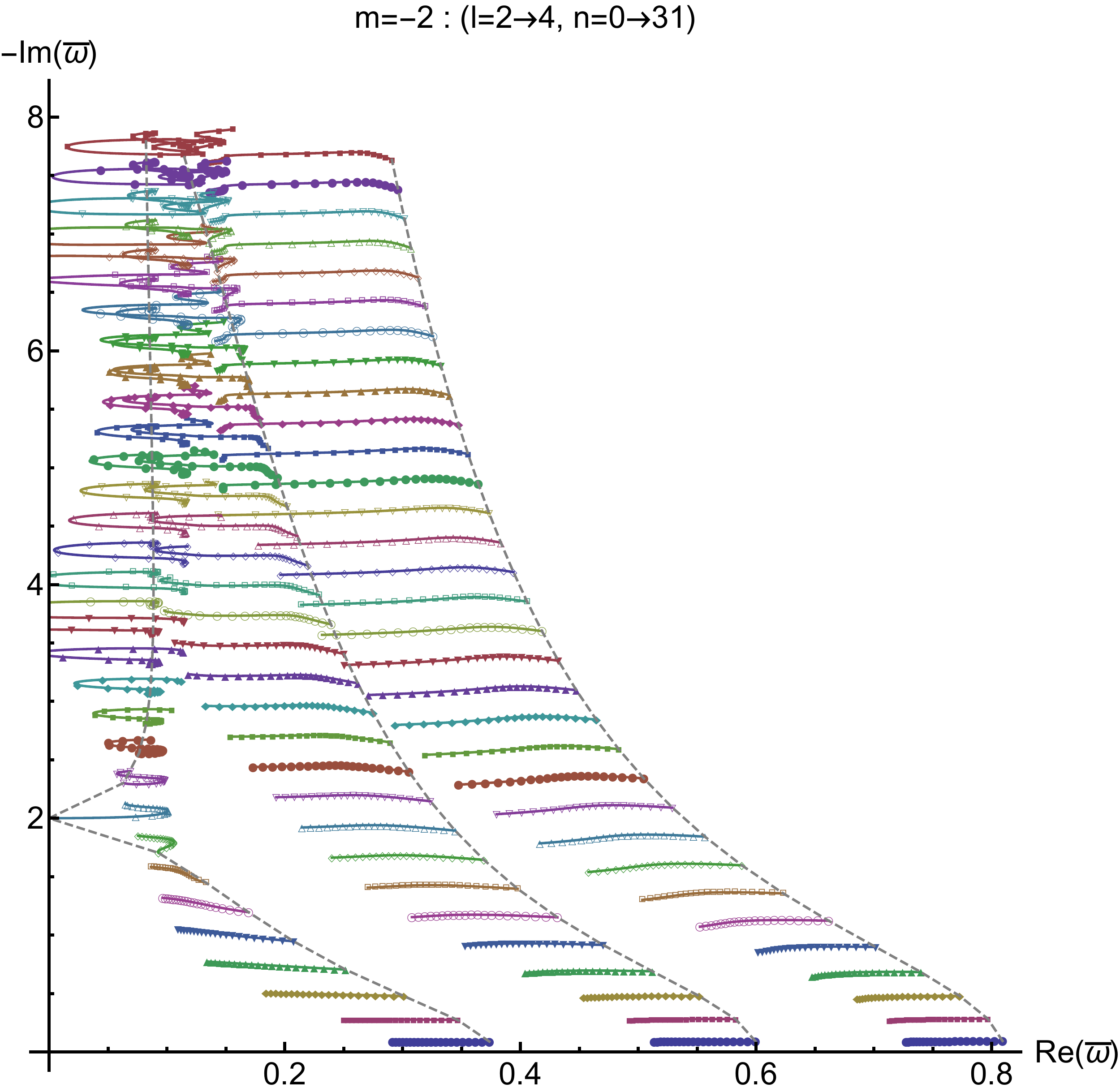}
\caption{\label{fig:mn2l2-4n00-31} Kerr QNM mode sequences for $m=-2$.
  See Fig.~\ref{fig:mn4l4n00-31} for a full description.  In this
  case, we plot the $\ell=2$, $\ell=3$ and $\ell=4$ sequences.  Note
that several of the $\ell=2$ sequences terminate at and re-emerge
from the NIA near $\bar\omega\sim4$, and several of the $\ell=3$ sequences
do the same near $\bar\omega\sim7$.  This behavior will be examined in
more detail in Figs.~\ref{fig:mn2l2n11-19} and \ref{fig:mn2l3n24-31}.}
\end{figure}

\begin{figure}
\includegraphics[width=\linewidth,clip]{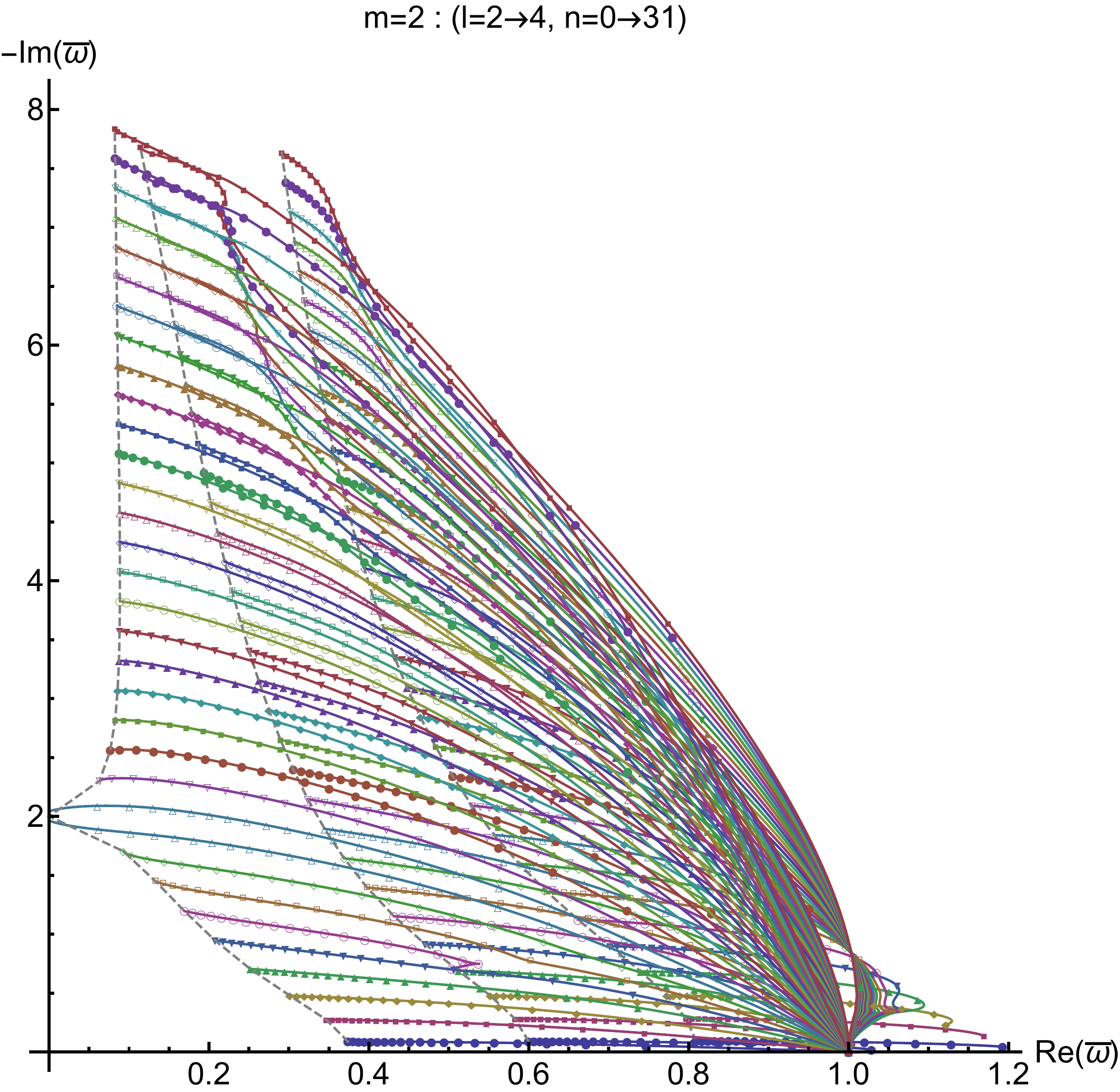}
\caption{\label{fig:mp2l2-4n00-31} Kerr QNM mode sequences for $m=2$.
  See Fig.~\ref{fig:mn4l4n00-31} for a full description.  In this
  case, we plot the $\ell=2$, $\ell=3$ and $\ell=4$ sequences.  Note
  that as $\bar{a}\rightarrow1$ many of the sequences approach an
  accumulation point at $\bar\omega=m/2$.}
\end{figure}

\begin{figure}
\includegraphics[width=\linewidth,clip]{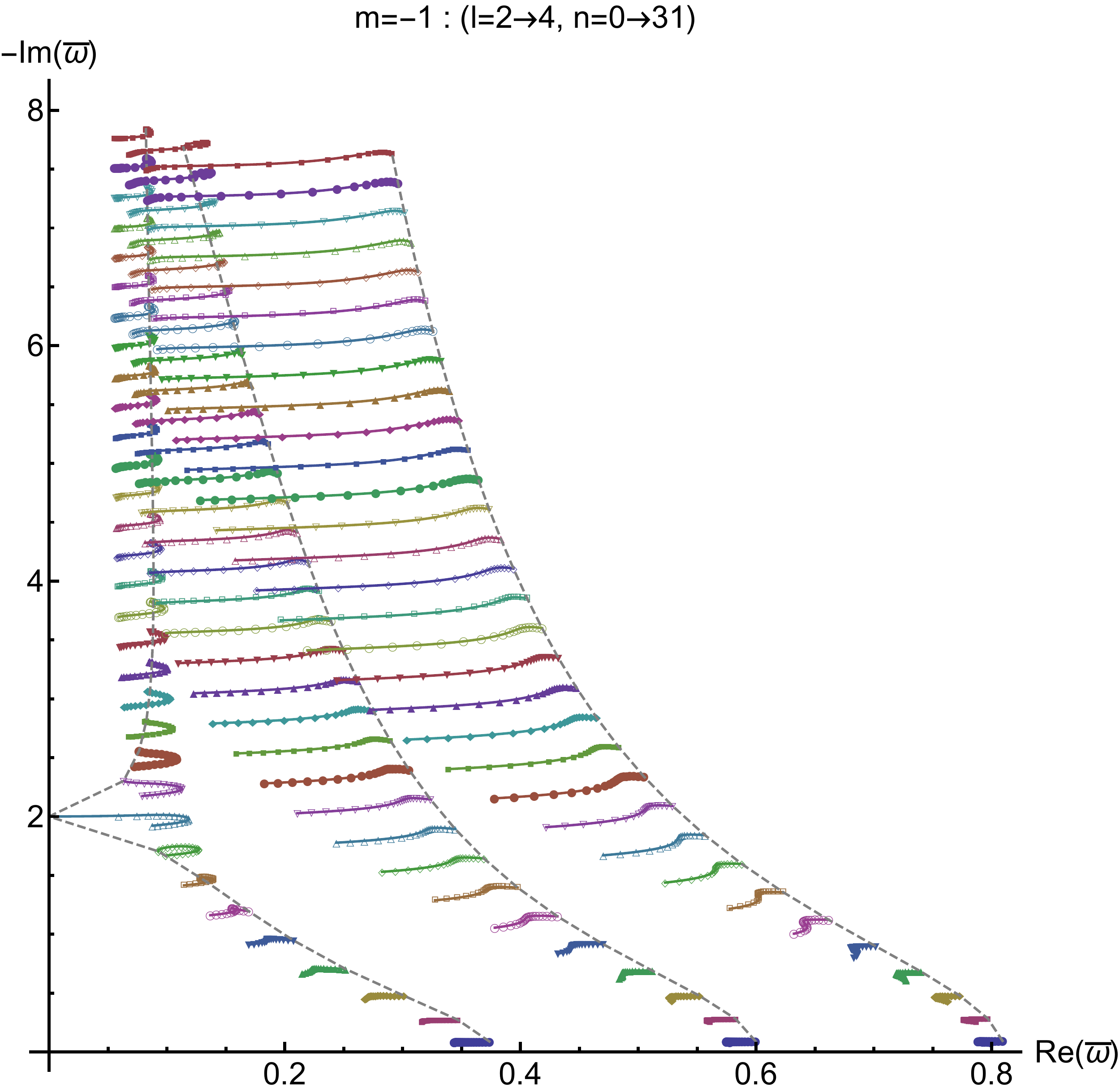}
\caption{\label{fig:mn1l2-4n00-31} Kerr QNM mode sequences for $m=-1$.
  See Fig.~\ref{fig:mn4l4n00-31} for a full description.  In this
  case, we plot the $\ell=2$, $\ell=3$ and $\ell=4$ sequences.}
\end{figure}

\begin{figure}
\includegraphics[width=\linewidth,clip]{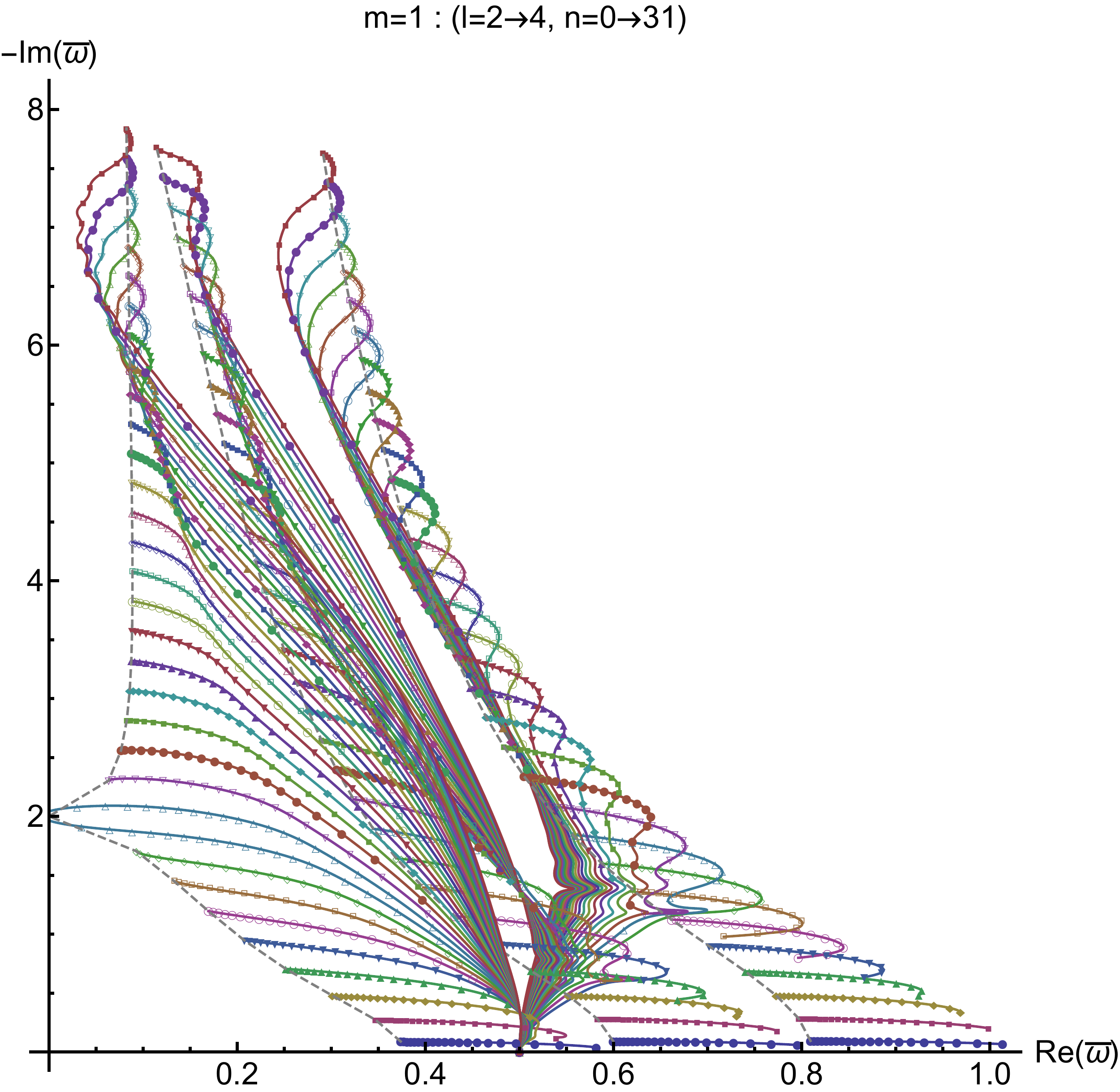}
\caption{\label{fig:mp1l2-4n00-31} Kerr QNM mode sequences for $m=1$.
  See Fig.~\ref{fig:mn4l4n00-31} for a full description.  In this
  case, we plot the $\ell=2$, $\ell=3$ and $\ell=4$ sequences.  Note
  that as $\bar{a}\rightarrow1$ many of the sequences approach an
  accumulation point at $\bar\omega=m/2$.}
\end{figure}

Figures~\ref{fig:mn2l2-4n00-31} and \ref{fig:mp2l2-4n00-31} display
sequences for $m=-2$ and $m=2$, while Figs.~\ref{fig:mn1l2-4n00-31}
and \ref{fig:mp1l2-4n00-31} display sequences for $m=-1$ and $m=1$.
In each of these figures, sequences with $\ell=2$, $3$, and $4$ are
displayed.  For $\ell=2$, the modes at the Schwarzschild limit (again
connected by a gray dashed line), show a new feature.  At $n=8$ the
mode becomes purely imaginary with $\bar\omega=\bar\Omega_2=-2i$.  For
$m\le0$, the $n=8$ sequences terminate at this frequency in the
Schwarzschild limit.  However, for $m>0$ (see
Figs.~\ref{fig:mp2l2-4n00-31} and \ref{fig:mp1l2-4n00-31}) {\em two}
sequences in each figure approach the NIA near
$\bar\omega=\bar\Omega_2$, and these sequences begin at
non-vanishing values of $\bar{a}$.  These sequences are all clearly
associated with the $n=8$ overtone.  Distinct sequences that should be
labeled by the same values for $\{\ell,m,n\}$ are referred to as
``overtone multiplets'' and are distinguished by a subscript on the
overtone index ($n=8_0$ and $8_1$ in this case).  We will discuss
these sequences further below.


\begin{figure}
\includegraphics[width=\linewidth,clip]{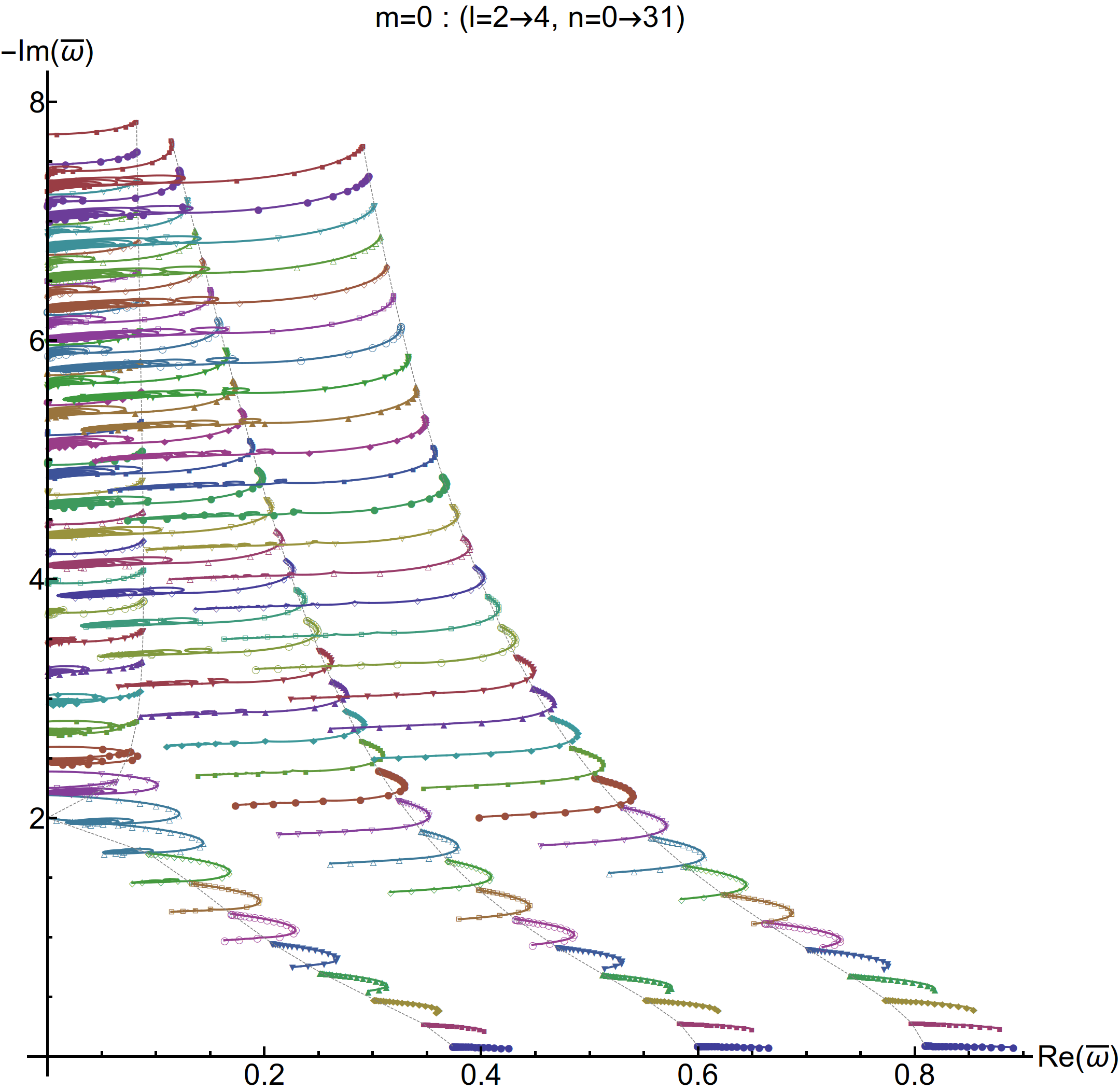}
\caption{\label{fig:m0l2-4n00-31} Overview of Kerr QNM mode sequences
  for $m=0$.  See Fig.~\ref{fig:mn4l4n00-31} for a full description.
  In this case, we plot the $\ell=2$, $\ell=3$ and $\ell=4$ sequences.
  The complex behavior near the NIA will be examined in more detail in
  Figs.~\ref{fig:m0l2-4n08-15}--\ref{fig:m0l4n25-31}}
\end{figure}

Finally, Fig.~\ref{fig:m0l2-4n00-31} displays an overview of the
sequences for $m=0$.  It too shows sequences with $\ell=2$, $3$, and
$4$.  This is a very dense figure, especially near the NIA, and we
will consider it in more manageable pieces below, but we see again the
same dashed gray line connecting modes at the Schwarzschild limit for
$\ell=2$ which becomes purely imaginary for $n=8$.

\subsection{Modes with $\ell=2$ and $n=8$}
\label{sec:l2n8modes}
\begin{figure}
\includegraphics[width=\linewidth,clip]{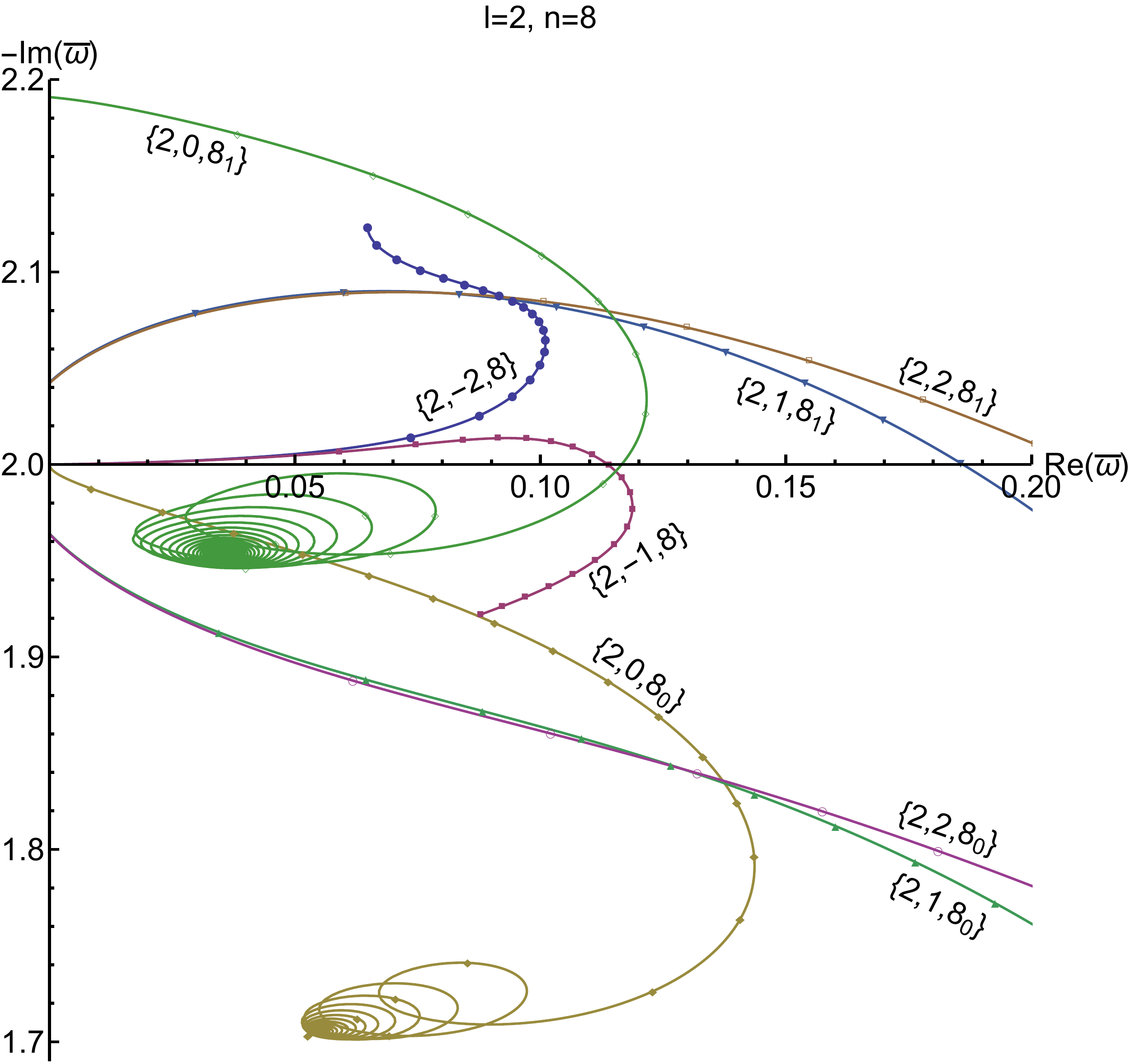}
\caption{\label{fig:malll2n8} Detail view near the NIA of Kerr QNM
  mode sequences for $\ell=2$ and $n=8$.  See
  Fig.~\ref{fig:mn4l4n00-31} for additional description.  Each
  sequence is labeled by it values for $\{\ell,m,n\}$.  Note that the
  $m=0$, $1$, and $2$ sequences are overtone multiplets as described
  in the text.  We will show in Sec.~\ref{sec:modes_on_NIA}, that the
  4 sequences with $m=1$ and $2$ must terminate without a mode on the
  NIA.  The $\{2,0,8_1\}$ emerges from the NIA at is simultaneously a
  QNM and a TTM${}_L$ with a non-vanishing value of $\bar{a}$.  The
  remaining 3 sequences branch out from $\bar\omega=-2i$ which is also
  simultaneously a QNM and a TTM${}_L$ with $\bar{a}=0$.}
\end{figure}

QNMs with $\ell=2$ and $n=8$ were the first computed gravitational
modes with frequencies seen to approach the
NIA\cite{leaver-1985,onozawa-1997,berticardoso-2003}.  All of the
gravitational QNM sequences for this case are plotted in
Fig.~\ref{fig:malll2n8}.  Whether or not these sequences could extend
to include modes with frequencies precisely on the NIA is a question
which remained controversial and poorly understood until it was
rigorously resolved by Maassen van den Brink\cite{van_den_brink-2000}
for modes in the Schwarzschild limit.  None-the-less, subsequent works
(including our own) did not fully embrace his
findings\cite{berticardoso-2003,Fiziev-2011,cook-zalutskiy-2014}.
However, upon gaining a fuller understanding of his approach, we now
agree with Maassen van den Brink's findings.  Regarding the existence
of QNMs with frequencies precisely on the NIA, he finds that the
algebraically special modes with frequencies
$\bar\omega=\bar\Omega_\ell$, where
\begin{equation}\label{eq:alg-spec-sch}
  M\Omega_\ell=\bar\Omega_\ell \equiv -\frac{i}{12}(\ell-1)\ell(\ell+1)(\ell+2),
\end{equation}
{\em are} QNMs.  More precisely, the $s=-2$ modes are simultaneously
QNMs and TTM${}_L$s, while there is no mode in the Schwarzschild limit
for $s=2$.  See Ref.\cite{van_den_brink-2000} and \cite{berti-QNM-2009}
for further details and comments.  

Furthermore, Maassen van den Brink\cite{van_den_brink-2000} finds that
a set of sequences of modes approach $\bar\Omega_\ell$ as
$\bar{a}\rightarrow0$.  For modes with ${\rm Re}(\bar\omega)\ge0$,
this set of sequences includes the modes with $m\le0$.  In
Fig.~\ref{fig:malll2n8}, where $\ell=2$, we see that the $\{2,-2,8\}$,
$\{2,-1,8\}$, and $\{2,0,8_0\}$ sequences appear to agree with this
predicted behavior.  In fact we have shown\cite{cook-zalutskiy-2014}
that they agree quantitatively, to high accuracy, with Maassen van den
Brink's predictions.  For $m>0$, he suggested several possible
behaviors, none of which precisely agree with what is seen
numerically\cite{van_den_brink-2000,van_den_brink-2003,berticardoso-2003,cook-zalutskiy-2014}.
As seen in Fig.~\ref{fig:malll2n8}, the $m=1$ and $2$ sequences exist
as overtone multiplets.  The $\{2,1,8_0\}$ and $\{2,2,8_0\}$ sequences
approach the NIA at a point slightly below $\bar\Omega_2$, while the
$\{2,1,8_1\}$ and $\{2,2,8_1\}$ sequences approach the NIA at a point
slightly above $\bar\Omega_2$.  This behavior was first seen in
Ref.\cite{berticardoso-2003} and confirmed in
Ref.\cite{cook-zalutskiy-2014} where the sequences were extended to
the neighborhood of the NIA with $\bar{a}$ becoming small, but
remaining finite.  See Table~\ref{tab:NIAnoQNM} for numerical values
of $\bar\omega$ and $\bar{a}$ adjacent to the NIA.

Since that work, we have found that the $m=0$ sequence has an overtone
multiplet partner.  Labeled as $\{2,0,8_1\}$ in
Fig.~\ref{fig:malll2n8}, we see that it has a spiraling shape similar
to that of its partner $\{2,0,8_0\}$.  However, similar to the
$\{2,1,8_{0,1}\}$ and $\{2,2,8_{0,1}\}$ multiplets, this sequence
approaches the NIA at a point displaced from $\bar\Omega_2$ and starts
at a non-zero value of $\bar{a}$.  See Table~\ref{tab:NIAdata2QNM} for
numerical values of $\bar\omega$ and $\bar{a}$ adjacent to
the NIA.

\subsection{Other modes approaching the NIA}
\label{sec:num_modes_on_NIA}

\begin{figure}
\includegraphics[width=\linewidth,clip]{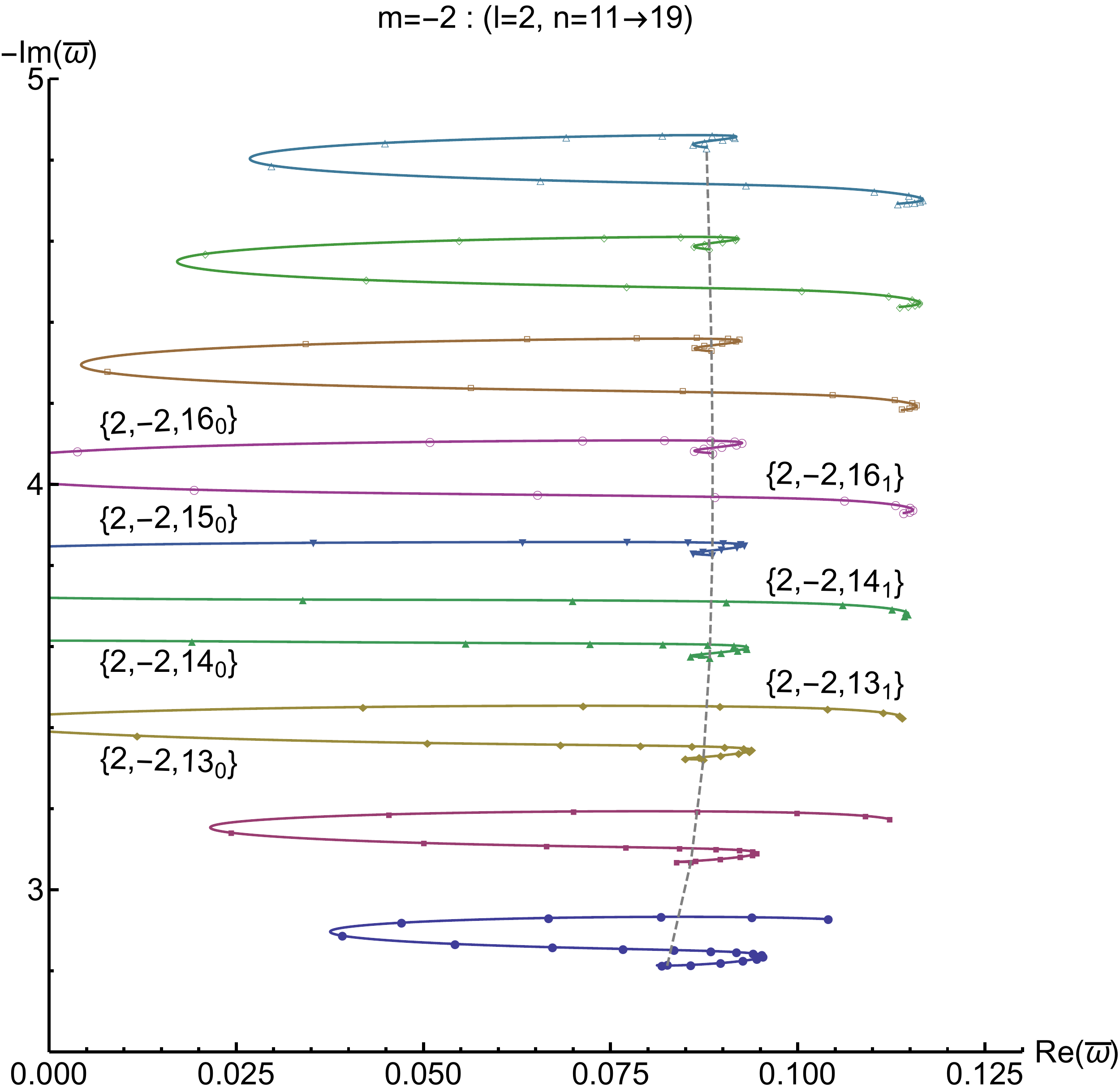}
\caption{\label{fig:mn2l2n11-19} Detail view near the NIA of Kerr QNM
  mode sequences for $\ell=2$, $m=-2$ and $11\le n\le19$.  See
  Fig.~\ref{fig:mn4l4n00-31} for additional description.  Each
  sequence is labeled by $\{\ell,m,n\}$ is part of an overtone
  multiplet.  Sequences with a $n_0$ overtone terminate at the NIA,
  while those with a $n_1$ overtone re-emerge from the NIA.  We will
  show in Sec.~\ref{sec:modes_on_NIA}, that none of these sequences
  have a mode precisely on the NIA.}
\end{figure}

\begin{figure}
\includegraphics[width=\linewidth,clip]{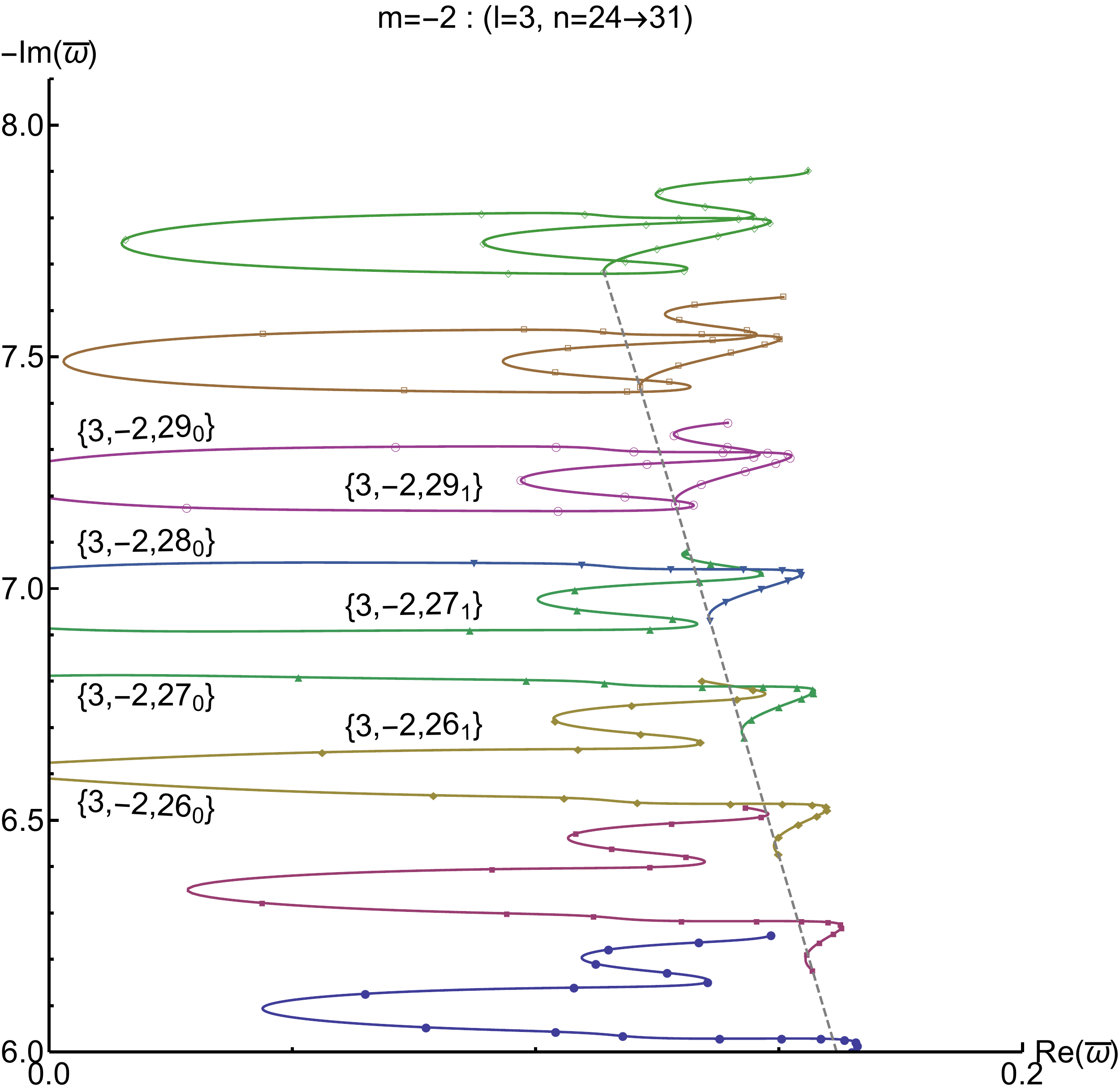}
\caption{\label{fig:mn2l3n24-31} Detail view near the NIA of Kerr QNM
  mode sequences for $\ell=3$, $m=-2$ and $24\le n\le31$.  See
  Fig.~\ref{fig:mn4l4n00-31} for additional description.  Each
  sequence is labeled by $\{\ell,m,n\}$ is part of an overtone
  multiplet.  Sequences with a $n_0$ overtone terminate at the NIA,
  while those with a $n_1$ overtone re-emerge from the NIA.  We will
  show in Sec.~\ref{sec:modes_on_NIA}, that none of these sequences
  have a mode precisely on the NIA.}
\end{figure}

As we computed solutions at larger values of $n$, we encountered many
other instances where sequences either approached or emerged from the
NIA.  In Fig.~\ref{fig:mn2l2-4n00-31}, we see several examples where
$m=-2$ modes exist in the neighborhood of the NIA.  For the case of
$\ell=2$ and $m=-2$, this behavior is seen more clearly in
Fig.~\ref{fig:mn2l2n11-19}.  Most of these sequences begin at the
Schwarzschild limit, $\bar{a}=0$, and smoothly move toward the
extremal limit of $\bar{a}=1$.  However, for overtones $13\le n\le16$
the sequences encounter the NIA.  With one exception, the sequences
skip a finite range of $\bar{a}$ where no modes are found and then
reemerge from the NIA and continue toward the extremal limit.  We also
consider such discontinuous sequences as overtone multiplets, labeling
the first segment with $n_0$ and the second with $n_1$.  See
Table~\ref{tab:NIAnoQNM} for numerical values of $\bar\omega$ and
$\bar{a}$ adjacent to the NIA.  The exceptional case is that of
$n=15$.  This sequence seems to have no second segment.  However,
examining the general behavior of the sequences in
Fig.~\ref{fig:mn2l2n11-19}, we see that the $\{2,-2,14_1\}$ could just
as easily have been labeled as $\{2,-2,15_1\}$, leaving $n=14$ with no
second segment.  In Fig.~\ref{fig:mn2l3n24-31} and
Table~\ref{tab:NIAnoQNM}, we see similar behavior in the $\ell=3$,
$m=-2$ modes for overtones $26\le n\le29$

\begin{table}
\begin{tabular}{llcl}
\hline\hline
\Chead{Mode} & \Chead{$\bar\omega$} && \Chead{$\bar{a}$} \\
$\{2,1,8_0\}$ & $5.38939\times10^{-10}-1.96407i$ && $0.00688189$ \\
$\{2,1,8_1\}$ & $2.56576\times10^{-10}-2.04259i$ && $0.0108327$ \\
$\{2,2,8_0\}$ & $8.05758\times10^{-10}-1.96384i$ && $0.00348258$ \\
$\{2,2,8_1\}$ & $3.27463\times10^{-8\ }-2.04223i$ && $0.00532788$ \\
\hline
$\{2,-2,13_0\}$ & $1.00256\times10^{-9\ }-3.38997i$ && $0.657472$ \\
$\{2,-2,13_1\}$ & $8.33053\times10^{-10}-3.43236i$ && $0.669473$ \\
$\{2,-2,14_0\}$ & $6.34428\times10^{-10}-3.61439i$ && $0.611751$ \\
$\{2,-2,14_1\}$ & $1.11866\times10^{-9\ }-3.72003i$ && $0.629591$ \\
$\{2,-2,15_0\}$ & $3.82197\times10^{-10}-3.84682i$ && $0.577994$ \\
$\{2,-2,16_0\}$ & $9.20034\times10^{-10}-4.07750i$ && $0.552996$ \\
$\{2,-2,16_1\}$ & $8.31778\times10^{-11}-4.00154i$ && $0.587497$ \\
\hline
$\{3,-2,26_0\}$ & $3.93481\times10^{-9\ }-6.59006i$ && $0.585204$ \\
$\{3,-2,26_1\}$ & $5.75685\times10^{-11}-6.62413i$ && $0.586625$ \\
$\{3,-2,27_0\}$ & $3.28135\times10^{-9\ }-6.81175i$ && $0.563660$ \\
$\{3,-2,27_1\}$ & $5.22956\times10^{-9\ }-6.91389i$ && $0.565234$ \\
$\{3,-2,28_0\}$ & $2.75891\times10^{-9\ }-7.04384i$ && $0.545236$ \\
$\{3,-2,29_0\}$ & $1.55414\times10^{-10}-7.27474i$ && $0.529576$ \\
$\{3,-2,29_1\}$ & $4.35556\times10^{-10}-7.19523i$ && $0.543328$ \\
\hline\hline
\end{tabular}
\caption{\label{tab:NIAnoQNM} Numerical solution for QNMs closest to
  NIA at beginning or end of selected mode sequences.  The first 4
  entries correspond to sequences in Fig.~\ref{fig:malll2n8}.  The
  next 7 entries correspond to sequences in
  Fig.~\ref{fig:mn2l2n11-19}.  The final 7 entries correspond to
  sequences in Fig.~\ref{fig:mn2l3n24-31}.  We will show in
  Sec.~\ref{sec:modes_on_NIA} that none of these sequences can have a
  mode precisely on the NIA.  However, our numerical results suggest
  that we can get arbitrarily close.}
\end{table}

\begin{figure}
\includegraphics[width=\linewidth,clip]{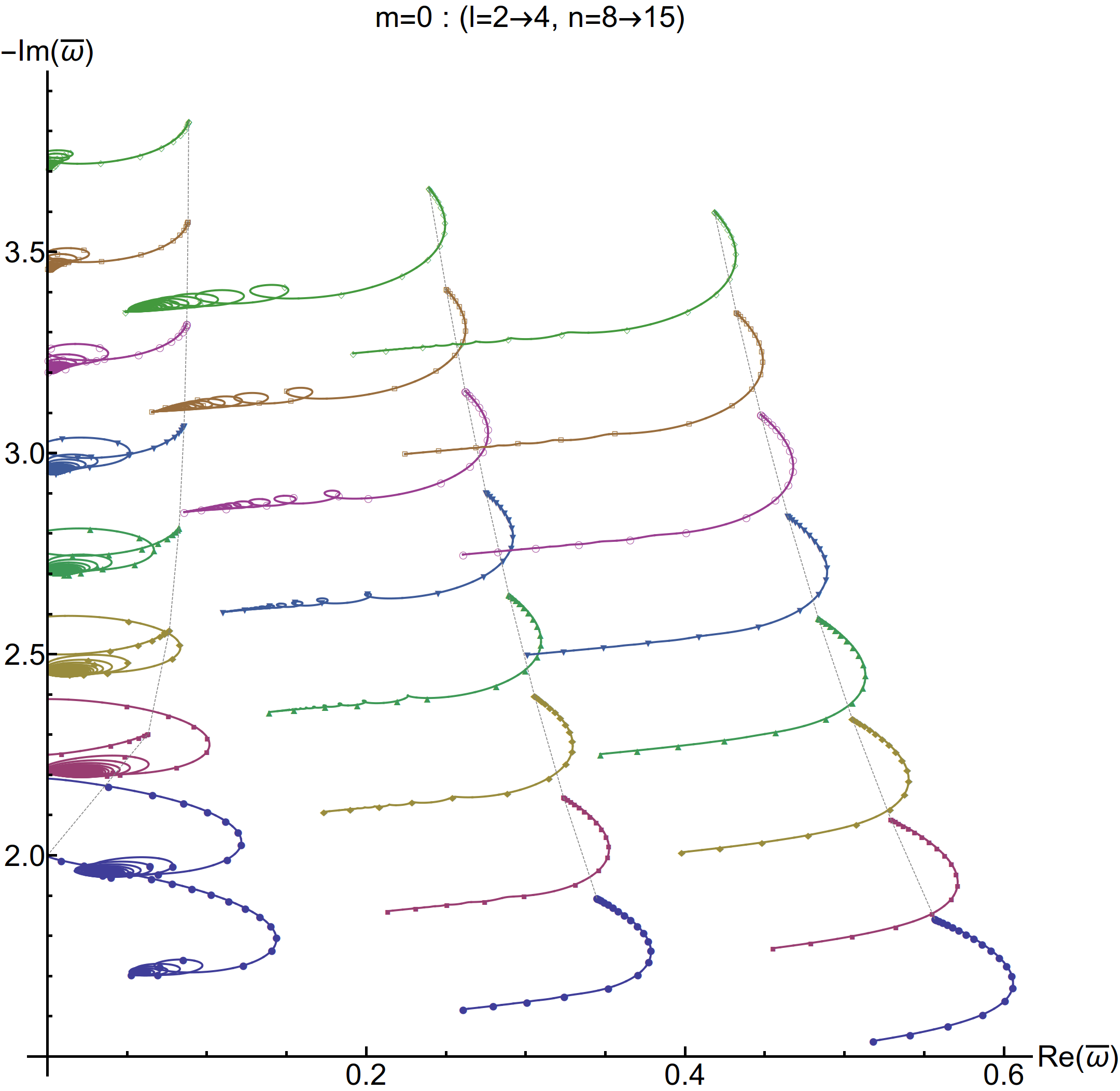}
\caption{\label{fig:m0l2-4n08-15} Detail view near the NIA of Kerr QNM
  mode sequences for $m=0$ and $8\le n\le15$.  See
  Fig.~\ref{fig:mn4l4n00-31} for additional description.  Sequences
  for $\ell=2$, $3$, and $4$ are shown.  All of the $\ell=2$ sequences
  are overtone multiplets.  The $\ell=2$ sequences also show looping
  behavior where many of the loops have numerous points of tangency
  with the NIA.}
\end{figure}

As seen in Fig.~\ref{fig:m0l2-4n00-31}, the vast majority of sequences
that approach the NIA have $m=0$.  Figure~\ref{fig:m0l2-4n08-15} shows
an expanded view containing only $8\le n\le15$.  For $\ell=2$, the
first two spiraling sequences near the bottom of the plot are
$\{2,0,8_0\}$ and $\{2,0,8_1\}$ discusses already in
Sec.~\ref{sec:l2n8modes}.  For $n=9$, we see a new behavior in the
$\ell=2$, $m=0$ sequences.  Starting at the Schwarzschild limit, the
$n=9$ sequence moves toward and terminates at the NIA at
$\bar\omega=-2.25i$ and a finite value of $\bar{a}$.  Following a
short interval in $\bar{a}$ with no modes, the sequence re-emerges
near $\bar\omega=-2.39i$ and loops back towards the NIA as $\bar{a}$
increases.  However, instead of terminating again at the NIA, the
sequence appears to touch the NIA at a point of tangency, continuing
to loop around again.  The upper right plot in
Fig.~\ref{fig:m0l2n09-26} shows the behavior of this mode in
isolation.  This sequence continues through a total of $7$ points of
tangency before subsequent loops pull away from the NIA.
Table~\ref{tab:NIAdata2QNM} lists values for $\bar\omega$ and
$\bar{a}$ at which the sequence terminates or emerges from the NIA,
while Table~\ref{tab:NIAdataLoops} lists these values for the $7$
points where the sequence becomes tangent to the NIA.  The data points
listed in this table are the particular numerical solutions in the
sequence that are closest to the NIA for each loop.  Numerical
solutions using Leaver's method\cite{leaver-1985,cook-zalutskiy-2014}
cannot be obtained precisely on the NIA.  However, quadratic
interpolation confirms the points of tangency to very high precision.

\begin{figure}
\includegraphics[width=\linewidth,clip]{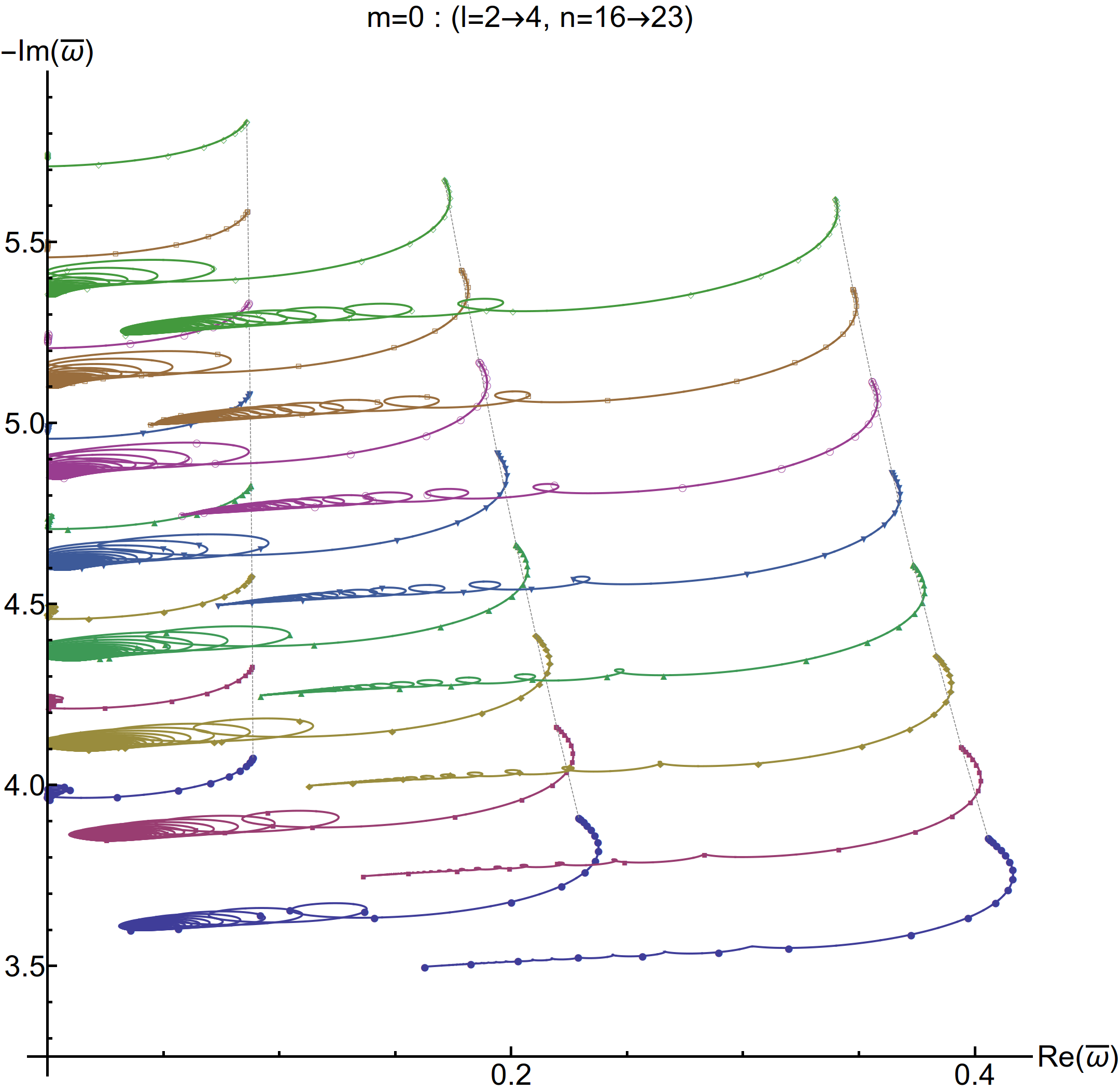}
\caption{\label{fig:m0l2-4n16-23} Detail view near the NIA of Kerr QNM
  mode sequences for $m=0$ and $16\le n\le23$.  See
  Fig.~\ref{fig:mn4l4n00-31} for additional description.  Sequences
  for $\ell=2$, $3$, and $4$ are shown.  All of the $\ell=2$ sequences
  are overtone multiplets.  Many of the $\ell=2$ and $3$ sequences
  show looping behavior where many of the loops have numerous
  points of tangency with the NIA.}
\end{figure}

\begin{figure}
\includegraphics[width=\linewidth,clip]{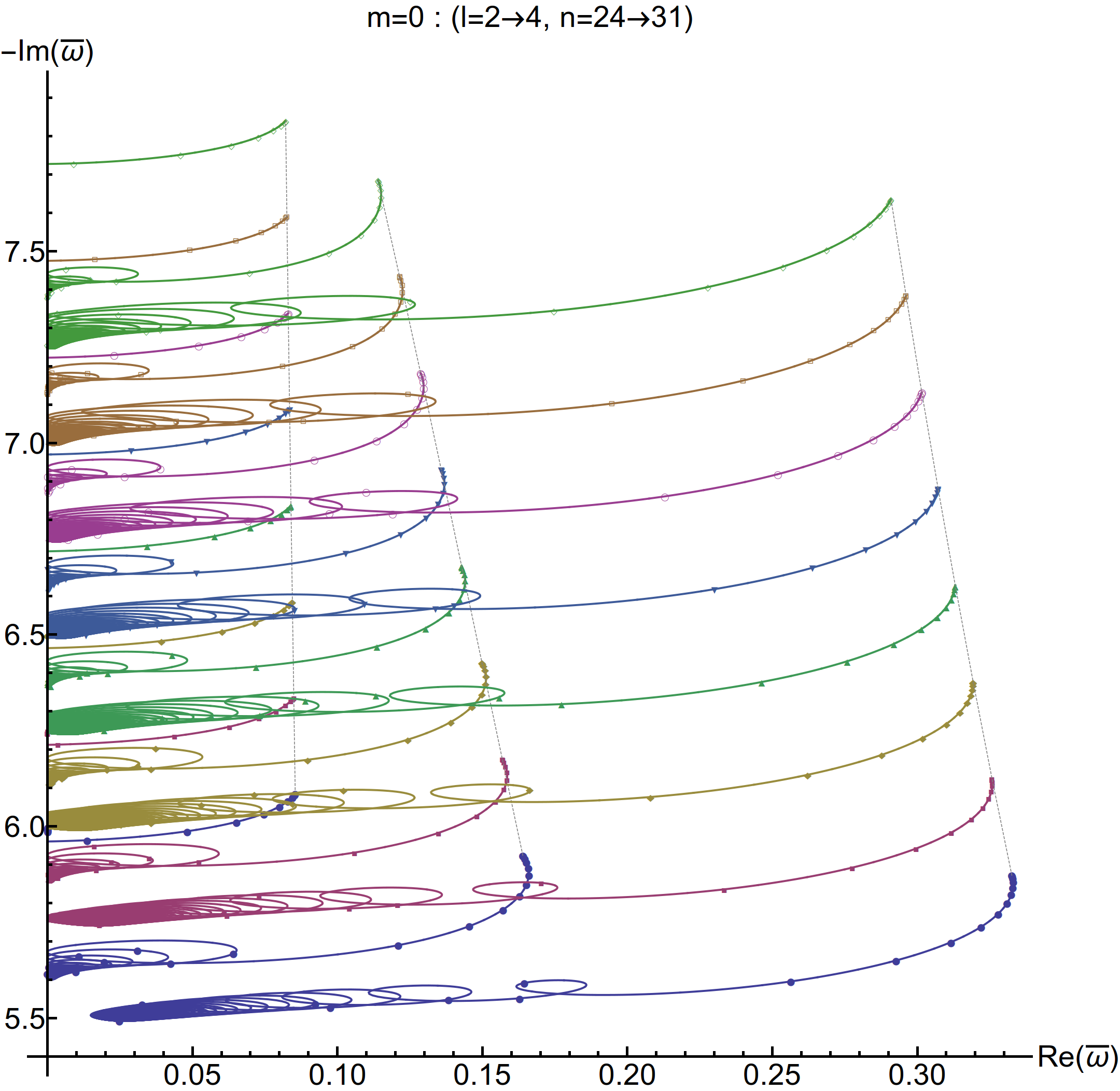}
\caption{\label{fig:m0l2-4n24-31} Detail view near the NIA of Kerr QNM
  mode sequences for $m=0$ and $24\le n\le31$.  See
  Fig.~\ref{fig:mn4l4n00-31} for additional description.  Sequences
  for $\ell=2$, $3$, and $4$ are shown.  Only the $n\le26$, $\ell=2$
  sequences are overtone multiplets.  Many of the sequences show
  looping behavior where many of the loops have numerous points of
  tangency with the NIA.}
\end{figure}

\begin{figure}
\begin{tabular}{cc}
\includegraphics[width=0.5\linewidth,clip]{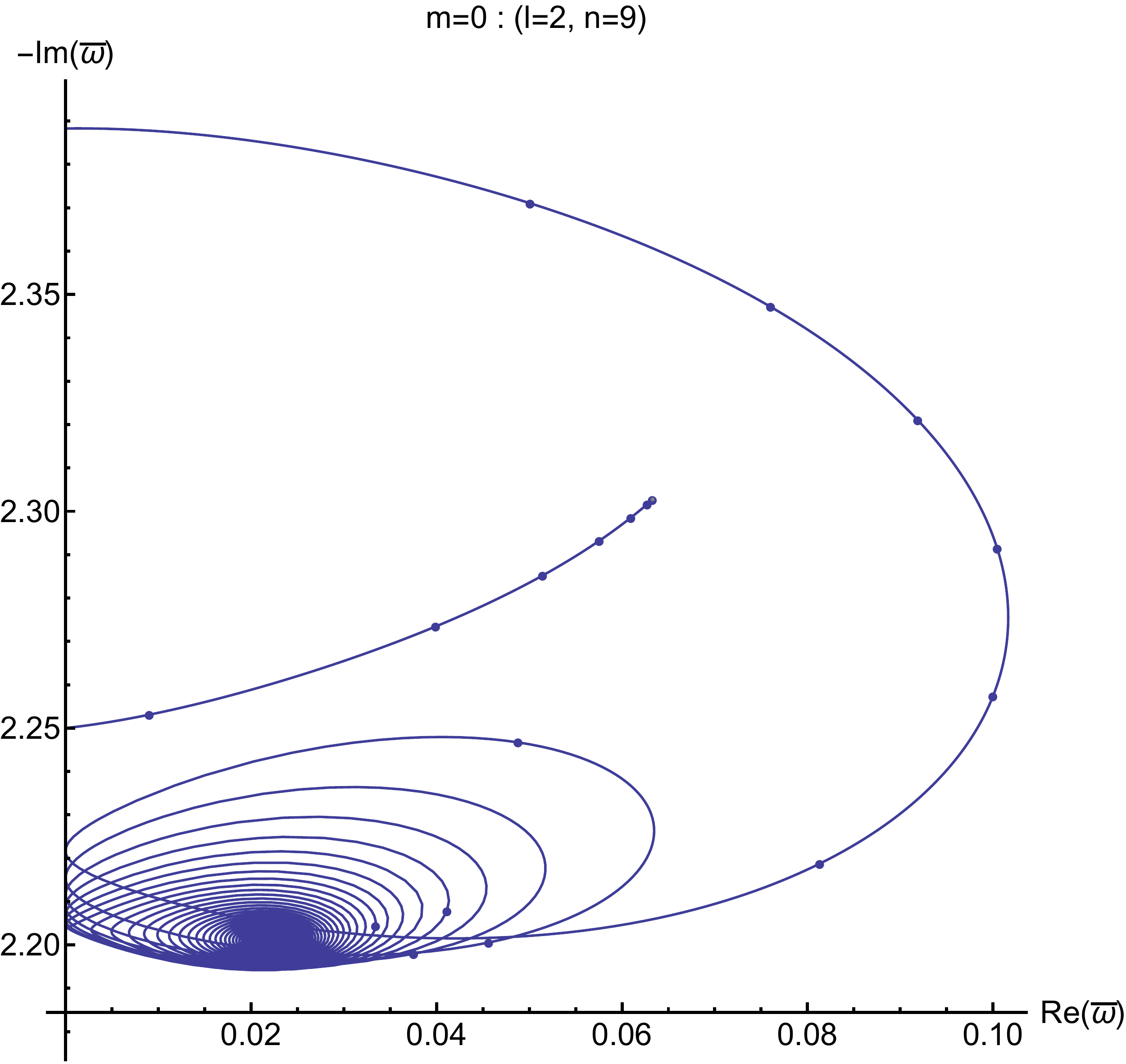} &
\includegraphics[width=0.5\linewidth,clip]{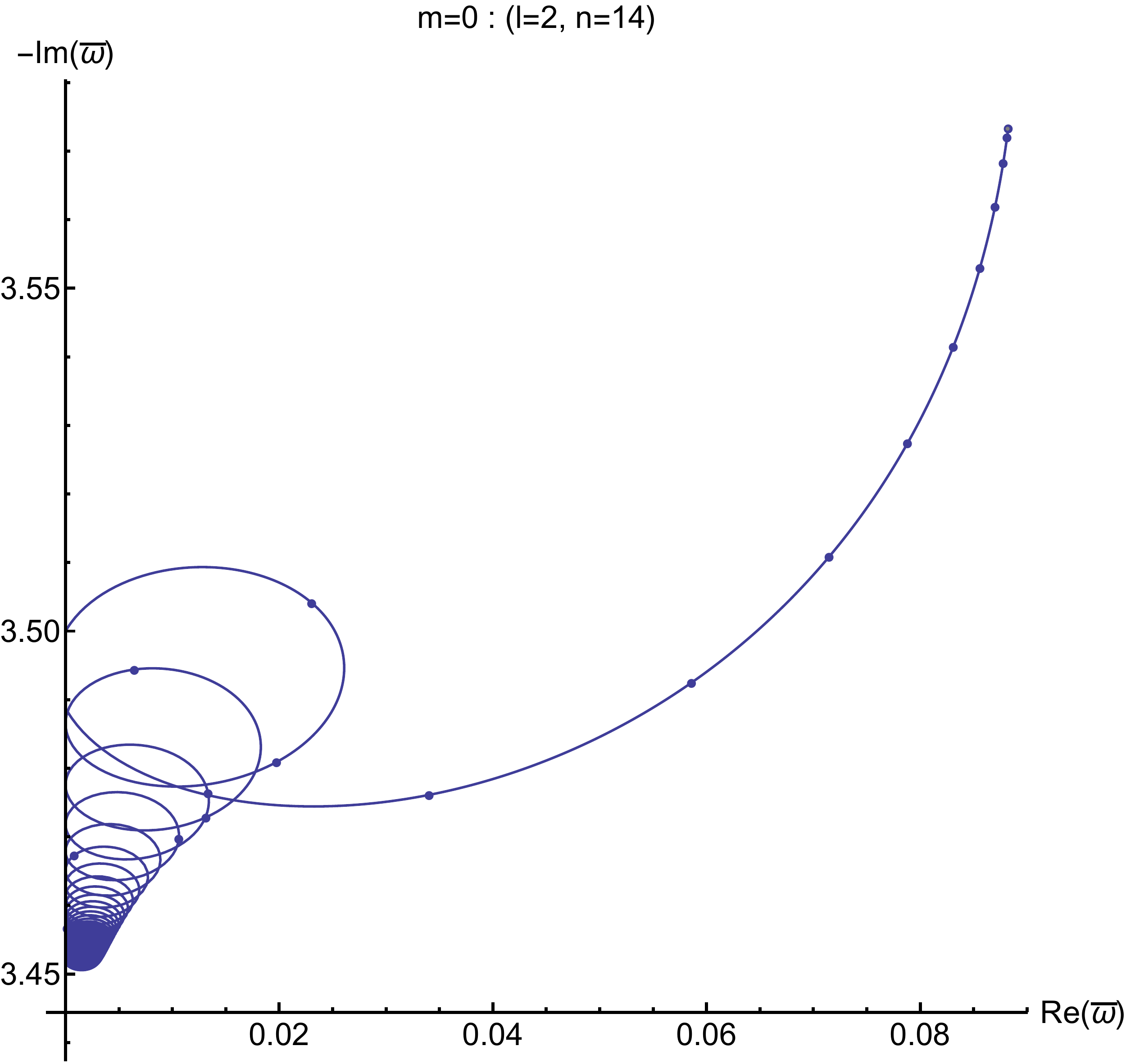} \\
\includegraphics[width=0.5\linewidth,clip]{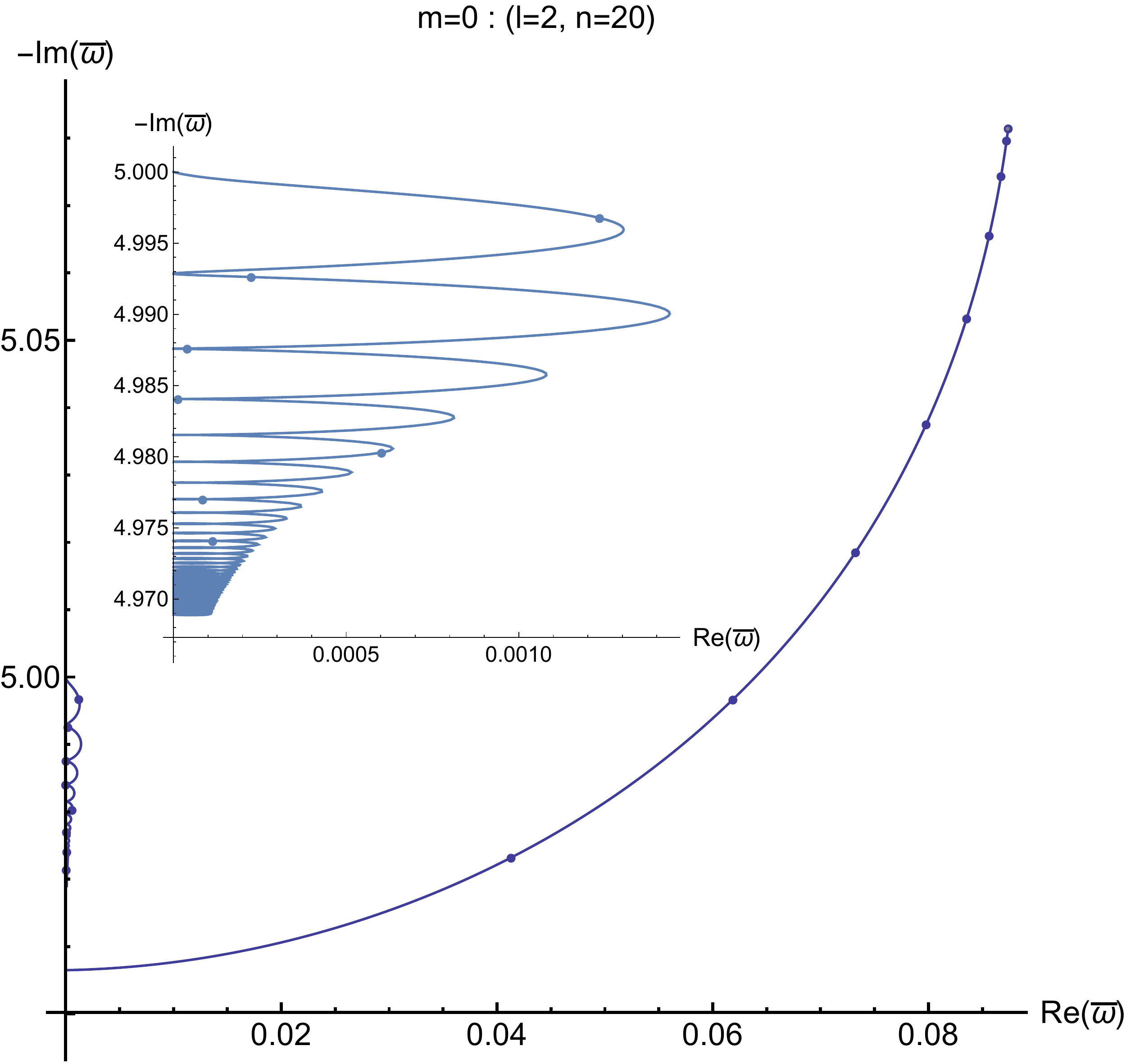} &
\includegraphics[width=0.5\linewidth,clip]{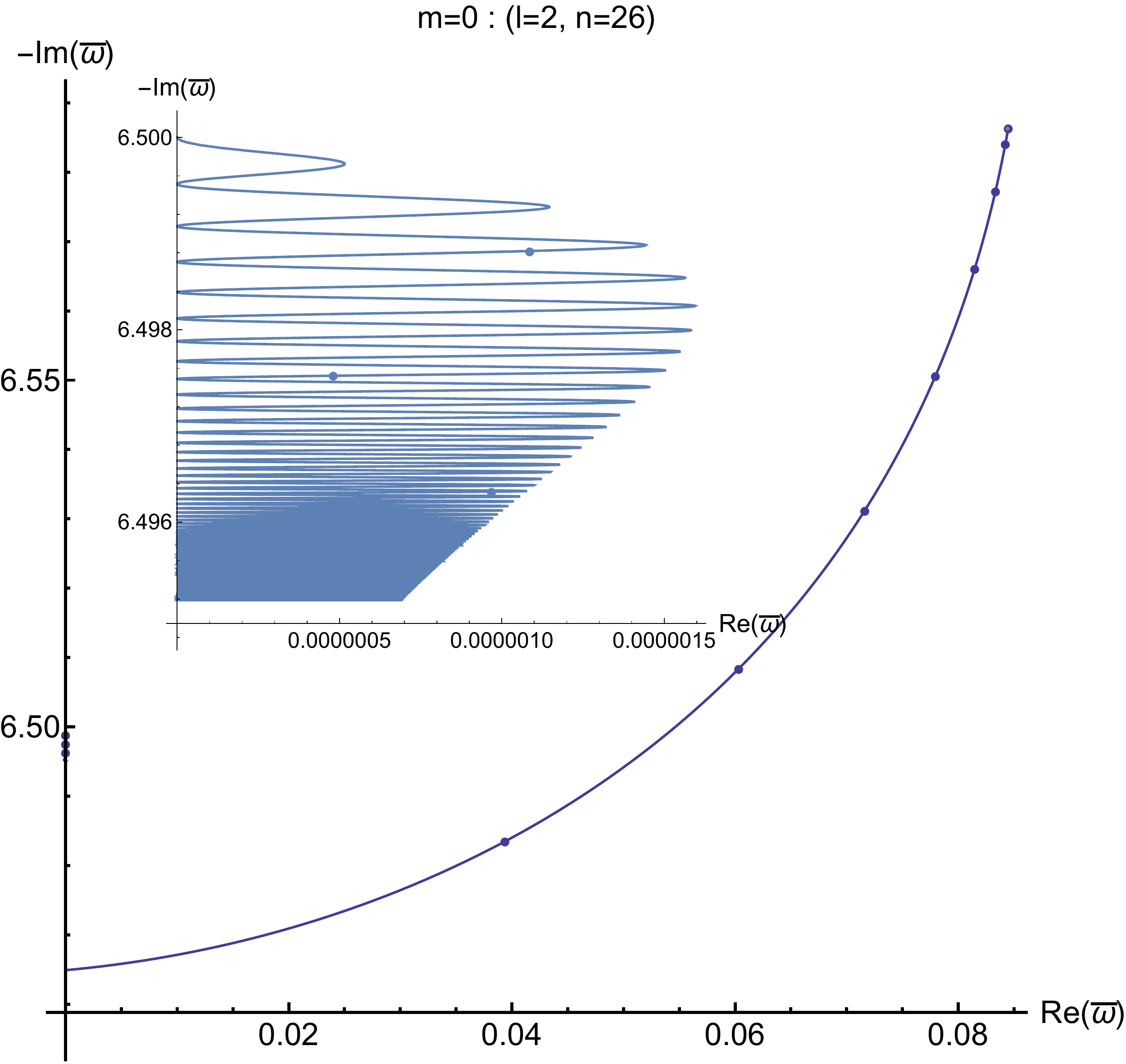}
\end{tabular}
\caption{\label{fig:m0l2n09-26} Examples of $\ell=2$, $m=0$ modes that
  appear to touch the NIA.  Shown are the $n=9$, $14$, $20$, and $26$
  sequences.  Each has a sequence segment that begins at the
  Schwarzschild limit and appears to terminate on the NIA.  Each
  sequence re-emerges from the NIA and then repeatedly loops back with
  many points of tangency with the NIA.  The lower two panels include
  insets that show an enlarged view of the second segment.}
\end{figure}

For the $\ell=2$, $m=0$ sequences, a similar behavior is seen for all
the sequences with $9\le n\le26$.  All of these sequences are
considered overtone multiplets.  Figures~\ref{fig:m0l2-4n16-23} and
\ref{fig:m0l2-4n24-31} respectively show expanded views containing
$16\le n\le23$ and $24\le n\le31$.  The $n_0$ segment begins at the
Schwarzschild limit and terminates at the NIA.  The $n_1$ segments
re-emerges from the NIA and performs numerous loops, many touching the
NIA at points of tangency.  As $n$ increases, we see that the distance
that the loops range away from the NIA decreases.  In fact for
$n\agt18$, the $n_1$ segment is nearly indistinguishable from the NIA.
The lower two plot in Fig.~\ref{fig:m0l2n09-26} show inset plots
giving expanded views of the $20_1$ and $26_1$ segments.

There are several interesting things to note about the $n_1$ segments
of these sequences.  First, as $n$ increases, the number of loops that
contain a point of tangency with the NIA increases rapidly.
Table~\ref{tab:NIAdataLoopsCount} shows the number of points of tangency
for each sequence for $9\le n\le17$.  For $18\le n\le26$ we have not
yet extended the sequence far enough to find the last loop to touch
the NIA.  For each sequence, we also list the values of $\bar\omega$
and $\bar{a}$ for the interpolated point of tangency of the last loop
that touches the NIA (or the last we have computed).  Note that for
$n\ge17$, the number of loops exceeds $2000$ for each sequence.
Using adaptive sequencing, we have fully resolved every loop with
sufficient accuracy and precision that we can locate each point
of tangency to better than $1\%$ of the spacing $\Delta\bar\omega$
between adjacent points of tangency.

For $\ell=2$ and $n>26$, the behavior of the sequences changes.  While
each sequence begins at the Schwarzschild limit and terminates at the
NIA, we no longer find a second, looping segment for these sequences.
We have carried out extensive searches for a second segment for $n=27$
with no evidence for any modes with $0.4<\bar{a}<1$.  While failing to
find such modes does not prove they do not exist, we are confident in
the ability of our numerical methods to find them were they to exist.
Furthermore, as we will discuss in Sec.~\ref{sec:Heun_polynomials}, we
have additional reasons to believe that sequences with $27\le n\le73$
have no second segment.  This has to do with an additional curious
behavior seen in the value of $\bar\omega$ at the NIA for one segment
in each of the $\ell=2$, $m=0$ overtone multiplets.  As can be seen in
Table~\ref{tab:NIAdata2QNM} and Fig.~\ref{fig:m0l2n09-26}, for $9\le
n\le13$ the $n_1$ segments emerge from the NIA at $\bar\omega=-(n/4)i$
to very high accuracy, while for the $14\le n\le26$ segments, the
$n_0$ segment terminates at the NIA at $\bar\omega=-(n/4)i$.  Note
that the $8_0$ sequence also emerges from the NIA at the corresponding
value of $\bar\omega=-2i$.

\begin{figure}
\begin{tabular}{cc}
\includegraphics[width=0.5\linewidth,clip]{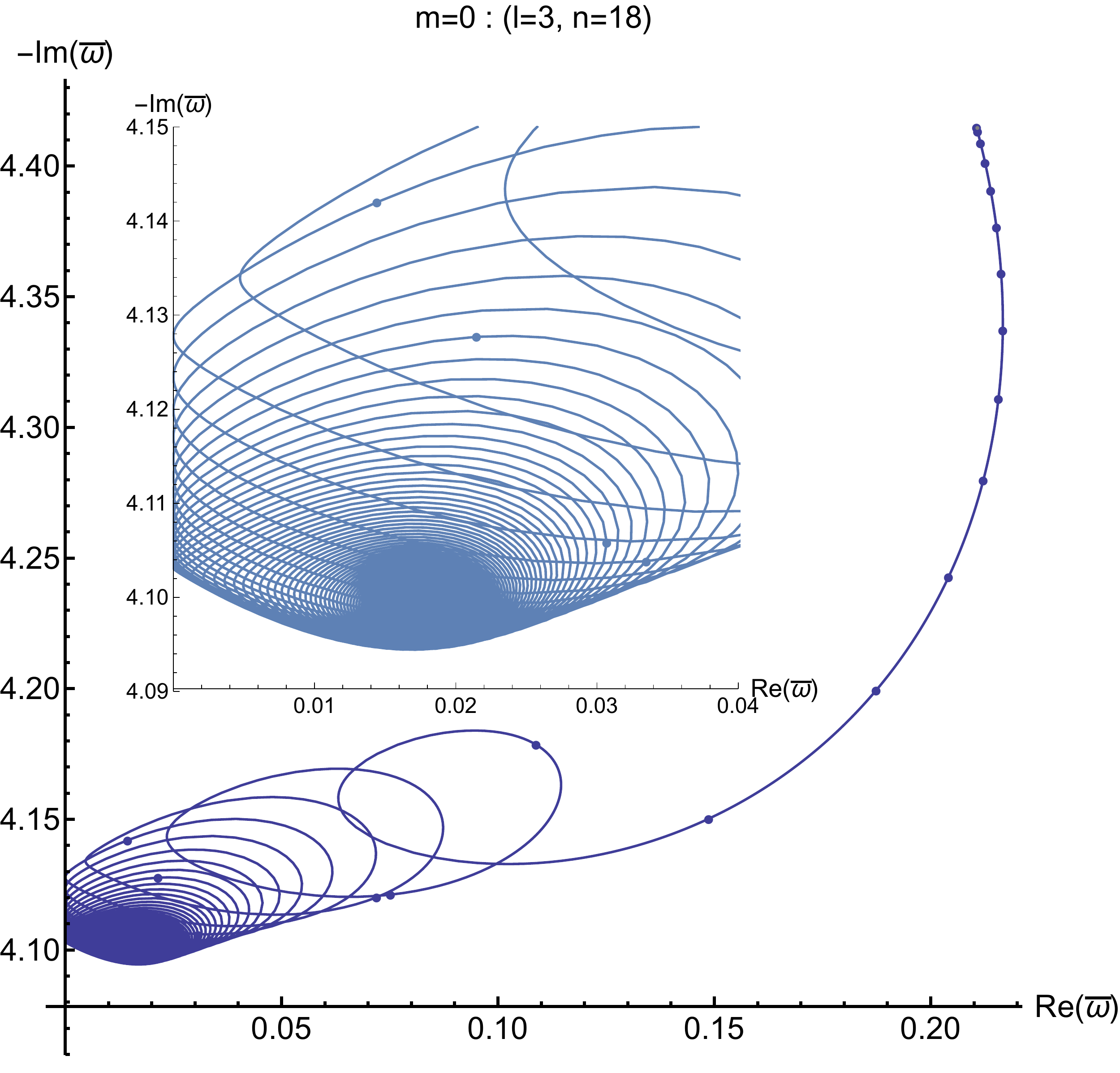} &
\includegraphics[width=0.5\linewidth,clip]{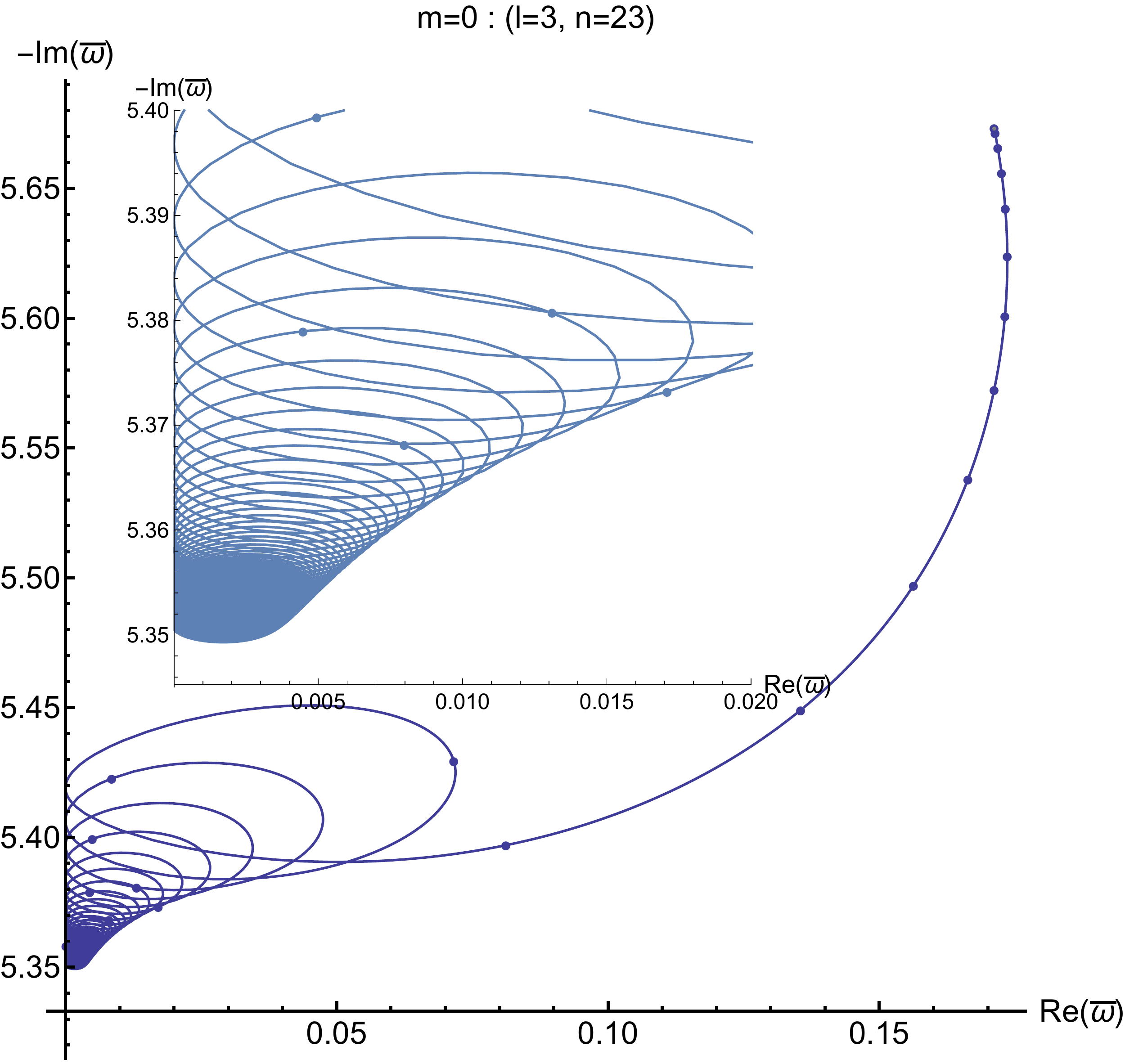} \\
\includegraphics[width=0.5\linewidth,clip]{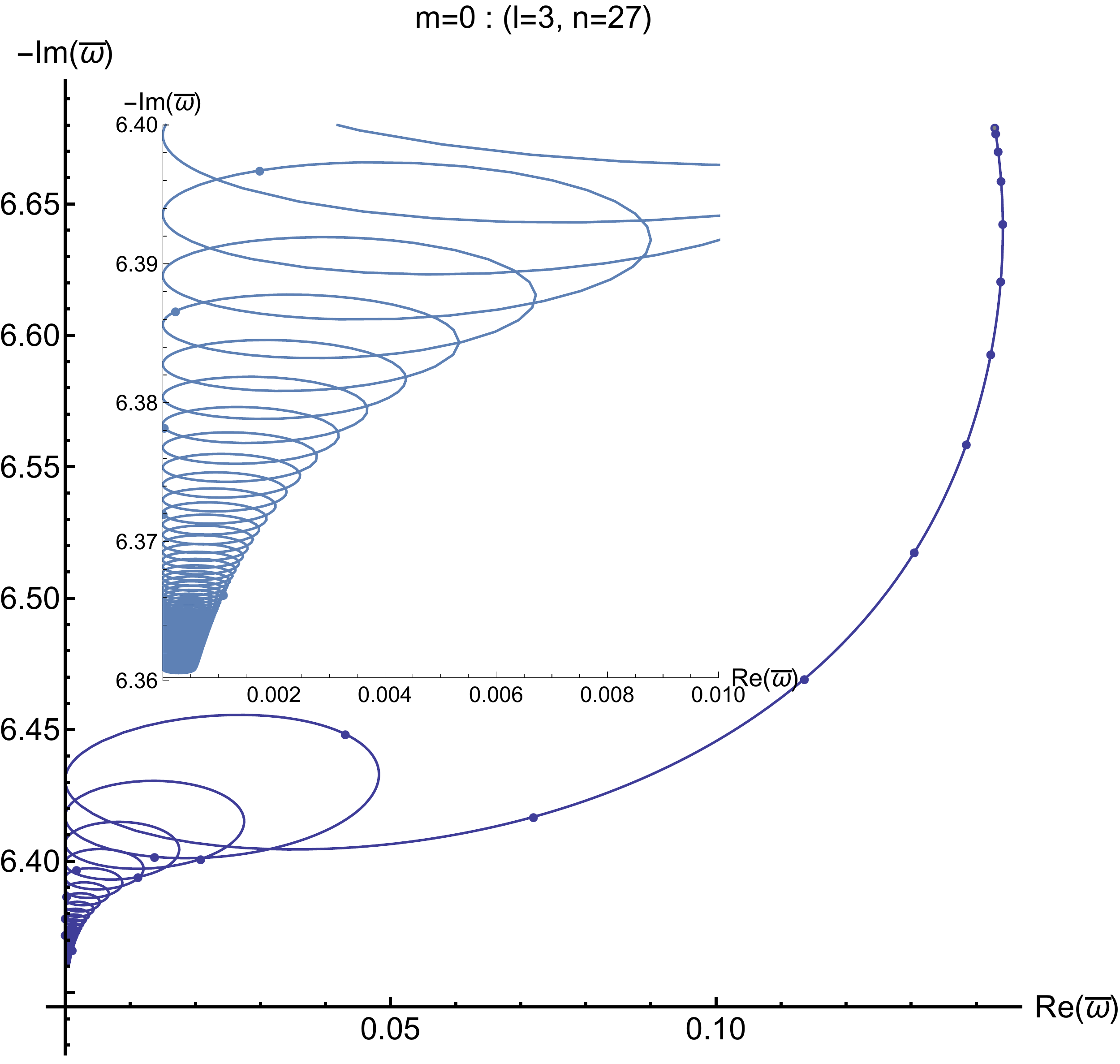} &
\includegraphics[width=0.5\linewidth,clip]{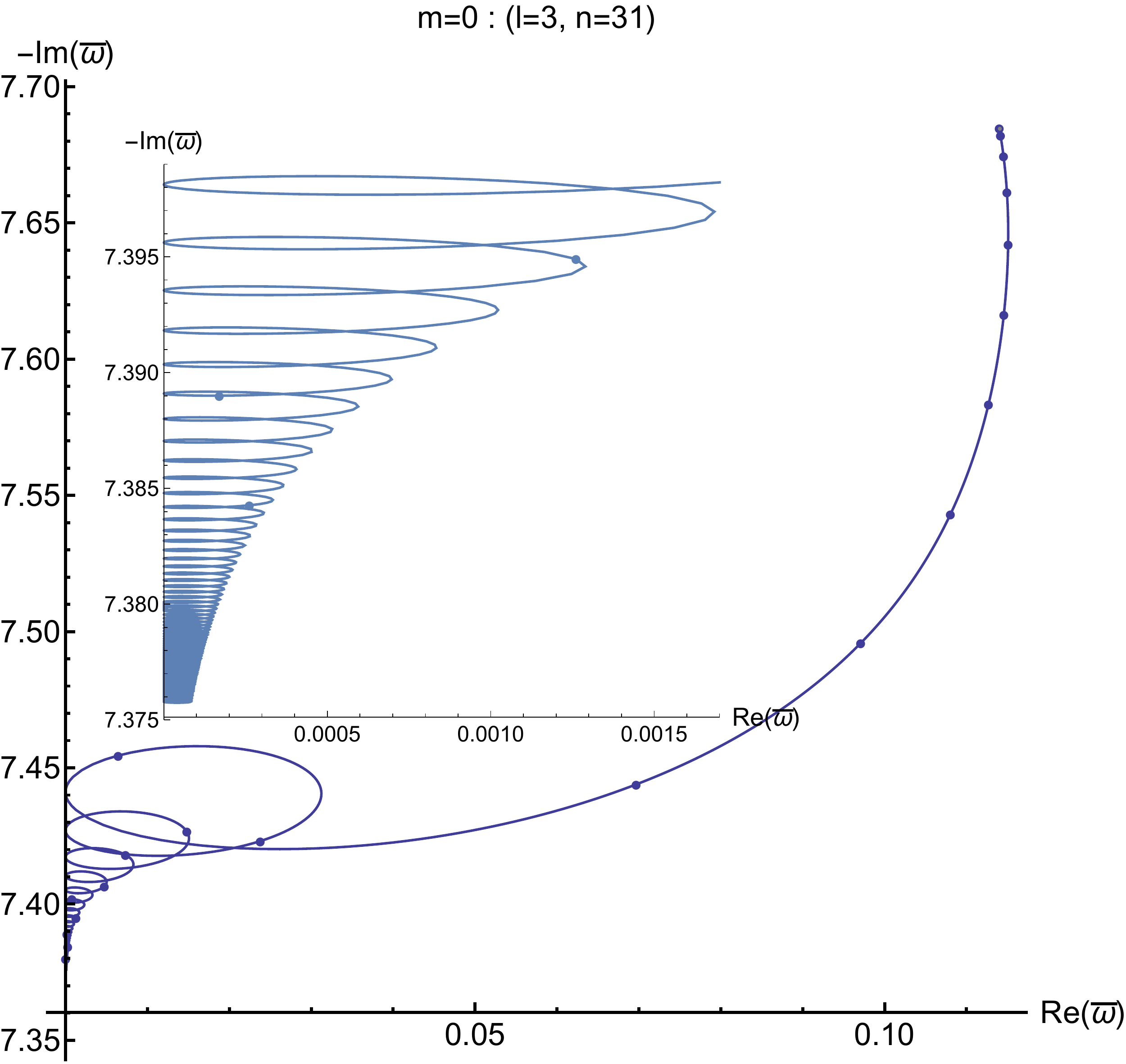}
\end{tabular}
\caption{\label{fig:m0l3n18-31} Examples of $\ell=3$, $m=0$ modes that
  appear to touch the NIA.  Shown are the $n=18$, $23$, $27$, and $31$
  sequences.  Each has a sequence begins at the Schwarzschild limit
  and then begin to exhibit loops.  Many of the loops have points of
  tangency with the NIA.  Each panel includes an inset that
  shows an enlarged view of the sequence near the NIA.}
\end{figure}

\begin{figure}
\begin{tabular}{cc}
\includegraphics[width=0.5\linewidth,clip]{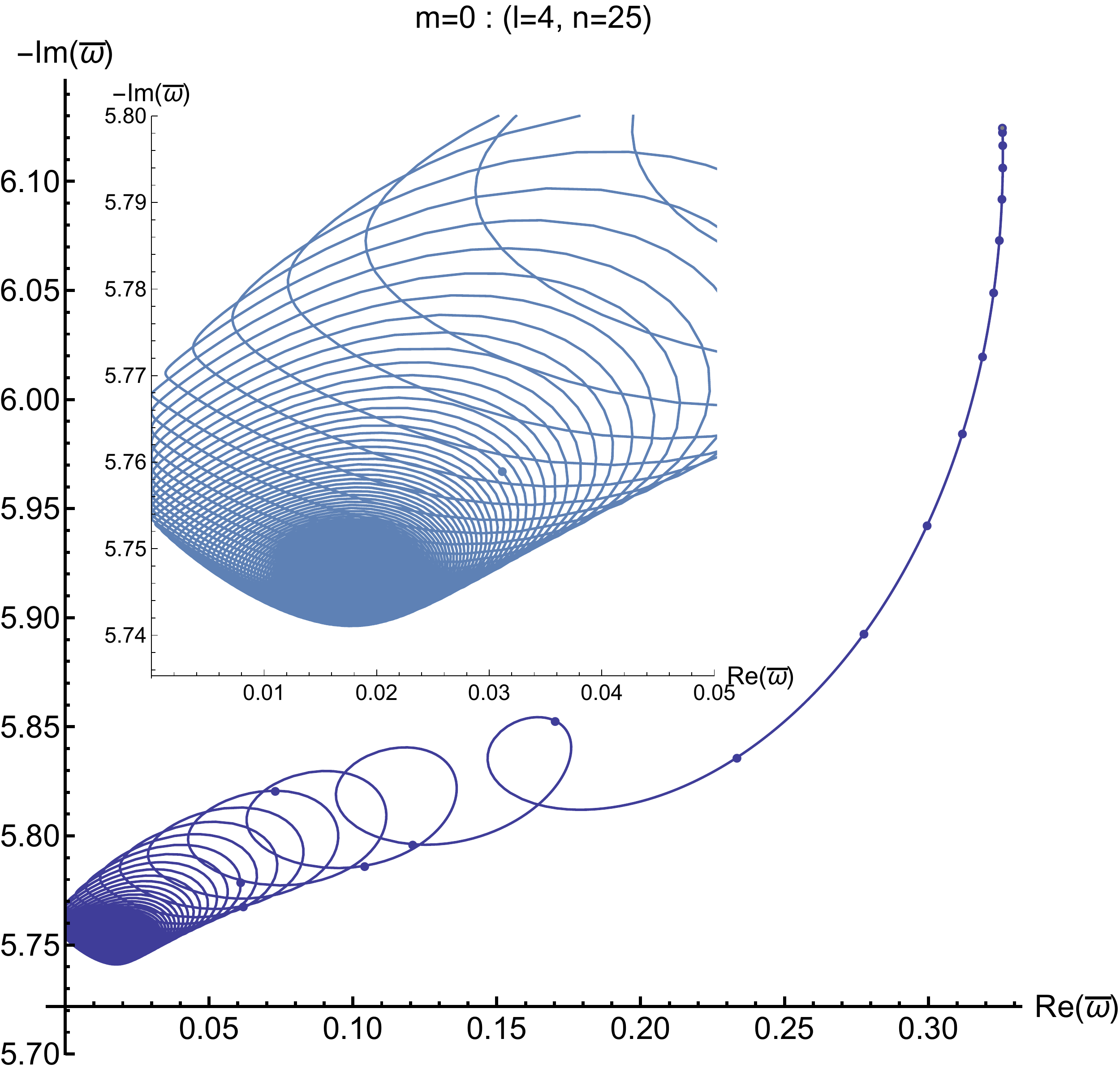} &
\includegraphics[width=0.5\linewidth,clip]{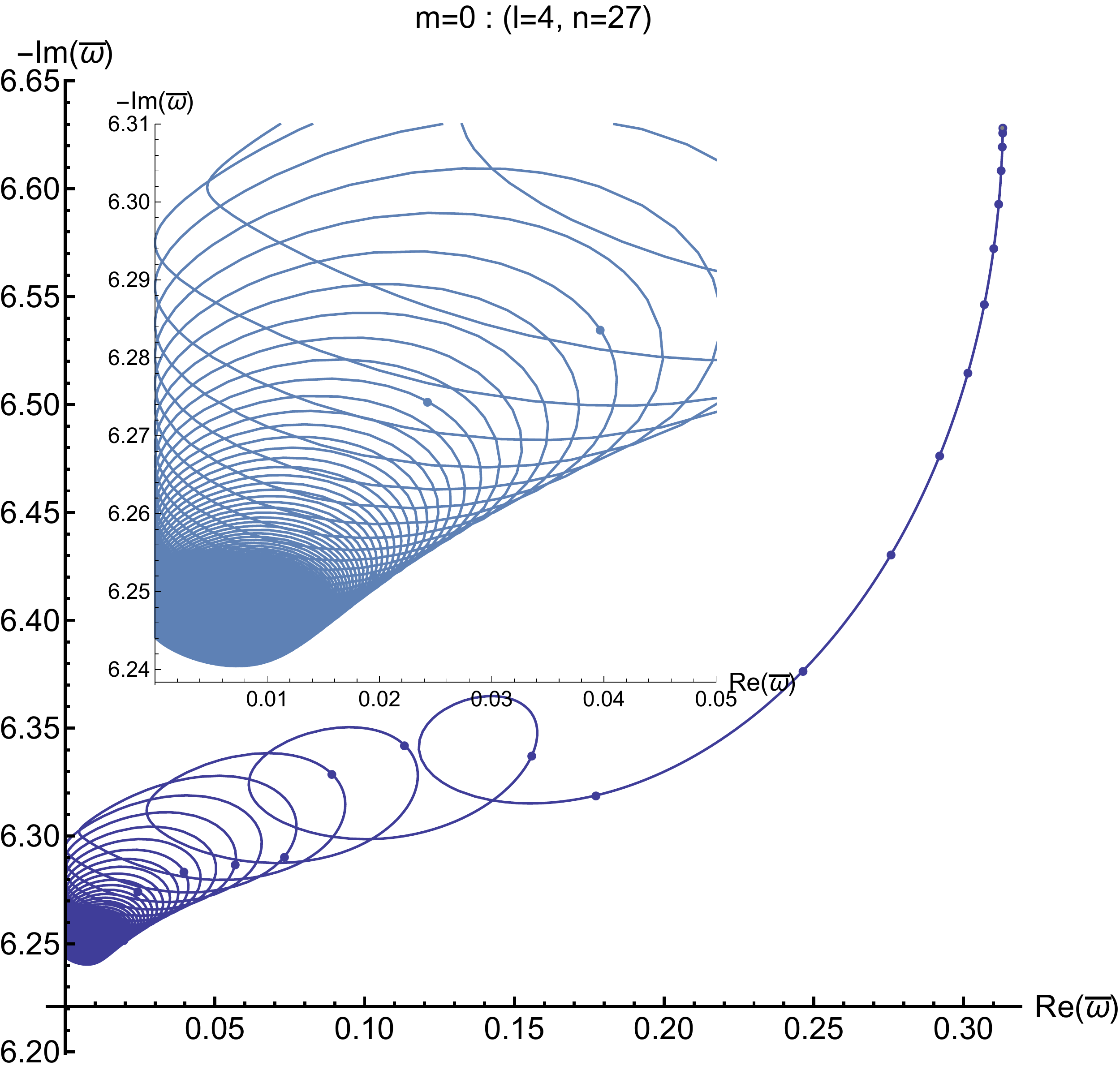} \\
\includegraphics[width=0.5\linewidth,clip]{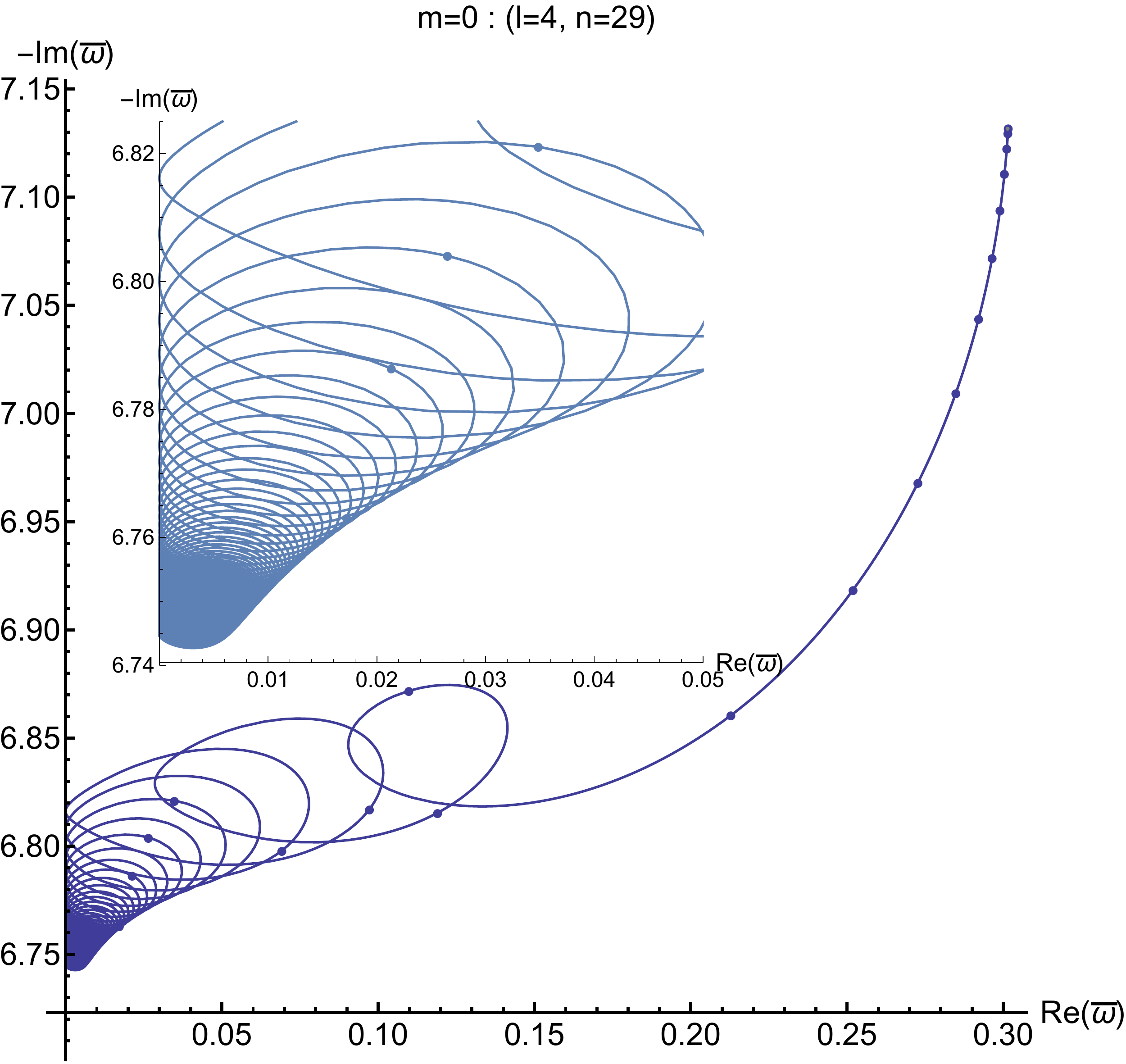} &
\includegraphics[width=0.5\linewidth,clip]{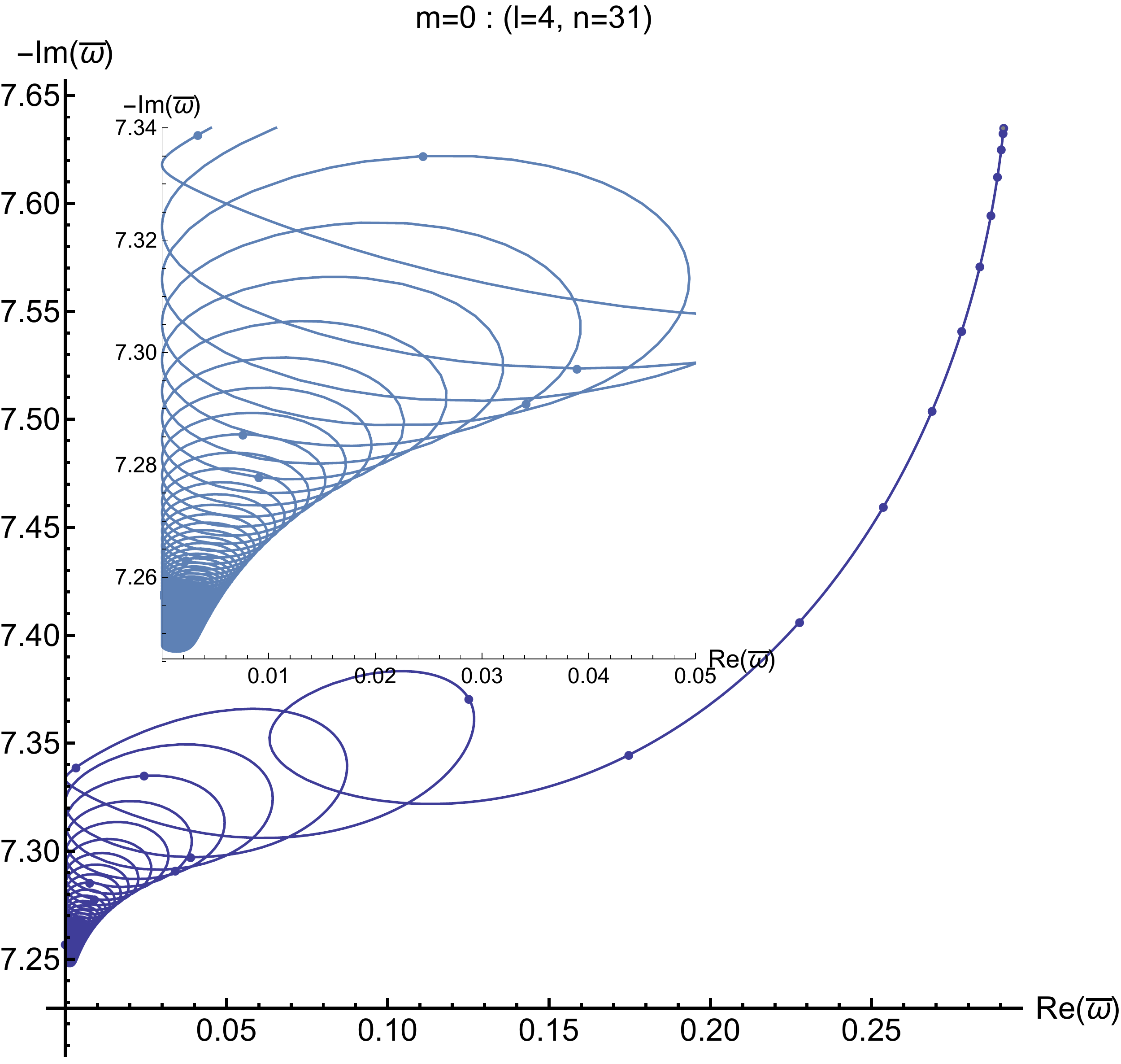}
\end{tabular}
\caption{\label{fig:m0l4n25-31} Examples of $\ell=4$, $m=0$ modes that
  appear to touch the NIA.  Shown are the $n=25$, $27$, $29$, and $31$
  sequences.  Each has a sequence begins at the Schwarzschild limit
  and then begin to exhibit loops.  Many of the loops have points of
  tangency with the NIA.  Each panel includes an inset that
  shows an enlarged view of the sequence near the NIA.}
\end{figure}

In addition to the various $\ell=2$, $m=0$ sequences which show
looping behavior with many points of tangency to the NIA, the $\ell=3$
and $4$, $m=0$ sequences also show a similar behavior starting with
$n=18$ for $\ell=3$ and at $n=25$ for $\ell=4$.
Tables\ref{tab:NIAdataLoops} and \ref{tab:NIAdataLoopsCount} also
contain relevant data for the cases of $\ell=3$ and $4$.  The one
significant difference compared to the $\ell=2$ case is that, so far,
these sequences do not terminate at or emerge from the NIA.  That is,
these are all single-segment sequences.
Figures~\ref{fig:m0l2-4n08-15}--\ref{fig:m0l2-4n24-31} show expanded
views including the $\ell=3$ and $4$ sequences, and
Figs.~\ref{fig:m0l3n18-31} and \ref{fig:m0l4n25-31} show isolated
sequences with loops touching the NIA.

\begin{table}
\begin{tabular}{llclcc}
\hline\hline
\Chead{Mode} & \Chead{$\bar\omega$} && \Chead{$\bar{a}$} && $N$ \\
$\{2,0,8_1\}$ & $8.26933\times10^{-10}-2.19086i$ && $0.315947$ && 9 \\
$\{2,0,9_0\}$ & $9.85696\times10^{-10}-2.25000i$ && $0.305661$ && 9 \\
$\{2,0,9_1\}$ & $1.44669\times10^{-9\ }-2.38829i$ && $0.404696$ && 10 \\
$\{2,0,10_0\}$ & $2.56367\times10^{-10}-2.50000i$ && $0.391144$ && 10 \\
$\{2,0,10_1\}$ & $6.40376\times10^{-10}-2.59349i$ && $0.451434$ && 11 \\
$\{2,0,11_0\}$ & $9.02419\times10^{-10}-2.75000i$ && $0.438874$ && 11 \\
$\{2,0,11_1\}$ & $6.35257\times10^{-10}-2.80677i$ && $0.476868$ && 12 \\
$\{2,0,12_0\}$ & $6.57536\times10^{-10}-3.00000i$ && $0.469040$ && 12 \\
$\{2,0,12_1\}$ & $6.29767\times10^{-10}-3.02770i$ && $0.489584$ && 13 \\
$\{2,0,13_0\}$ & $2.82277\times10^{-11}-3.25000i$ && $0.489617$ && 13 \\
$\{2,0,13_1\}$ & $1.19986\times10^{-10}-3.25539i$ && $0.494193$ && 14\\
$\{2,0,14_0\}$ & $2.95683\times10^{-10}-3.48876i$ && $0.493501$ && 15\\
$\{2,0,14_1\}$ & $2.09637\times10^{-10}-3.50000i$ && $0.504617$ && 14 \\
$\{2,0,15_0\}$ & $5.24731\times10^{-10}-3.72672i$ && $0.489333$ && 16 \\
$\{2,0,15_1\}$ & $2.47748\times10^{-10}-3.75000i$ && $0.516351$ && 15 \\
$\{2,0,16_0\}$ & $4.77568\times10^{-10}-3.96828i$ && $0.482912$ && 17 \\
$\{2,0,16_1\}$ & $1.98281\times10^{-10}-4.00000i$ && $0.526296$ && 16 \\
$\{2,0,17_0\}$ & $4.02699\times10^{-10}-4.21263i$ && $0.475069$ && 18 \\
$\{2,0,17_1\}$ & $6.66620\times10^{-11}-4.25000i$ && $0.535491$ && 17 \\
$\{2,0,18_0\}$ & $6.61545\times10^{-10}-4.45911i$ && $0.466375$ && 19 \\
$\{2,0,18_1\}$ & $7.68713\times10^{-11}-4.50000i$ && $0.544763$ && 18 \\
$\{2,0,19_0\}$ & $5.31223\times10^{-10}-4.70719i$ && $0.457216$ && 20 \\
$\{2,0,19_1\}$ & $3.31607\times10^{-12}-4.75000i$ && $0.554862$ && 19 \\
$\{2,0,20_0\}$ & $9.01498\times10^{-10}-4.95648i$ && $0.447860$ && 21 \\
$\{2,0,20_1\}$ & $1.77914\times10^{-11}-5.00000i$ && $0.566592$ && 20 \\
$\{2,0,21_0\}$ & $1.98444\times10^{-10}-5.20668i$ && $0.438486$ && 22 \\
$\{2,0,21_1\}$ & $1.35783\times10^{-11}-5.25000i$ && $0.580946$ && 21 \\
$\{2,0,22_0\}$ & $1.16769\times10^{-9\ }-5.45755i$ && $0.429213$ && 23 \\
$\{2,0,22_1\}$ & $4.51259\times10^{-12}-5.50000i$ && $0.599334$ && 22 \\
$\{2,0,23_0\}$ & $3.28288\times10^{-10}-5.70893i$ && $0.420123$ && 24 \\
$\{2,0,23_1\}$ & $2.35358\times10^{-12}-5.75000i$ && $0.624019$ && 23 \\
$\{2,0,24_0\}$ & $4.68041\times10^{-12}-5.96069i$ && $0.411266$ && 25 \\
$\{2,0,24_1\}$ & $4.06023\times10^{-15}-6.00000i$ && $0.659109$ && 24 \\
$\{2,0,25_0\}$ & $1.80872\times10^{-10}-6.21271i$ && $0.402672$ && 26 \\
$\{2,0,25_1\}$ & $1.03997\times10^{-13}-6.25000i$ && $0.713250$ && 25 \\
$\{2,0,26_0\}$ & $1.00792\times10^{-10}-6.46493i$ && $0.394359$ && 27 \\
$\{2,0,26_1\}$ & $5.55008\times10^{-16}-6.50000i$ && $0.808299$ && 26 \\
$\{2,0,27\}$ & $3.52849\times10^{-10}-6.71728i$ && $0.386334$ && 28 \\
$\{2,0,28\}$ & $9.30984\times10^{-10}-6.96973i$ && $0.378598$ && 29 \\
$\{2,0,29\}$ & $1.06736\times10^{-9\ }-7.22223i$ && $0.371147$ && 30 \\
$\{2,0,30\}$ & $4.28210\times10^{-11}-7.47477i$ && $0.363976$ && 31 \\
$\{2,0,31\}$ & $5.96821\times10^{-10}-7.72732i$ && $0.357074$ && 32 \\
\hline\hline
\end{tabular}
\caption{\label{tab:NIAdata2QNM} Numerical solution for QNMs closest
  to NIA at beginning or end of selected mode sequences.  The entries
  correspond to $\ell=2$ sequences plotted in
  Figs.~\ref{fig:m0l2-4n08-15}--\ref{fig:m0l2n09-26}.  We will show in
  Sec.~\ref{sec:modes_on_NIA} that all of these sequences terminate
  at, or emerge from a mode precisely on the NIA.  More than half of
  these are simultaneously QNMs and TTM${}_L$s, however the
  $9_0$--$13_0$ and $14_1$--$26_1$ are simply QNMs. The column labeled
  by $N$ contains the value for $N_\pm$ for either
  Eq.~(\ref{eq:omega_plus}) or (\ref{eq:omega_minus}).}
\end{table}

\begin{table}
\begin{tabular}{lcclcl}
\hline\hline
\Chead{Mode} & \# && \Chead{$\bar\omega$} && \Chead{$\bar{a}$} \\
$\{2,0,9_1\}$ & 1 && $1.60237\times10^{-9\ }-2.22167i$ && $0.735354$ \\
 & 2 && $9.58918\times10^{-10}-2.21540i$ && $0.810761$ \\
 & 3 && $5.47607\times10^{-10}-2.21151i$ && $0.856778$ \\
 & 4 && $2.40980\times10^{-10}-2.20889i$ && $0.887379$ \\
 & 5 && $2.74679\times10^{-10}-2.20705i$ && $0.908916$ \\
 & 6 && $1.45896\times10^{-10}-2.20569i$ && $0.924713$ \\
 & 7 && $4.29091\times10^{-11}-2.20466i$ && $0.936673$ \\
\hline
$\{3,0,18\}$ & 1 && $3.05492\times10^{-13}-4.12779i$ && $0.799512$ \\
 & 2 && $2.70918\times10^{-13}-4.12339i$ && $0.829180$ \\
 & 3 && $9.56952\times10^{-12}-4.11988i$ && $0.852443$ \\
 & 4 && $1.44533\times10^{-13}-4.11704i$ && $0.871092$ \\
 & 5 && $1.38208\times10^{-11}-4.11471i$ && $0.886310$ \\
 & 6 && $1.26720\times10^{-12}-4.11277i$ && $0.898914$ \\
 & 7 && $1.25936\times10^{-11}-4.11115i$ && $0.909486$ \\
 & 8 && $3.00991\times10^{-16}-4.10976i$ && $0.918448$ \\
 & 9 && $4.83160\times10^{-14}-4.10858i$ && $0.926119$ \\
 & 10 && $1.19638\times10^{-13}-4.10756i$ && $0.932740$ \\
 & 11 && $3.64452\times10^{-13}-4.10667i$ && $0.938496$ \\
 & 12 && $3.97142\times10^{-15}-4.10589i$ && $0.943533$ \\
 & 13 && $5.83628\times10^{-13}-4.10520i$ && $0.947970$ \\
 & 14 && $2.86913\times10^{-12}-4.10460i$ && $0.951897$ \\
 & 15 && $2.47378\times10^{-13}-4.10406i$ && $0.955392$ \\
 & 16 && $5.94897\times10^{-13}-4.10357i$ && $0.958515$ \\
 & 17 && $1.35759\times10^{-14}-4.10314i$ && $0.961319$ \\
\hline
 $\{4,0,25\}$ & 1 && $1.40943\times10^{-12}-5.76781i$ && $0.881851$ \\
 & 2 && $8.01725\times10^{-12}-5.76574i$ && $0.891625$ \\
 & 3 && $9.00602\times10^{-12}-5.76391i$ && $0.900203$ \\
 & 4 && $4.68881\times10^{-12}-5.76230i$ && $0.907778$ \\
 & 5 && $7.51494\times10^{-13}-5.76085i$ && $0.914503$ \\
 & 6 && $1.16406\times10^{-11}-5.75957i$ && $0.920504$ \\
 & 7 && $1.58023\times10^{-11}-5.75841i$ && $0.925884$ \\
 & 8 && $3.18629\times10^{-12}-5.75737i$ && $0.930726$ \\
 & 9 && $5.53437\times10^{-12}-5.75643i$ && $0.935101$ \\
 & 10 && $1.13290\times10^{-12}-5.75557i$ && $0.939068$ \\
 & 11 && $7.15779\times10^{-13}-5.75479i$ && $0.942678$ \\
 & 12 && $2.17820\times10^{-12}-5.75408i$ && $0.945972$ \\
 & 13 && $3.19872\times10^{-13}-5.75343i$ && $0.948987$ \\
\hline\hline
\end{tabular}
\caption{\label{tab:NIAdataLoops} Numerical solution for QNMs closest
  to NIA for selected sequences with loops that have points of
  tangency with the NIA.  The first 7 entries correspond to the 7
  points of tangency in the first $\ell=2$ sequence to have such
  points of tangency (see upper-left panel of
  Fig.~\ref{fig:m0l2n09-26}).  The next 17 entries are for the
  corresponding first $\ell=3$ sequences to have points of tangency to
  the NIA (see upper-left panel of Fig.~\ref{fig:m0l3n18-31}).  The
  final 13 entries are for the corresponding first $\ell=4$ sequences
  to have points of tangency to the NIA (see upper-left panel of
  Fig.~\ref{fig:m0l4n25-31}).  We will show in
  Sec.~\ref{sec:modes_on_NIA} that no QNM or TTM mode exists at any
  point of tangency.}
\end{table}

\begin{table}
\begin{tabular}{lcclcl}
\hline\hline
\Chead{Mode} & $N_L$ && \Chead{$\bar\omega$} && \Chead{$\bar{a}$} \\
     $\{2,0,9_1\}$ & 7 && $-2.20466311i$ && $0.9366728030$ \\
   $\{2,0,10_1\}$ & 20 && $-2.450680438i$ && $0.982207022698$ \\
   $\{2,0,11_1\}$ & 41 && $-2.699696290i$ && $0.993545856885$ \\
   $\{2,0,12_1\}$ & 81 && $-2.9496977199i$ && $0.997755524347$ \\
  $\{2,0,13_1\}$ & 154 && $-3.2004198500i$ && $0.999215241283$ \\
  $\{2,0,14_1\}$ & 292 && $-3.45173825872i$ && $0.999735368677$ \\
  $\{2,0,15_1\}$ & 555 && $-3.70358925546i$ && $0.999913629889$ \\
 $\{2,0,16_1\}$ & 1058 && $-3.95591591329i$ && $0.999972512866$ \\
 $\{2,0,17_1\}$ & 2036 && $-4.20866196834i$ && $0.999991533852$ \\
$\{2,0,18_1\}$ & $>2506$ && $-4.4617737635i$ && $0.999993710751$ \\
$\{2,0,19_1\}$ & $>2826$ && $-4.71520182427i$ && $0.999994473567$ \\
$\{2,0,20_1\}$ & $>2186$ && $-4.9689036556i$ && $0.999989779470$ \\
$\{2,0,21_1\}$ & $>2363$ && $-5.22284221839i$ && $0.999990332460$ \\
$\{2,0,22_1\}$ & $>2590$ && $-5.47698706788i$ && $0.999991145753$ \\
$\{2,0,23_1\}$ & $>2860$ && $-5.73131247094i$ && $0.999992049325$ \\
$\{2,0,24_1\}$ & $>2105$ && $-5.98579710816i$ && $0.999984056118$ \\
$\{2,0,25_1\}$ & $>2487$ && $-6.24042211241i$ && $0.999987583136$ \\
$\{2,0,26_1\}$ & $>2929$ && $-6.49517245393i$ && $0.999990311150$ \\
\hline
     $\{3,0,18\}$ & 17 && $-4.1031415544i$ && $0.961319209948$ \\
     $\{3,0,19\}$ & 40 && $-4.3492316497i$ && $0.986072890543$ \\
     $\{3,0,20\}$ & 75 && $-4.5981480445i$ && $0.994372141386$ \\
    $\{3,0,21\}$ & 131 && $-4.84824846927i$ && $0.997708161178$ \\
    $\{3,0,22\}$ & 225 && $-5.09911272304i$ && $0.999072207558$ \\
    $\{3,0,23\}$ & 383 && $-5.35056428456i$ && $0.999633513815$ \\
    $\{3,0,24\}$ & 652 && $-5.60251180138i$ && $0.999857946825$ \\
   $\{3,0,25\}$ & 1112 && $-5.85489118443i$ && $0.999945856002$ \\
   $\{3,0,26\}$ & 1837 && $-6.10765020533i$ && $0.999978219330$ \\
$\{3,0,27\}$ & $>1421$ && $-6.3607457778i$ && $0.999960719027$ \\
$\{3,0,28\}$ & $>1694$ && $-6.6141303333i$ && $0.999970035027$ \\
$\{3,0,29\}$ & $>1697$ && $-6.8677728413i$ && $0.999967824364$ \\
$\{3,0,30\}$ & $>1825$ && $-7.1216409598i$ && $0.999970063305$ \\
$\{3,0,31\}$ & $>1971$ && $-7.37570857244i$ && $0.999972448891$ \\
\hline
  $\{4,0,25\}$ & 13 && $-5.753433018i$ && $0.948986760638$ \\
  $\{4,0,26\}$ & 47 && $-5.995998951i$ && $0.983358646036$ \\
  $\{4,0,27\}$ & 88 && $-6.244148245i$ && $0.993102653036$ \\
 $\{4,0,28\}$ & 151 && $-6.4939321820i$ && $0.997042993390$ \\
 $\{4,0,29\}$ & 249 && $-6.7446195562i$ && $0.998718278044$ \\
 $\{4,0,30\}$ & 404 && $-6.9959315212i$ && $0.999450207454$ \\
 $\{4,0,31\}$ & 657 && $-7.2477418438i$ && $0.999769051122$ \\
\hline\hline
\end{tabular}
\caption{\label{tab:NIAdataLoopsCount} We list all of the $m=0$ sequences
  with $\ell=2$--$4$ and $n\le31$ which have loops with points of
  tangency to the NIA.  The second column, $N_L$, lists the number of
  such points of tangency for each sequence.  If the number is
  preceded by $>$ then it represents the number we have computed so
  far, but that we have not yet found the last point of tangency for
  that sequence.  The third column, $\bar\omega$, list the
  interpolated value for the frequency at the last point of tangency.
  The last column gives the corresponding interpolated value for
  $\bar{a}$.}
\end{table}

\section{Modes on the NIA}
\label{sec:modes_on_NIA}

In Sec.~\ref{sec:numerical_results}, we have presented significant
evidence that many QNM exist with frequencies that are arbitrarily
close to the NIA, but as we mention there, numerical solutions for
QNMs cannot be obtained for modes precisely on the NIA when using
Leaver's method.  While Maassen van den Brink\cite{van_den_brink-2000}
has answered the question of the existence of QNMs on the NIA for the
special frequencies of $\bar\omega=\bar\Omega_\ell$ when $\bar{a}=0$, we
now have what seems to be a countably infinite number of modes with
$\bar{a}\ne0$ and frequencies on the NIA, and we do not know if these
are QNMs.

Here, we will outline the framework needed to answer this question.
We begin by looking at Leaver's method and show why it cannot be used
to compute QNMs of the Kerr geometry on the NIA.  We will do this
using the framework of solutions of the confluent Heun equation
because the theory of confluent Heun polynomials allows us to develop
a method for locating potential QNMs with frequencies on the NIA.
Finally, we will examine these modes with frequencies on the NIA and
describe how we determine whether or not they are QNMs.

\subsection{Solutions of the confluent Heun equation}
\label{sec:sol_of_CHE} 

The confluent Heun equation is a second-order linear ordinary
differential equation, obtained from the Heun equation when one
regular singular point is lost by confluence with another and the point
at infinity becomes irregular\cite{Heun-eqn}.  Written in {\em
  nonsymmetrical canonical form}, the confluent Heun equation reads
\begin{equation}\label{eq:Heun_NSCF}
\frac{d^2H(z)}{dz^2} 
+ \left(4p + \frac\gamma{z}+\frac\delta{z-1}\right)\frac{dH(z)}{dz}
+ \frac{4\alpha pz -\sigma}{z(z-1)}H(z)=0,
\end{equation}
where the remaining two regular singular points are at $z=0,1$ and the
irregular singular point is at $z=\infty$.  It is defined by five 
parameters: $p$, $\alpha$, $\gamma$, $\delta$, and $\sigma$.

Frobenius solutions local to each of the three
singular points can be defined in terms of two functions\cite{Heun-eqn},
\begin{align}
\label{eq:local_a_sol_series}
 Hc^{(a)}(p,\alpha,\gamma,\delta,\sigma;z)&= \sum_{k=0}^\infty{c^{(a)}_kz^k}, \\
\label{eq:local_r_sol_series}
 Hc^{(r)}(p,\alpha,\gamma,\delta,\sigma;z)&= \sum_{k=0}^\infty{c^{(r)}_kz^{-\alpha-k}}.
\end{align}
The local solution $Hc^{(a)}(p,\alpha,\gamma,\delta,\sigma;z)$ is defined by
the three-term recurrence relation
\begin{subequations}
\label{eq:local_a_defs}
\begin{align}
\label{eq:local_a_3term}
  0 =& f^{(a)}_kc^{(a)}_{k+1}+ g^{(a)}_kc^{(a)}_k\!+ h^{(a)}_kc^{(a)}_{k-1}\ :
   \begin{array}{l}c^{(a)}_{-1}=0,\\ c^{(a)}_0=1,\end{array} \\
\label{eq:local_a_g}
  g^{(a)}_k =& k(k-4p+\gamma+\delta-1)-\sigma, \\
\label{eq:local_a_f}
  f^{(a)}_k =& -(k+1)(k+\gamma), \\
\label{eq:local_a_h}
  h^{(a)}_k =& 4p(k+\alpha-1).
\end{align}
\end{subequations}
At the regular singular point at $z=0$, the characteristic exponents
(roots of the indicial equation) are $\{0,1-\gamma\}$ so the local
solutions have the leading behavior
\begin{subequations}
\label{eq:local_sol_0}
\begin{align}
  \lim_{z\to0}H(z)\sim1& \quad\mbox{or}\quad z^{1-\gamma},
\intertext{and the two solutions local to $z=0$ are given by}
\label{eq:local_sol_z0a}
  Hc^{(a)}(p,\alpha,\gamma,\delta,\sigma;z&), \\
\label{eq:local_sol_z0b}
  z^{1-\gamma}Hc^{(a)}(p,\alpha+1-\gamma,&2-\gamma,\delta, \\
 &\sigma+(1-\gamma)(4p-\delta);z). \nonumber
\end{align}
\end{subequations}
For the regular singular point at $z=1$, the characteristic exponents
are $\{0,1-\delta\}$ so the local solutions have the leading behavior
\begin{subequations}
\label{eq:local_sol_1}
\begin{align}
  \lim_{z\to1}H(z)\sim1& \quad\mbox{or}\quad (z-1)^{1-\delta},
\intertext{and the two solutions local to $z=1$ are given by}
\label{eq:local_sol_z1a}
  Hc^{(a)}(-p,\alpha,\delta,&\gamma,\sigma-4p\alpha;1-z) \\
\label{eq:local_sol_z1b}
  (z-1)^{1-\delta}Hc^{(a)}(-p,\alpha+&1-\delta,2-\delta,\gamma, \\
   \sigma-(&1-\delta)\gamma-4p(\alpha+1-\delta);1-z). \nonumber
\end{align}
\end{subequations}

The solution $Hc^{(r)}(p,\alpha,\gamma,\delta,\sigma;z)$ is defined in a
similar way:
\begin{subequations}
\label{eq:local_r_defs}
\begin{align}
\label{eq:local_r_3term}
  0 =& f^{(r)}_kc^{(r)}_{k+1}+ g^{(r)}_kc^{(r)}_k+ h^{(r)}_kc^{(r)}_{k-1}\ :
   \begin{array}{l}c^{(r)}_{-1}=0,\\ c^{(r)}_0=1,\end{array} \\
\label{eq:local_r_g}
  g^{(r)}_k =& (k+\alpha)(k+4p+\alpha-\gamma-\delta+1)-\sigma, \\
\label{eq:local_r_f}
  f^{(r)}_k =& -4p(k+1), \\
\label{eq:local_r_h}
  h^{(r)}_k =& -(k+\alpha-1)(k+\alpha-\gamma).
\end{align}
\end{subequations}
For the irregular singular point at $z=\infty$, the local solutions
have the leading behavior
\begin{subequations}
\label{eq:local_sol_inf}
\begin{align}
  \lim_{z\to\infty}H(z)\sim z^{-\alpha}& \quad\mbox{or}\quad 
  e^{-4pz}z^{\alpha-\gamma-\delta}.
\intertext{and the two solutions local to $z=\infty$ are given by}
\label{eq:local_sol_zinfa}
  Hc^{(r)}(p,\alpha,\gamma,\delta,\sigma;&z) \\
\label{eq:local_sol_zinfb}
  e^{-4pz}Hc^{(r)}(-p,-\alpha+\gamma&+\delta,\gamma,\delta,\sigma-4p\gamma;z).
\end{align}
\end{subequations}

If a solution is simultaneously a Frobenius solution for two adjacent
singular points, then the solution is called a {\em confluent Heun
  function}. In the special case that a solution is simultaneously a
Frobenius solution for all three singular points, then the solution is
a {\em confluent Heun polynomial}.  A polynomial solution requires
that the series solution terminates.  A necessary, but not sufficient
condition for this to occur is for the second parameter, $\alpha$, of
either Eq.~(\ref{eq:local_a_sol_series}) or
(\ref{eq:local_r_sol_series}) to be a nonpositive integer, $-q$,
resulting in $h^{(a,r)}_{q+1}=0$.

If we think of the recurrence relations in Eqs.~(\ref{eq:local_a_3term})
and (\ref{eq:local_r_3term}) as infinite-dimensional tridiagonal
systems, then if $h^{(a,r)}_{q+1}=0$, we can think of the tridiagonal
coefficient matrix 
\begin{equation}\label{eq:tridiag_cond}
  \left[\begin{array}{ccccc|cccc}
  g_0 & f_0 & 0 & \cdots & 0 & 0 & 0 & 0 & \cdots \\
  h_1 & g_1 & f_1 & \ddots & 0 & 0 & 0 & 0 & \cdots \\
  0 & h_2 & g_2 & f_2 & \ddots & 0 & 0 & 0 & \cdots \\
  0 & 0 & \ddots & \ddots & \ddots & \ddots & \ddots & \ddots & \ddots \\
  0 & 0 & \ddots & h_q & g_q & f_q & 0 & 0 & \cdots \\
\hline
  0 & 0 & \cdots & 0 & 0 & g_{q+1} & f_{q+1} & 0 & \cdots \\
  0 & 0 & \cdots & 0 & 0 & h_{q+2} & g_{q+2} & f_{q+2} & \ddots \\
  \vdots & \vdots & \cdots & \vdots & \vdots & \ddots & \ddots & \ddots & \ddots \\
  \end{array}\right]
\end{equation}
in block form.  The two blocks on the diagonal are both tridiagonal.
The upper-right block has only one non-zero element, $f_q$.  The
lower-left block is all zeros.  The vanishing of the determinant of
the upper-left block, referred to as the $\Delta_{q+1}=0$ condition,
is the necessary and sufficient condition that $c^{a,r}_{q+1}=0$,
which guarantees that the series will terminate\cite{Heun-eqn}.

\subsection{The radial Teukolsky equation and Leaver's method}
\label{sec:radial-teukolsky-Leaver}
The radial Teukolsky equation, Eq.~(\ref{eqn:radialR:Diff_Eqn}), has
regular singular points at the inner and outer horizons.  These are 
located at the roots, $r_\pm$, of $\Delta=0$:
\begin{equation}
  r_\pm \equiv M\pm\sqrt{M^2-a^2}.
\end{equation}
The outer or event horizon is labeled by $r_+$ and the inner or Cauchy
horizon by $r_\minus$.  $r=\infty$ is an irregular singular point.
We define the following dimensionless variables:
\begin{subequations}
\begin{align}
  \bar{r} &\equiv \frac{r}{M}, \\
  \bar{a} &\equiv \frac{a}{M}, \\
  \bar\omega &\equiv M\omega.
\end{align}
\end{subequations}
In terms of the dimensionless coordinate
\begin{equation}
  z \equiv \frac{r-r_\minus}{r_+-r_\minus} 
  = \frac{\bar{r}-\bar{r}_\minus}{\bar{r}_+-\bar{r}_\minus},
\end{equation}
the radial Teukolsky equation, (\ref{eqn:radialR:Diff_Eqn}), can be
placed into nonsymmetrical canonical form, (\ref{eq:Heun_NSCF}), by
making the transformation
\begin{equation}
R(r(z)) = z^\eta(z-1)^\xi e^{(\bar{r}_+-\bar{r}_\minus)\bar\zeta z} H(z).
\end{equation}
The parameters $\bar\zeta$, $\xi$, and $\eta$ must be of the form
\begin{subequations}\label{eq:Teukolsky_Heun_parameters}
\begin{align}
  \bar\zeta&=\pm i\bar\omega\equiv\bar\zeta_\pm, \\
  \xi&=\frac{-s\pm(s + 2i\sigma_+)}2\equiv\xi_\pm, \\
  \eta&=\frac{-s\pm(s - 2i\sigma_\minus)}2\equiv\eta_\pm,
\end{align}
\end{subequations}
where
\begin{equation}\label{eqn:sigmapm-def}
 \sigma_\pm\equiv\frac{2\bar\omega\bar{r}_\pm - m\bar{a}}{\bar{r}_+-\bar{r}_\minus}.
\end{equation}
See Ref.\cite{cook-zalutskiy-2014,Fiziev-2009b} for a complete
discussion.  The parameters $\bar\zeta$, $\xi$, and $\eta$ can each
take on one of two values allowing for a total of eight ways to
achieve nonsymmetrical canonical form, but in each case, the five
parameters defining the confluent Heun equation are given by
\begin{subequations}
\begin{align}
  p &= (\bar{r}_+-\bar{r}_\minus)\frac{\bar\zeta}2 \\
  \alpha &= 1+s+\xi+\eta - 2\bar\zeta + s\frac{i\bar\omega}{\bar\zeta} \\
  \gamma &= 1+s+2\eta \\
  \delta &= 1+s+2\xi \\
  \sigma &= \scA{s}{\ell{m}}{\bar{a}\bar\omega} + \bar{a}^2\bar\omega^2 
   - 8\bar\omega^2 + p(2\alpha+\gamma-\delta) \\
& \mbox{}\hspace{0.75in}
   +\left(1+s-\frac{\gamma+\delta}2\right)\left(s+\frac{\gamma+\delta}2\right).
\nonumber
\end{align}
\end{subequations}

When looking for QNMs, in most cases we are looking for a {\em
  confluent Heun function} which is simultaneously a local solution at
$z=1$ and at $z=\infty$. In this case, it is convenient to choose
$\bar\zeta=\bar\zeta_+$ and $\xi=\xi_\minus$.  The choice of
$\xi_\minus$ means that the solution local to $z=1$, given by
Eq.~(\ref{eq:local_sol_z1a}), represents the desired boundary
condition of no waves emerging from the black hole.  The choice of
$\bar\zeta_+$ means that the solution local to $z=\infty$, given by
Eq.~(\ref{eq:local_sol_zinfa}), represents the desired boundary
condition of no waves coming in at infinity.  The choice of $\eta$ is
associated with $z=0$ and is not important yet.

The desired {\em confluent Heun functions} are readily found using
``Leaver's method''\cite{leaver-1985,leaver-1986} which consists of
removing the asymptotic behavior via $H(z)=z^{-\alpha}\bar{R}(z)$ and
rescaling the radial coordinate as $z\to\frac{z-1}{z}$ so the relevant
domain is $0\le z\le1$.  The solution is expanded as
$\bar{R}(z)=\sum_{n=0}^\infty{a_nz^n}$, resulting in a new three-term
recurrence relation for the coefficients $a_n$: 
\begin{subequations}
\begin{align}
\label{eq:Leaver_rad_2-term}
  0 &= a_0\beta_0 + a_1\alpha_0, \\ 
\label{eq:Leaver_rad_3-term}
  0 &= a_{n+1}\alpha_n + a_n\beta_n + a_{n-1}\gamma_n.
\end{align}
\end{subequations} 
See Ref.\cite{cook-zalutskiy-2014} for the values of coefficients
$\alpha_n$, $\beta_n$, and $\gamma_n$.  This new series has a radius
of convergence of one and more precisely, with
$r_n\equiv\frac{a_{n+1}}{a_n}$,
\begin{equation}\label{eqn:a_ratio_expansion}
   \lim_{n\to\infty} r_n = 1 + \frac{u_1}{\sqrt{n}}
      + \frac{u_2}{n} + \frac{u_3}{n^{\frac32}} +\cdots,
\end{equation}
and
\begin{equation}\label{eq:asymptotic_a}
  \lim_{n\to\infty}a_n \propto n^{u_2} e^{2u_1\sqrt{n}}.
\end{equation}
See Ref.\cite{cook-zalutskiy-2014} for additional details.  The two
parameters, $u_1$ and $u_2$ take on the values
\begin{subequations}
\begin{align}
  u_1 &=\pm\sqrt{-4p} = \pm\sqrt{-2i(\bar{r}_+-\bar{r}_\minus)\bar\omega}, \\
  u_2 &= -\frac14(8p-4\alpha+2\gamma+2\delta+3)
\end{align}
\end{subequations}
There will be two independent series solutions to the recurrence
relation, and they are distinguished by the two possible sign choices
for $u_1$.  The QNM solution we seek will be a {\em minimal} solution
we denote by $a_n\to f_n$, and we label the other set of coefficients
by $a_n\to g_n$.  A minimal solution has the property that
$\lim_{n\to\infty}{\frac{f_n}{g_n}}=0$.  For $u_1(\bar\omega)$, the
branch cut is along the negative imaginary axis and the minimal
solution corresponds to the sign choice that gives
$\rm{Re}(u_1(\bar\omega))<0$.  So long as $\rm{Re}(\bar\omega)\ne0$ or
$\rm{Im}(\bar\omega)>0$, this gives
$\lim_{n\to\infty}{\frac{f_n}{g_n}}\sim
e^{-4|\rm{Re}(u_1)|\sqrt{n}}=0$ and a minimal solution will
exist\footnote{Note that the discussion of the sign choice for $u_1$
  in Ref.~\cite{cook-zalutskiy-2014} contains an error.}.  The ratio
$r_n$ can be written as a continued fraction in terms of the
coefficients of the recurrence relation for the $a_n$.
\begin{subequations}\label{eq:Leaver_ratio}
\begin{align}\label{eq:Leaver_ratio_recur}
  r_n &= \frac{-\gamma_{n+1}}{\beta_{n+1}+ \alpha_{n+1}r_{n+1}}
  \qquad n=0,1,2,\ldots, \\
\label{eq:Leaver_ratio_CF}
  &= \frac{-\gamma_{n+1}}{\beta_{n+1}-} 
    \frac{\alpha_{n+1}\gamma_{n+2}}{\beta_{n+2}-}
    \frac{\alpha_{n+2}\gamma_{n+3}}{\beta_{n+3}-}\ldots.
\end{align}
\end{subequations}
The key property of this recurrence relation is given by Pincherle's
theorem\cite{Gautschi-1967}.
\begin{theorem}[Pincherle]\label{th:Pincherle}
  The continued fraction $r_0$ converges if and only
  if the recurrence relation Eq.~(\ref{eq:Leaver_rad_3-term}) possesses
  a minimal solution $a_n=f_n$, with $f_0\ne0$.  In case of
  convergence, moreover, one has $\frac{f_{n+1}}{f_n}=r_n$ with
  $n=0,1,2,\ldots$ provided $f_n\ne0$.
\end{theorem}
Since the continued fraction $r_0$ must converge to a specific value
given by Eq.~(\ref{eq:Leaver_rad_2-term}), the QNM solutions are found
at those frequencies $\bar\omega$ where $r_0$ does converge to this
required value.

But, it is very important to recognize that for $\bar\omega$ on the
NIA, $\rm{Re}(u_1(\bar\omega))=0$.  Since $u_1$ is purely imaginary on
the NIA, $\lim_{n\to\infty}\frac{f_n}{g_n}$ becomes
oscillatory\footnote{Note that in the ratio $\frac{f_n}{g_n}$, the
  factors of $n^{u_2}$ in Eq.~(\ref{eq:asymptotic_a}) cancel.} and
{\em a minimal solution cannot exist unless the infinite series
  solution terminates}.  Thus, any QNM solution on the NIA must be of
the form of a {\em confluent Heun polynomial}.  Furthermore, {\em the
  continued fraction cannot be used to determine the QNM frequencies
  $\bar\omega$ on the NIA}.

\subsection{Polynomial solutions}
\label{sec:Heun_polynomials}
Because confluent Heun polynomials are simultaneous Frobenius
solutions of all three singular points, and the radial Teukolsky
equation can be put in the form of the confluent Heun equation in
eight different ways depending of the choice of the parameters
$\{\bar\zeta,\xi,\eta\}$, the same confluent Heun polynomial can be
computed in several different ways.  Additional discussion of this can
be found in Ref.~\cite{cook-zalutskiy-2014} where the examples of
TTM${}_L$ and TTM${}_R$ polynomial solutions were examined in detail.
Here, we will describe how to find QNM solutions.

Consider the boundary condition at the event horizon, $z=1$.  We must
ensure that no waves propagate out from the horizon.  The two local
solutions at the horizon are
$\lim_{z\rightarrow1}R(z)\sim(z-1)^{-s-i\sigma_+}$, which represents
waves traveling into the horizon, and
$\lim_{z\rightarrow1}R(z)\sim(z-1)^{i\sigma_+}$, which represents
waves traveling out from the horizon.  If we choose $\xi=\xi_\minus$,
then the local confluent Heun solution of Eq.~(\ref{eq:local_sol_z1a})
will achieve the first local behavior and, in general, the second
behavior cannot be part of the series solution.  The necessary, but not
sufficient, condition for the series of this local solution to
terminate is for its second parameter $\alpha$ to be a non-positive
integer $\alpha=-q$.

Now, let us assume that we also satisfy the remaining necessary and
sufficient condition, $\Delta_{q+1}=0$.  We will postpone the details
of how we do that.  For now, we assume we have a polynomial solution
that should satisfy the boundary condition for a QNM at the event
horizon.  How do we know if this solution satisfies the necessary
boundary condition at infinity?  The key is to recall that a confluent
Heun polynomial solution is {\em simultaneously} a polynomial solution
at all three singular points.  Equations~(\ref{eq:local_sol_zinfa})
and (\ref{eq:local_sol_zinfb}) represent the two local solutions at
the outer boundary, $z=\infty$.  It is easy to see that the second
parameter of Eq.~(\ref{eq:local_sol_zinfa}) is also $\alpha$, so if we
use the same set of parameters (including $\alpha=-q$) as we used to
obtain the polynomial solution above, then we are guaranteed that
Eq.~(\ref{eq:local_sol_zinfa}) will yield the same polynomial
solutions as we obtained via Eq.~(\ref{eq:local_sol_z1a}) above which
has the desired QNM behavior at the event horizon.  Now, however,
using Eq.~(\ref{eq:local_sol_zinfa}) allows us to understand the
behavior of the solution at the outer boundary.  At the outer
boundary, $z=\infty$, we must ensure that no waves enter from
infinity.  The two local behaviors at infinity are
$\lim_{z\rightarrow\infty}R(z)\sim
z^{-1-2s+2i\bar\omega}e^{i(\bar{r}_+-\bar{r}_\minus)\bar\omega{z}}$,
which represents waves traveling out at infinity, and
$\lim_{z\rightarrow\infty}R(z)\sim
z^{-1-2i\bar\omega}e^{-i(\bar{r}_+-\bar{r}_\minus)\bar\omega{z}}$,
which represents waves traveling in from infinity.  Using
Eq.~(\ref{eq:local_sol_zinfa}), the parameter choice
$\bar\zeta=\bar\zeta_+$ will achieve the first local behavior and, in
general, the second behavior cannot be part of the series solution.

In order to satisfy the $\Delta_{q+1}=0$ condition, we must construct
the $(q+1)$-dimensional upper-left block of (\ref{eq:tridiag_cond}).
With our choices of $\xi=\xi_\minus$ and $\bar\zeta=\bar\zeta_+$, we
could use the coefficients from the recurrence relations associated
with either Eq.~(\ref{eq:local_sol_z1a}) or (\ref{eq:local_sol_zinfa})
to construct this matrix.  However, we have not yet fixed the choice
for $\eta$.

At the Cauchy horizon, $z=0$, the two local solutions are
$\lim_{z\rightarrow0}R(z)\sim z^{-s+i\sigma_\minus}$ and
$\lim_{z\rightarrow0}R(z)\sim z^{-i\sigma_\minus}$.  The first
behavior is associate with the local solution of
Eq.~(\ref{eq:local_sol_z0a}) if we choose $\eta=\eta_+$.  Notice that
the second parameter of Eq.~(\ref{eq:local_sol_z0a}) is again
$\alpha$, so the same choice of parameters as above will yield the
same polynomial solution satisfying the QNM boundary conditions {\em
  and} the first local behavior at $z=0$ mentioned above.  Thus, if we
choose the parameter set $\{\bar\zeta_+,\xi_\minus,\eta_+\}$ and let
$q=-\alpha$ be a non-negative integer, then the matrices constructed
from Eqs.~(\ref{eq:local_sol_z0a}), (\ref{eq:local_sol_z1a}), or
(\ref{eq:local_sol_zinfa}) will yield the {\em same confluent Heun
  polynomial solution} if the $\Delta_{q+1}=0$ condition is
satisfied\footnote{Satisfying the $\Delta_{q+1}=0$ for any one of the
  three matrices guarantees the condition will be satisfied by the
  other two matrices at the same value of $\bar{a}$}.

With the parameter set $\{\bar\zeta_+,\xi_\minus,\eta_+\}$, the
condition that $\alpha=-q$ can be rewritten as
\begin{equation}
   q+s+1\equiv N_+ = 2i\left[\frac{2\bar\omega-m\bar{a}}{\bar{r}_+-\bar{r}_\minus}+\bar\omega\right],
\end{equation}
where $N_+\ge s+1$ will be either an integer or a half-odd integer
depending on $s$.  This can be rewritten as a constraint on the values
of $\bar\omega$ that can potentially be associated with a confluent
Heun polynomial solution
\begin{equation}\label{eq:omega_plus}
\bar\omega=\bar\omega_+ \equiv
\frac{\bar{a}m-iN_+\sqrt{1-\bar{a}^2}}{2(1+\sqrt{1-\bar{a}^2})}.
\end{equation}

Now, consider the other possible choices for the parameter set
$\{\bar\zeta,\xi,\eta\}$.  Each pair of local solutions,
Eqs.~(\ref{eq:local_sol_0}), (\ref{eq:local_sol_1}), or
(\ref{eq:local_sol_inf}), are associated respectively with the
parameters $\eta$, $\xi$, and $\bar\zeta$, and each choice allows
one of the pairs of solutions to yield one of the two possible local
behaviors for $R(z)$.  For example, if we switch our choice for $\eta$
so that the parameter set is
$\{\bar\zeta_+,\xi_\minus,\eta_\minus\}$, we can consider solutions
with the same physical behaviors at all three singular points as
described above if we construct our coefficient matrix using the
recurrence relations associated with Eq.~(\ref{eq:local_sol_z0b}).  In
this case, the necessary condition for a polynomial solution is
$\alpha+1-\gamma=-q$, but in terms of the new parameter set this
yields exactly the same constraint that $\bar\omega =\bar\omega_+$.

Alternatively, if we use the parameter set
$\{\bar\zeta_+,\xi_\minus,\eta_\minus\}$ but construct our
coefficient matrix using the recurrence relations associated with
Eq.~(\ref{eq:local_sol_z0a}) (or vi Eq.~(\ref{eq:local_sol_z1a}) or
(\ref{eq:local_sol_zinfa})), then we are considering a second possible
set of QNM solutions where the local behavior at the Cauchy horizon
has changed.  In this case, the condition that $\alpha=-q$ can be
rewritten as
\begin{equation}
  q+1\equiv N_\minus = 4i\bar\omega,
\end{equation}
where $N_\minus\ge1$ is an integer, and we can rewrite this condition
as the constraint
\begin{equation}\label{eq:omega_minus}
\bar\omega=\bar\omega_\minus \equiv
-i\frac{N_\minus}4.
\end{equation}

In total, each of the six local solutions combines with two of the
eight possible choices for the parameter set
$\{\bar\zeta,\xi,\eta\}$ to correspond to one or the other of the
two possible polynomial QNM solutions distinguished by the two
possible local behaviors at the Cauchy horizon.  These possibilities
are summarized by
\begin{equation}\label{eq:QNM_boundary_set}
  -q= \left\{\begin{array}{cc}
            \alpha & (\bar\zeta_+,\xi_\minus,\eta_\pm), \\
            \alpha+1-\delta & (\bar\zeta_+,\xi_+,\eta_\pm), \\
            -\alpha+\gamma+\delta & (\bar\zeta_+,\xi_\minus,\eta_\pm), \\
            \alpha+1-\gamma & (\bar\zeta_+,\xi_\minus,\eta_\mp).
  \end{array}\right. 
\end{equation}
In each case, the upper sign choice for $\eta$ corresponds to a
solution with the local behavior of $R(z)\sim z^{-s+i\sigma_\minus}$
and the necessary condition that $\bar\omega=\bar\omega_+$, while the
lower sign choice corresponds to $R(z)\sim z^{-i\sigma_\minus}$ and
the necessary condition that $\bar\omega=\bar\omega_\minus$.  A
similar analysis for the TTM${}_L$ and TTM${}_R$ cases can be found in
Ref.\cite{cook-zalutskiy-2014}.

As foreshadowed in Sec.~\ref{sec:num_modes_on_NIA}, we note that
several of the sequences approaching the NIA (see
Table~\ref{tab:NIAdata2QNM}) do so at a frequency
$\bar\omega=\bar\omega_\minus$.  Moreover, all of the $m=0$ sequences
we have examined which begin at, terminate at, or become tangent to
the NIA satisfy either $\bar\omega=\bar\omega_+$ or
$\bar\omega=\bar\omega_\minus$.  For the $\bar\omega_+$ case, this
requires agreement with both $\bar\omega$ and $\bar{a}$.  More than
half of the modes in Table~\ref{tab:NIAdata2QNM}, and all of the modes
in Tables~\ref{tab:NIAdataLoops} and \ref{tab:NIAdataLoopsCount}
satisfy the constraint $\bar\omega=\bar\omega_+$.  In
Table~\ref{tab:NIAdata2QNM}, the value of $N_\pm$ is listed in the
last column.

The fact that so many $m=0$ sequences are approaching the NIA at
precisely the frequencies constrained by $\bar\omega_+$ and
$\bar\omega_\minus$ gives us confidence that we are approaching
polynomial solutions on the NIA.  However, these conditions are only
necessary, not sufficient, for the existence of polynomial solutions.
To guarantee that we have found confluent Heun polynomial solutions,
we must also solve the $\Delta_{q+1}=0$ condition.

\subsubsection{Solving the $\Delta_{q+1}=0$ condition}
\label{sec:Deltaq-cond}
Mode frequencies, $\bar\omega$, of the coupled Teukolsky equations
depend only on $s$, $m$, and $\bar{a}$, and effectively $\ell$ through
the choice of a particular eigenvalue $\scA{s}{\ell
  m}{\bar{a}\bar\omega}$ from our solutions to the angular Teukolsky
equation, Eq.~(\ref{eqn:swSF_DiffEqn}).  As we saw for general QNMs,
given fixed $s$, $\ell$, and $m$, the mode frequencies form sequences
parameterized by $\bar{a}$.  For the case of confluent Heun
polynomials, the modes are again parameterized solely by $\bar{a}$.
In this case, the $\Delta_{q+1}=0$ condition effectively replaces the
radial Teukolsky equation, but there is an additional constraint.

To construct the coefficient matrix for the $\Delta_{q+1}=0$
condition, we choose one of the local Heun solutions along with the
set of parameters $\{\bar\zeta,\xi,\eta\}$.  How does the additional
constraint affect our solutions?  For the case of TTMs, as discussed
in detail in Sec.~III.B of Ref.~\cite{cook-zalutskiy-2014}, this
condition only fixes an integer value for $q$ which fixes the size of
the coefficient matrix and the order of the polynomial solutions.  For
the TTMs, this condition does not directly constrain the mode
$\bar\omega$.  The $\Delta_{q+1}=0$ condition then yields an algebraic
equation which turns out to be the square of the Starobinsky
constant\cite{chandra-1984}.  This can be solved, together with the
angular Teukolsky equation, to yield the mode frequency $\bar\omega$
as a continuous function of $\bar{a}$.  See Fig.~24 of
Ref.~\cite{cook-zalutskiy-2014} for plots of the various $\ell=2$ and
$\ell=3$ TTM mode sequences.

When considering polynomial QNM solutions, we have seen that the
additional constraint fixes the mode frequency as a function of
$\bar{a}$ and a new parameter $N_\pm$ (see Eqs.~(\ref{eq:omega_plus})
and (\ref{eq:omega_minus})).  The $\Delta_{q+1}=0$ condition no longer
has the freedom to pick $\bar\omega$ to satisfy the condition for each
value of $\bar{a}$.  Instead, we must search for values of $\bar{a}$
at which the $\Delta_{q+1}=0$ condition is satisfied, subject to the
constraint that $\bar\omega=\bar\omega_\pm(N_\pm,\bar{a})$.
Typically, we can only expect this to yield isolated, discrete
solutions instead of a continuum.

While the $\Delta_{q+1}=0$ condition could be easily reduced to an
algebraic equation for the TTM cases\cite{cook-zalutskiy-2014}, this
becomes too difficult for QNMs because the size of the matrix can get
arbitrarily large depending on the value of $N_\pm$.  Therefore, we
construct the determinant of the coefficient matrix numerically and use
root-finding methods to locate the zeros of the determinant.  To obtain
a numerical value for the determinant, we must choose values for $s$,
$m$, ${}_sA_{\ell m}$, $N_\pm$ and $\bar{a}$.  To search for roots, we
first fix $s$, $m$, and $N_\pm$.  The determinant is then considered
a function of $\bar{a}$ with ${}_sA_{\ell m}$ chosen as follows for 
each value of $\bar{a}$.

${}_sA_{\ell m}$ is an eigenvalue of the spin-weighted spheroidal
differential equation given in Eq.~(\ref{eqn:swSF_DiffEqn}).  With $s$
and $m$ fixed, the value of the oblateness parameter
$c=\bar{a}\bar\omega$ is computed using the the current value of
$\bar{a}$ with $\bar\omega$ set via Eq.~(\ref{eq:omega_plus}) or
(\ref{eq:omega_minus}) as appropriate together with the fixed value of
$N_\pm$.  With all its parameters fixed, Eq.~(\ref{eqn:swSF_DiffEqn})
is solved using the spectral method described in
Ref.\cite{cook-zalutskiy-2014}.  The solution yields an ordered set of
eigenvalues that are labeled by $\ell$.  During any given search, we
fix which element of the set of eigenvalues to use.  For example, for
$m=0$ and $s=-2$, the first eigenvalue is labeled $\ell=2$, the second
by $\ell=3$, and so on.  The labeling of the eigenvalue is not
absolute.  Sequences of solutions, can cross depending on what
criteria are used to define the sequences and so the labeling of
solutions is not unique.  When comparing solutions, we must be
careful to compare the values of ${}_sA_{\ell m}$ and not simply
$(\ell,m)$ index pairs.

The matrix of coefficients can be quite large.  For the case of
$\bar\omega=\bar\omega_+$, the matrix is $(N_+-s)\times(N_+-s)$.  For
the case of $\bar\omega=\bar\omega_-$, the matrix is $N_\minus\times
N_\minus$.  Since $N_\pm$ can get large, we must be concerned about
the matrix being ill-conditioned.  We also find that the value of the
determinant can vary over many orders of magnitude as $\bar{a}$ varies
between $0$ and $1$.  To ensure that numerical problems are not
significant, we have used two methods for computing the determinant.
The first is simply to compute the determinant directly.  But, we also
use singular-value decomposition to decide when the matrix has a
vanishing determinant.

In singular value decomposition, a matrix ${\bf M}$ is decomposed
as ${\bf M}={\bf U}\cdot{\bf D}\cdot{\bf V}^\dagger$ where ${\bf U}$
and ${\bf V}$ are unitary matrices and ${\bf D}$ is a diagonal
matrix whose elements are real and non-negative.  A much better
behaved proxy for the determinant is to look for roots of
\begin{equation}
  \det({\bf U}\cdot{\bf V}^*)\min({\rm diag}({\bf D})).
\end{equation}
Using only the minimum diagonal element from ${\bf D}$ keeps the
function from varying so dramatically in magnitude.  It is important
to keep the determinant of ${\bf U}\cdot{\bf V}^*$.  Even though it
has unit magnitude, this term contains all of the phase information.
When we look for polynomial solutions with $\bar\omega=\bar\omega_+$
and let $m\ne0$, the determinant yields a complex number.  When $m=0$,
or when we consider $\bar\omega=\bar\omega_\minus$, the determinant
will be real. In either case, the phase information is important so
that we have a smooth function across the zeros.

For the case of polynomial solutions with $\bar\omega=\bar\omega_+$,
we choose the parameter set $\{\bar\zeta_+,\xi_\minus,\eta_+\}$ and
use the coefficients from the recurrence relation associated with
Eq.~(\ref{eq:local_sol_z0a}) to build the matrix, explicitly replacing
$\bar\omega$ with Eq.~(\ref{eq:omega_plus}).  We have carried out an
extensive search for polynomial solutions (roots of the determinant or
its proxy) for gravitational QNMs ($s=-2$).

\begin{figure}
\includegraphics[width=\linewidth,clip]{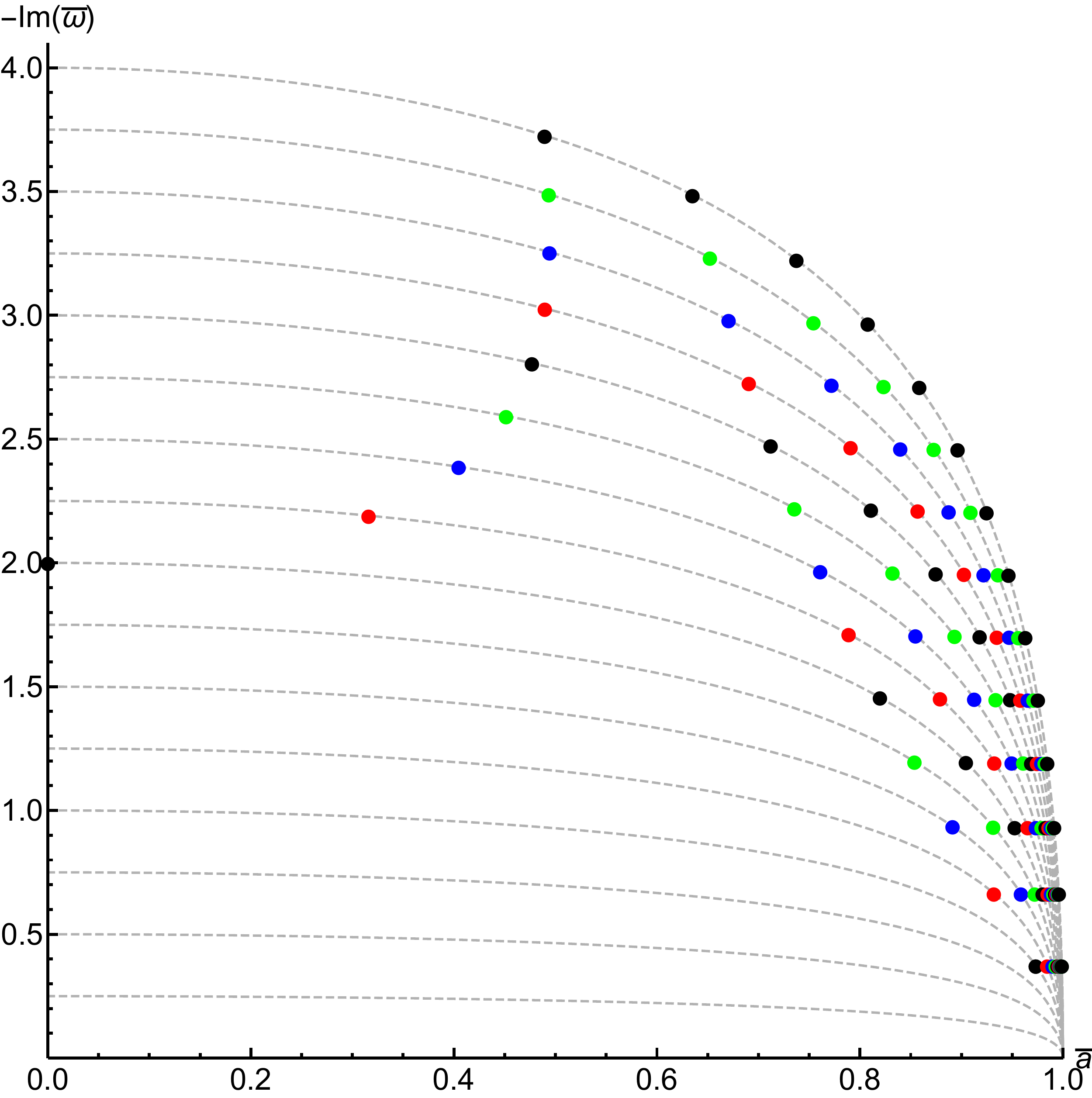}
\caption{\label{fig:l2omegapN4-16} Polynomial modes with
  $\bar\omega=\bar\omega_+$.  The dashed gray lines are the possible
  values of $\bar\omega_+(N_+,\bar{a})$ for $1\le N_+\le16$.  The
  marker denote the discrete points where the $\Delta_{q+1}=0$
  condition is satisfied.}
\end{figure}

First, we have found no evidence for polynomial solutions with
$m\ne0$.  For all cases examined, the determinant moves around the
complex plain as we vary $\bar{a}$, but never crosses the origin.
However, for $m=0$ we find what seems to be a countably infinite set
of polynomial solutions.  As expected, these solutions are not
continuous, but occur at discrete values of $\bar{a}$.  
Figure~\ref{fig:l2omegapN4-16} displays the solutions for $4\le
N_+\le16$.  Notice that this figure differs from previous figures in
that the horizontal axis measures $\bar{a}$ instead of ${\rm
  Re}(\bar\omega)$ (${\rm Re}(\bar\omega)=0$ for these solutions).  In
this figure, we have also plotted the constraint
$\bar\omega=\bar\omega_+(N_+,\bar{a})$ as dashed gray lines for $1\le
N_+\le16$.  The circular marks on each line denote the particular
values of $\bar{a}$ at which the $\Delta_{q+1}=0$ condition is
satisfied, marking a valid polynomial solution.
We find no
solutions for $N_+<4$.  There is one solution for $N_+=4$, two for
$N_+=5$, and so on to 4 solutions at $N_+=7$.  At $N_+=8$, we find not
5, but 6 solutions. The jump in the number of solutions corresponds to
the existence of a root for $\bar{a}=0$.  This is the mode
corresponding to the algebraically special solution with
$\bar\omega=\bar\Omega_2$.  Between $8\le N_+\le133$ we find $N_+-2$
solutions.  We have not yet searched beyond $N_+=133$.

\begin{figure}
\begin{tabular}{cc}
\includegraphics[width=0.5\linewidth,clip]{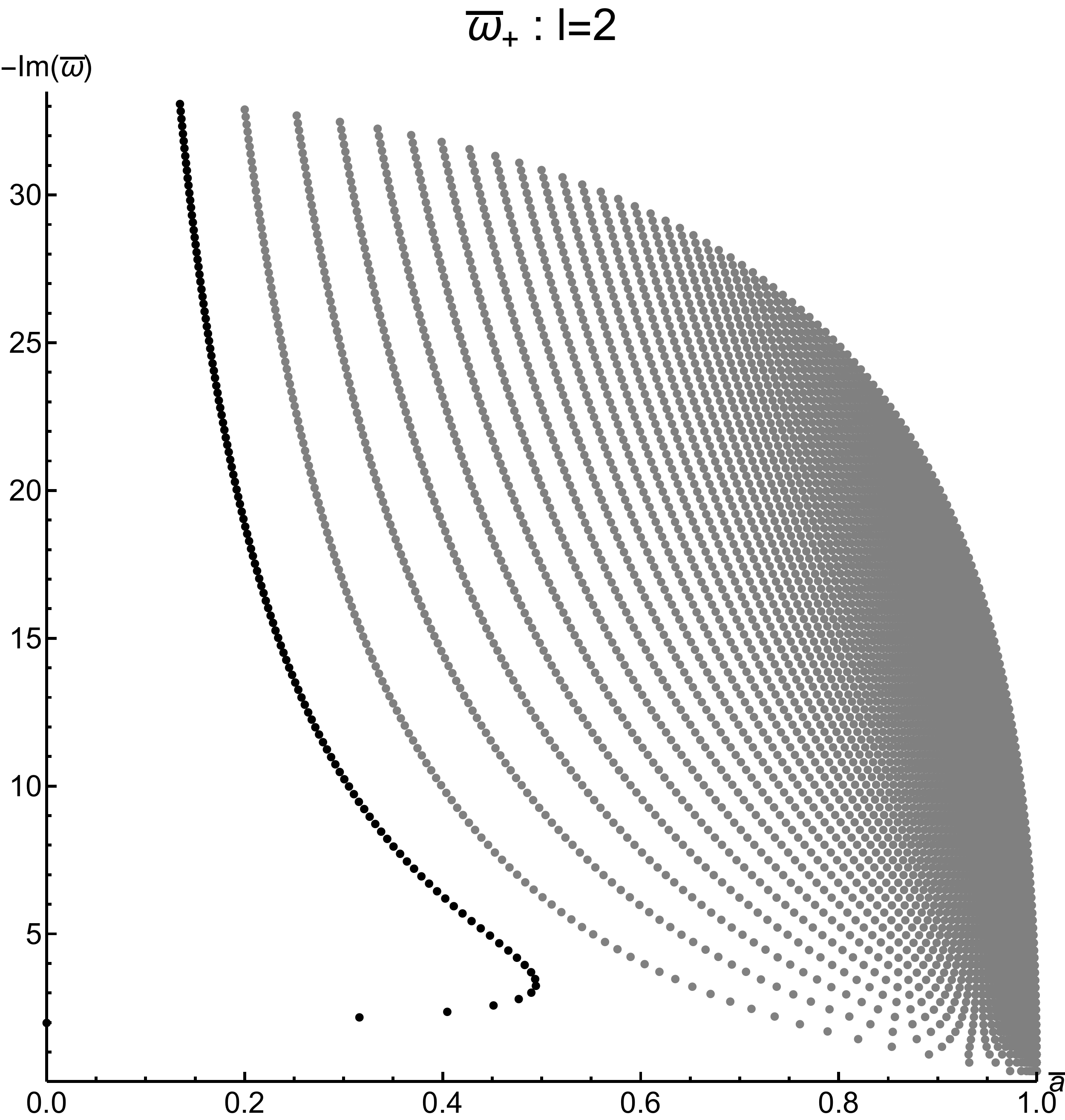} &
\includegraphics[width=0.5\linewidth,clip]{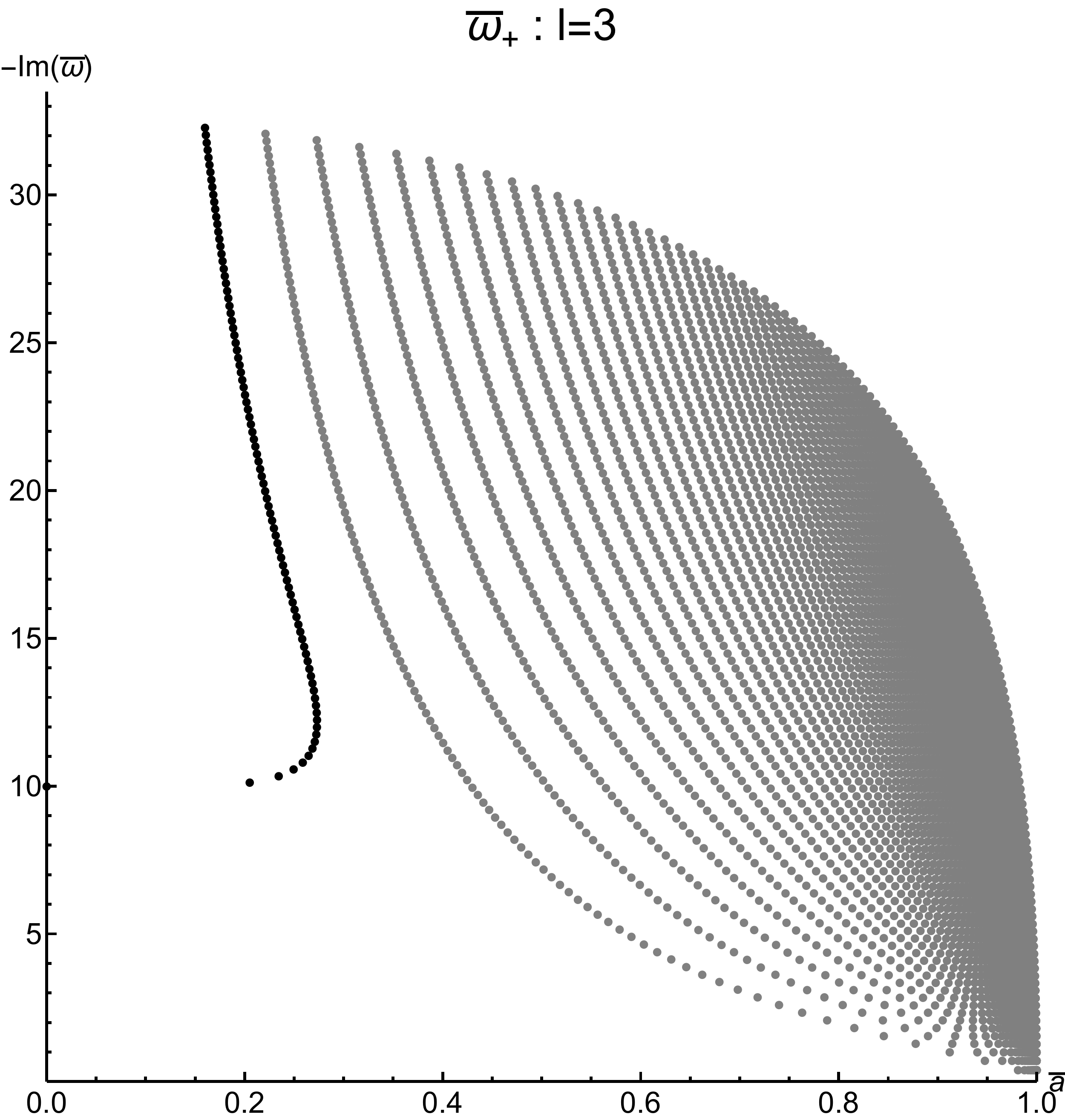} \\
\includegraphics[width=0.5\linewidth,clip]{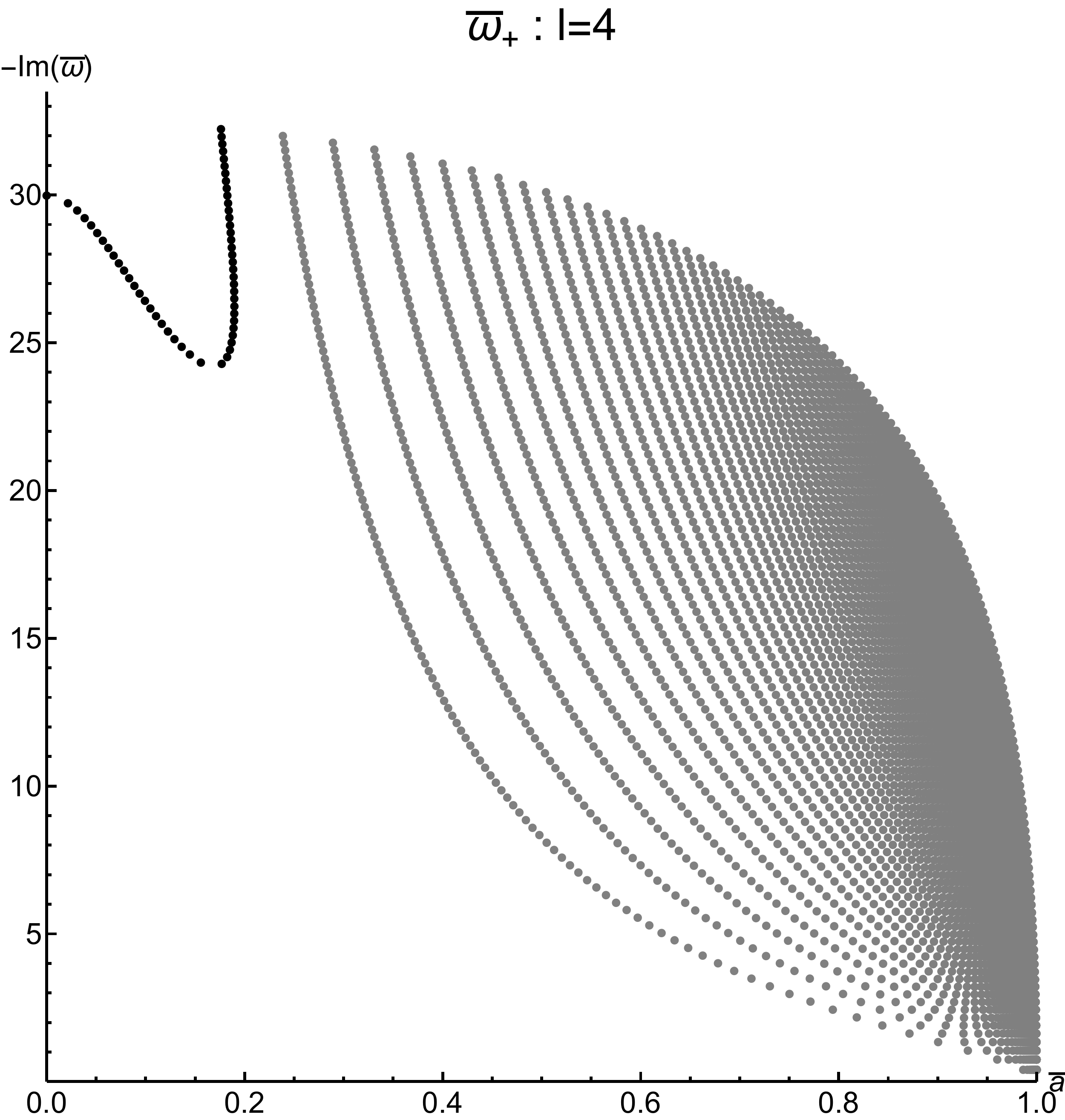} &
\end{tabular}
\caption{\label{fig:l234omegapN4-133} Polynomial modes with
  $\bar\omega=\bar\omega_+$.  The upper-left panel shows results for
  $\ell=2$, the upper-right panel shows $\ell=3$, and the lower-left
  $\ell=4$.  In each panel, the black dots are a subset of solutions
  that are shown in Sec.~\ref{sec:Generic_Anomalous_Miraculous} to be
  polynomial solutions that are {\em simultaneously QNMs and
    TTM${}_L$s}.  The gray dots are a subset of solutions that are
  shown to be {\em neither} QNM nor TTM.}
\end{figure}

We have also carried out similar searches for $\ell=3$ and $\ell=4$.
For $\ell=3$ the first solution is found for $N_+=5$, while for
$\ell=4$, the first solution is found for $N_+=6$.
Figure~\ref{fig:l234omegapN4-133} shows individual plots for $\ell=2$,
$3$, and $4$.  Each plot includes all of the solutions we have found
up to $N_+=133$.\footnote{We searched up to $N_+=132$ for $\ell=3$, and
  $N_+=131$ for $\ell=4$.}  We have omitted the lines denoting
$\bar\omega_+$ for clarity.

For the case of polynomial solutions with
$\bar\omega=\bar\omega_\minus$, we choose the parameter set
$\{\bar\zeta_+,\xi_\minus,\eta_-\}$ and use the coefficients from the
recurrence relation associated with Eq.~(\ref{eq:local_sol_z0a}) to
build the matrix, explicitly replacing $\bar\omega$ with
Eq.~(\ref{eq:omega_minus}).  We have again carried out an extensive
search for polynomial solutions for gravitational QNMs ($s=-2$).

\begin{figure}
\begin{tabular}{cc}
\includegraphics[width=0.5\linewidth,clip]{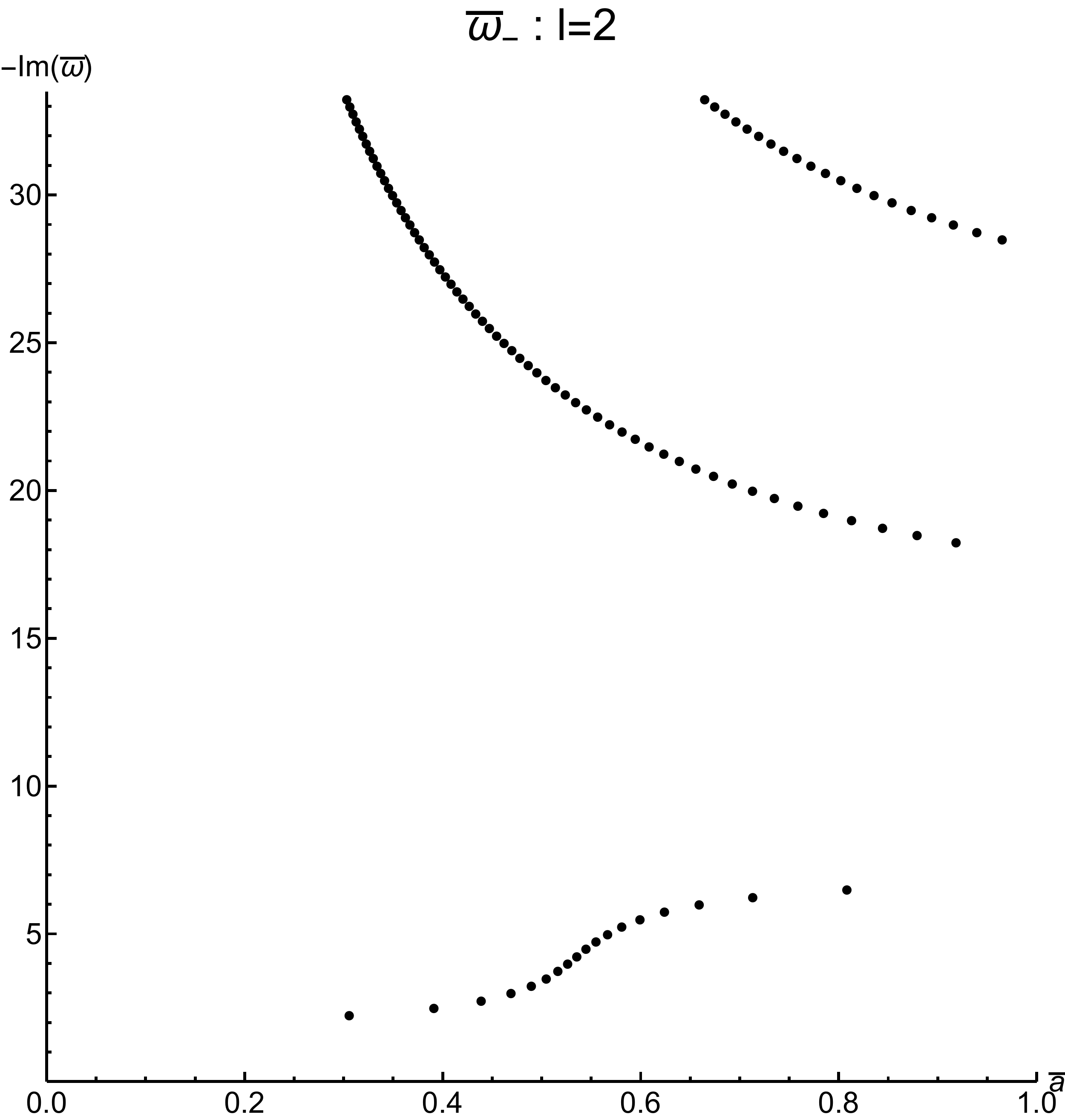} &
\includegraphics[width=0.5\linewidth,clip]{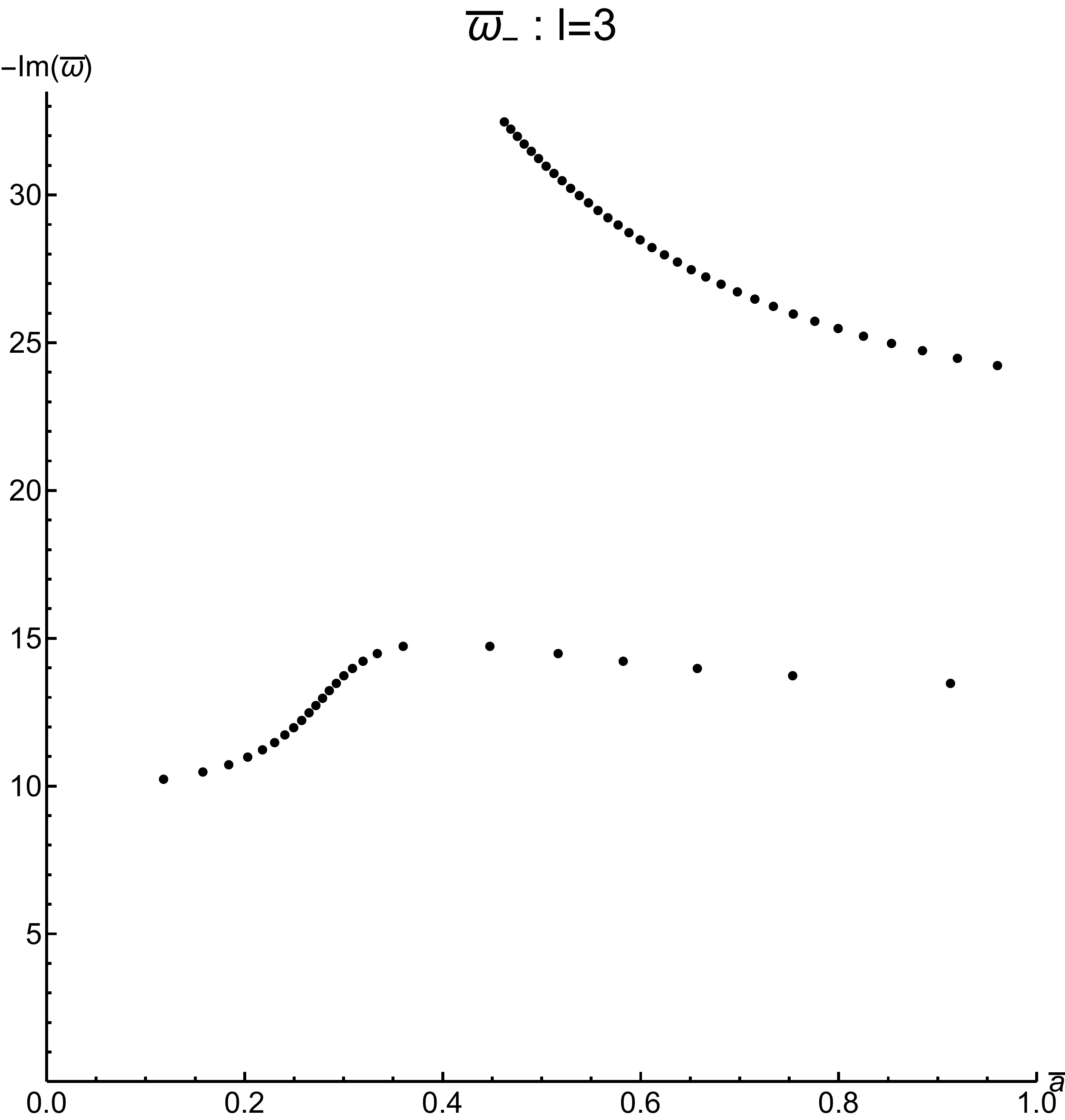} \\
\includegraphics[width=0.5\linewidth,clip]{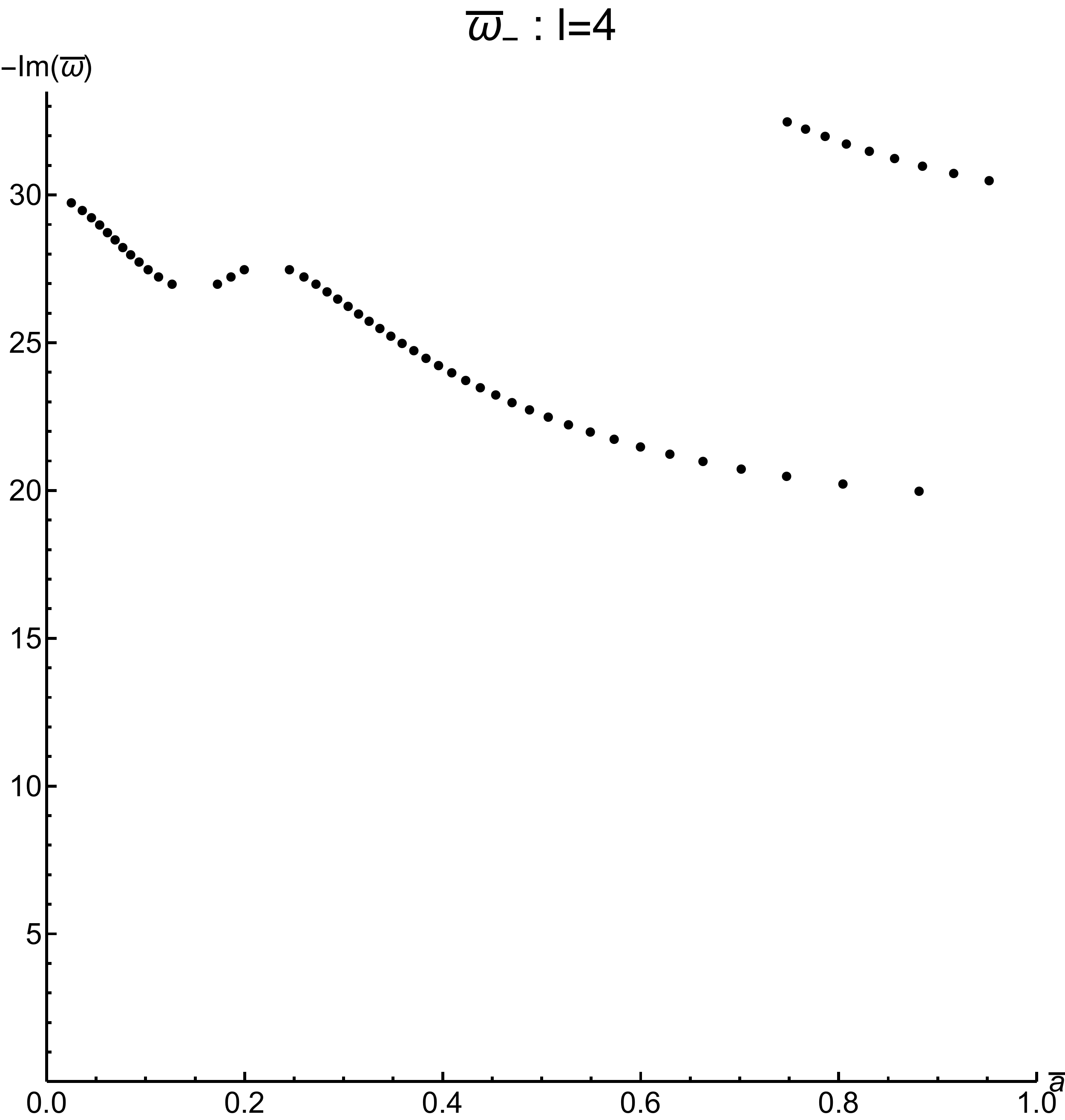} &
\end{tabular}
\caption{\label{fig:l234omegamN9-133} Polynomial modes with
  $\bar\omega=\bar\omega_\minus$.  The upper-left panel shows results
  for $\ell=2$, the upper-right panel shows $\ell=3$, and the
  lower-left $\ell=4$.  As shown in
  Sec.~\ref{sec:Generic_Anomalous_Miraculous}, all of these polynomial
  solutions are QNMs.}
\end{figure}

As before, we have found no evidence of polynomial solutions with
$m\ne0$.  For $m=0$ we do find solutions, but they appear to be less
numerous.  The solutions for $\ell=2$ are seen in the upper-left plot
of Fig.~\ref{fig:l234omegamN9-133}.  We find no roots for
$N_\minus<9$, and for $9\le N_\minus\le26$ we find a single polynomial
solution for each value of $N_\minus$.  Between $27\le N_\minus\le72$
there appear to be no roots, but for $N_\minus\ge73$ there seems to be
at least one polynomial solution for each $N_\minus$.  Starting at
$N_\minus=114$, we find a second polynomial solution for each
$N_\minus$.  While we have not extended our search beyond
$N_\minus=133$, it seem likely that the solutions persist indefinitely
as $N_\minus$ increases, and it would not be surprising to find even
more solutions for each $N_\minus$ as we move to larger $N_\minus$.
Figure~\ref{fig:l234omegamN9-133} also shows the result for $\ell=3$
and $4$.  In these cases, the counting of the number of roots for 
each $N_\minus$ is more complicated, but the behavior is clear from 
the figure.

Many, but not all of the polynomial solutions we have found correspond
to specific modes on the QNM sequences that exist arbitrarily close to
the NIA that we outlined in Sec.~\ref{sec:num_modes_on_NIA}.
Table~\ref{tab:NIAdata2QNM} displays the QNM solutions we have found
closest to the NIA for sequences that either appear to terminate on
the NIA or emerge from the NIA.  We noted that many of these seem to
coincide with the $\bar\omega=\bar\omega_\minus$ constraint.  Indeed,
every $\ell=2$, $m=0$ sequence that is an overtone multiplet has one
segment that corresponds precisely to one of the $\bar\omega_\minus$
polynomial solutions in the upper-left plot of
Fig.~\ref{fig:l234omegamN9-133} with $9\le N_\minus\le26$.  The fact
that we find no more polynomial solutions until $N_\minus=73$ is why
we are so confident that the $\ell=2$, $m=0$ QNM sequences we have
computed with $n>26$ are not overtone multiplets.  The beginning of
this set of polynomial solutions also heralds the beginning of this
set of overtone multiplets.  However, the reason is a little more
subtle, owing to the degeneracy between the $\bar\omega_+$ and
$\bar\omega_\minus$ when $N_+=N_\minus=8$ and $\bar{a}=0$.\footnote{In
  fact, as we will see in Sec~\ref{sec:Generic_Anomalous_Miraculous},
  this solution is ``anomalous'' at both $z=0$ and $z=1$}

The remaining entries from Table~\ref{tab:NIAdata2QNM} correspond
precisely to certain polynomial solutions in the upper-left plot of
Fig.~\ref{fig:l234omegapN4-133}.  The left-most set of solutions,
shown as black dots, starts with the the algebraically special
solution at $\bar\omega=-2i$ and $\bar{a}=0$.  The next 5 solutions
move to the right and up slightly, then the remaining solutions move
back leftward and up rapidly in the plot.  After the algebraically
special solution, the next 24 polynomial solutions correspond to the
remaining entries from Table~\ref{tab:NIAdata2QNM}.  These
$\bar\omega_+$ solutions, and the 18 $\bar\omega_\minus$ polynomial
solutions discussed above, correspond to all of the $\ell=2$, $m=0$
QNM sequences we have computed which either terminate at or emerge from
the NIA.

However, there are still several thousand instances where we have
found looping solutions that encounter the NIA at a point of tangency.
We have computed these for $\ell=2$, $3$, and $4$ (see
Figs.~\ref{fig:m0l2n09-26}, \ref{fig:m0l3n18-31}, and
\ref{fig:m0l4n25-31}; Tables~\ref{tab:NIAdataLoops} and
\ref{tab:NIAdataLoopsCount}).  While we cannot compute a solution at
the point of tangency using Leaver's method, we can use quadratic
interpolation to estimate the values of $\bar\omega$ and $\bar{a}$ at
the point of tangency.  In every case, the interpolated result is in
precise agreement with one of the $\bar\omega_+$ solutions shown as
gray dots in Fig.~\ref{fig:l234omegapN4-133}.

All three plots in Fig.~\ref{fig:l234omegapN4-133} contain a
``left-most set'' of solutions, shown as black dots, that starts at
$\bar{a}=0$ with $\bar\omega=\bar\Omega_\ell$.  They also contain a
large number of additional polynomial solutions that are grouped to
the right of this set.  Referring to Fig.~\ref{fig:l2omegapN4-16}, we
see that this set of solutions breaks into roughly horizontal groups
extending toward $\bar{a}=1$, with a spacing in $\bar\omega$ of
roughly $i/4$ between each grouping.  The first group starts with
$N_+=4$ and ${\rm Im}(\bar\omega)\sim-i/2$, the second with $N_+=5$
and ${\rm Im}(\bar\omega)\sim-3i/4$, and so on.  Through $N_+=133$ we
have found that once a horizontal grouping starts, it includes a
member for every subsequent value of $N_+$.  In
Fig.~\ref{fig:l2omegapN4-16}, the gray dashed lines represent
$\bar\omega_+(N_+,\bar{a})$ for $1\le N_+\le16$.  It is clear that as
we let $N_+$ increase, all of these curves will extend down to ${\rm
  Im}(\bar\omega)=0$ at $\bar{a}=1$.  It seems that each roughly
horizontal grouping of solutions forms a countably infinite set,
becoming arbitrarily dense as $\bar{a}\rightarrow1$.  Furthermore, the
plots in Fig.~\ref{fig:l234omegapN4-133} suggest that there is a
countably infinite number of such roughly horizontal groupings of
solutions, each itself countably infinite, for each value of $\ell$.
\begin{figure}
\includegraphics[width=\linewidth,clip]{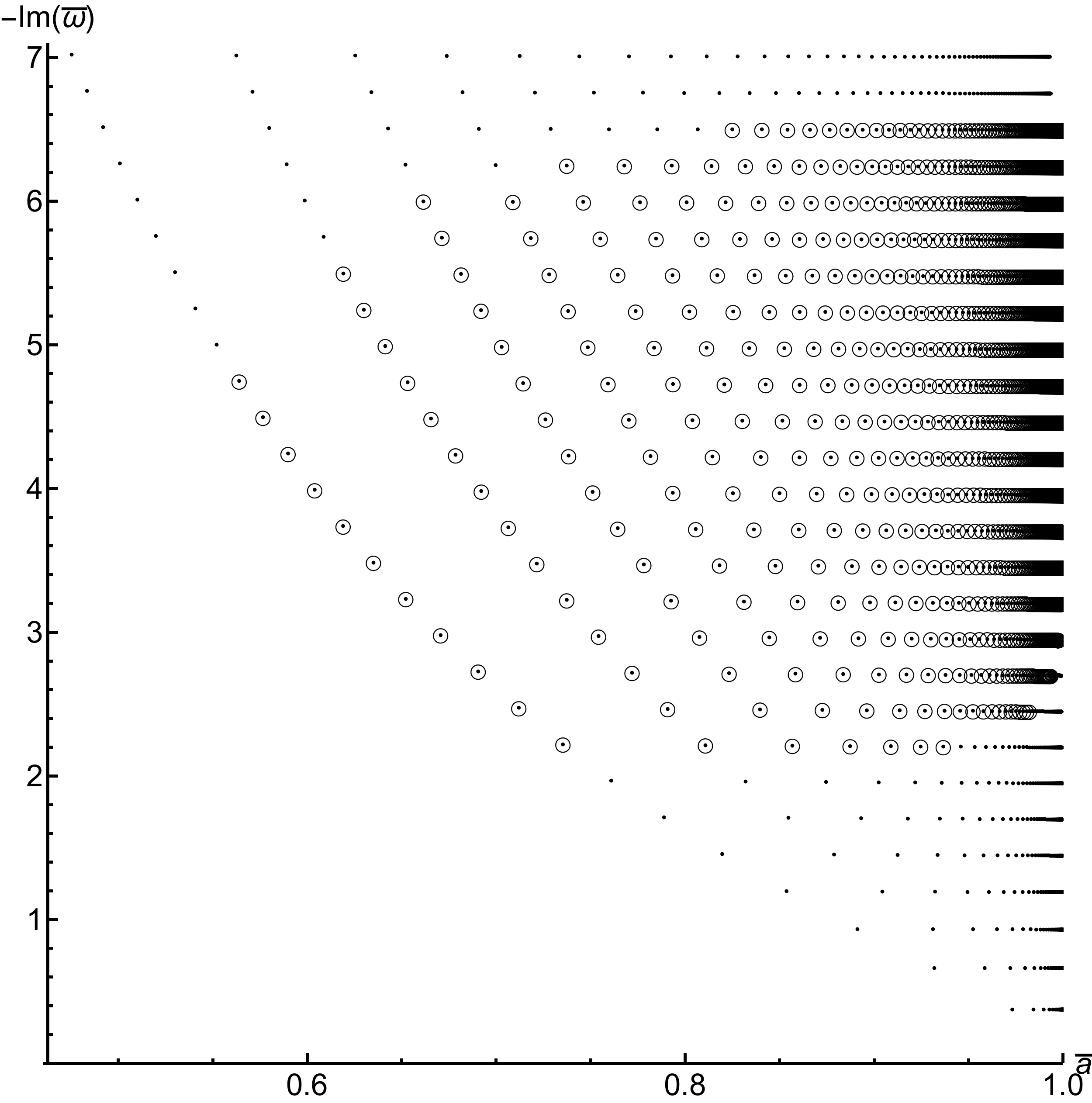}
\caption{\label{fig:l2omegap-cmp} Comparison of $\ell=2$ polynomial
  modes with QNMs interpolated to the NIA.  The open circles represent
  points where QNM sequences become tangent to the NIA.  The points of
  tangency are determined by interpolation.  The small solid dots are
  polynomial modes obeying $\bar\omega=\bar\omega_+$.  Any mismatch in
  the centering of dots and circles is a graphing artifact.  The
  maximum deviation has an absolute error of $\mathcal{O}(10^{-7})$.
  As shown in Sec.~\ref{sec:Generic_Anomalous_Miraculous}, {\em none}
  of these polynomial solutions are QNMs or TTMs.}
\end{figure}

We find that for each looping QNM sequence, each point of tangency
corresponds precisely to a point in one of the roughly horizontal
groupings.  Figure~\ref{fig:l2omegap-cmp} shows, for $\ell=2$, the
correspondence between all of the interpolated points of tangency and
the polynomial solutions.  For example, the seven points of tangency
in the $\{2,0,9_1\}$ sequence (see Fig.~\ref{fig:m0l2n09-26} and
Table~\ref{tab:NIAdataLoops}) correspond to the grouping that begins
with $N_+=9$.  The 20 points in the $\{2,0,10_1\}$ sequence correspond
to the next grouping beginning with $N_+=10$.  Since there are a
finite number of points of tangency on each sequence, and an infinite
number of polynomial solutions in each roughly horizontal grouping,
only a small number of these polynomial solutions correspond to points
along a QNM sequence.  Similar behavior is seen for $\ell=3$ and $4$.

\subsection{Generic, anomalous, and miraculous solutions}
\label{sec:Generic_Anomalous_Miraculous}
We have found two families ($\bar\omega_+$ and $\bar\omega_\minus$) of
confluent Heun polynomial solutions on the NIA.  The methods we have
used have been tailored to finding QNMs.  But we must be careful to
examine, for each regular singular point, the behavior of the roots of
the indicial equation to be certain that the solution actually
represents a QNM.

Our methods for finding solutions, both Leaver's method and the method
of confluent Heun polynomials, are based on power-series methods.  When
we consider the regular singular points at $z=0$ and $z=1$, standard
Frobenius theory tells us that, when the roots of the indicial equation
differ by an integer, only the local series solution corresponding to
the larger root is guaranteed to exist.  The second local solution
will usually include a $\log$ term.  However, it is possible for the
coefficient multiplying this $\log$ term to vanish.  Subtleties can 
also occur at irregular-singular points, but that is not important
in this work.

As seen in Eq.~(\ref{eq:local_sol_0}), the roots of the indicial
equation for $z=0$ are $\{0,1-\gamma\}$, and so we must be careful
whenever $\gamma$ is an integer.  The behavior of solutions at
the Cauchy horizon is associated with the $\eta$ parameter, so it
is not surprising that we can express $\gamma$ as
\begin{equation}
  \gamma=\gamma_\pm \equiv 1 + s + 2\eta_\pm = 1\pm s\mp2i\sigma_\minus.
\end{equation}
In general, $\sigma_\minus$ is complex and $\gamma$ cannot be an
integer.  In fact, so long as $2i\sigma_\minus$ is not an integer (or
a half-odd integer if $s$ is a half odd integer), then $\gamma$ cannot
be an integer.  This is the {\em generic} case.  If $\bar{a}=0$, then
$\sigma_\minus=0$ and $\gamma$ is an integer (or zero) when $s$ is an
integer.  This case has been examined by Maassen van den
Brink\cite{van_den_brink-2000}.  Furthermore, if $m=0$ and
$\bar\omega$ is purely imaginary, then $2i\sigma_\minus$ will be an
integer at that frequency for certain discrete values of $\bar{a}$.
And if $m\ne0$, special combinations of $\bar\omega$ and $\bar{a}$ can
allow $2i\sigma_\minus$ to be an integer.  However, since the behavior
is the solution at the Cauchy horizon is not relevant to determining
if a mode is a QNM, we will not investigate this further.

The behavior of the solution at the event horizon, $z=1$, is critical
for determining if a mode is a QNM.  As seen in
Eq.~(\ref{eq:local_sol_1}), the roots of the indicial equation for
$z=1$ are $\{0,1-\delta\}$, and so we must be careful whenever
$\delta$ is an integer.  The behavior of solutions at the event
horizon is associated with the $\xi$ parameter, so it is not
surprising that we can express $\delta$ as
\begin{equation}
  \delta=\delta_\pm \equiv 1 + s + 2\xi_\pm = 1\pm s\pm2i\sigma_+.
\end{equation}
In this case, solutions are {\em generic} so long as $2i\sigma_+$ is
not an integer (or a half-odd integer if $s$ is a half odd integer).
From Eq.~(\ref{eqn:sigmapm-def}), we see that $\sigma_+$ depends
on $\bar\omega$ and $\bar{a}$.  Rewriting this, we can find the
frequencies that yield non-generic solutions:
\begin{equation}
\bar\omega=
\frac{\bar{a}m-i(2i\sigma_+)\sqrt{1-\bar{a}^2}}{2(1+\sqrt{1-\bar{a}^2})}.
\end{equation}
Comparing this to Eq.~(\ref{eq:omega_plus}), we see that the equations
are identical with $2i\sigma_+=N_+$.  So, {\em any confluent Heun
  polynomial solution obeying the constraint that
  $\bar\omega=\bar\omega_+$ is non-generic at the event horizon} for
any choice of $s$.  Thus, for the majority of the gravitational,
$s=-2$, polynomial modes on the NIA which we have found, we cannot
immediately conclude that they are QNMs.  

Before we explore this further, we should consider the case of
polynomial solutions obeying the constraint that
$\bar\omega=\bar\omega_\minus$.  In this case, if $\bar{a}=0$ and if
$s$ is an integer, then the mode is non-generic.  Again, this case has
been examined by Maassen van den Brink\cite{van_den_brink-2000}.  If
$\bar{a}\ne0$ but $m=0$, then a mode can be non-generic if
$\frac{N_\minus(1+\sqrt{1-\bar{a}^2})}{2\sqrt{1-\bar{a}^2}}$ is an
integer when $s$ is an integer (or a half-odd integer when $s$ is a
half-odd integer).  All of the confluent Heun polynomial solutions on
the NIA that we have computed so far that obey the
$\bar\omega=\bar\omega_\minus$ constraint {\em do not satisfy this
  condition}.  While each such mode must be tested individually, so
far, all have proven to be generic and we can conclude that {\em all
  of the $\bar\omega=\bar\omega_\minus$ polynomial solutions} (see
  Fig.~\ref{fig:l234omegamN9-133}) {\em are QNMs}.

\subsubsection{Non-generic modes}
\label{lab:nongeneric}
Let us now consider the behavior at the event horizon ($z=1$) of
non-generic modes on the NIA.  All of these modes have
$\bar\omega=\bar\omega_+$ and are characterized by the parameter
$N_+=2i\sigma_+$ (see Fig.~\ref{fig:l234omegapN4-133}).  The two
possible local behaviors of these non-generic modes on the NIA are
either $(z-1)^{-s-N_+/2}$, which we would naively associate with modes
propagating into the horizon, and $(z-1)^{N_+/2}$, which we would
expect to be modes propagating out from the horizon.  Frobenius theory
tells us that two possibilities exist for non-generic solutions.  The
simplest case is that the only series solution (and it is a truncated
series for our polynomial solutions) has leading behavior associated
with the largest exponent.  For $N_+>-s$, this is $(z-1)^{N_+/2}$.
The other series solution (and it need not be truncated) will have
leading behavior $(z-1)^{-s-N_+/2}$, but will also include a term that
is proportional to $\ln(z-1)$ multiplied by the first solution.
Maassen van den Brink labels this non-generic case as
``anomalous''\cite{van_den_brink-2000} and argues that the second
solution including the $\ln(z-1)$ term cannot represent a mode
propagating into the horizon.  Surprisingly, he finds that the series
solution, which appears to correspond to a TTM${}_L$, simultaneously
represents a QNM and a TTM${}_L$.  In Fig.~\ref{fig:l234omegapN4-133},
all ``anomalous'' solutions are displayed with a black dot.  We will
return to discuss this in more detail below.

The second possible non-generic behavior occurs when the coefficient
multiplying the $\ln(z-1)$ happens to vanish.  Maassen van den Brink
labels this {\em doubly} non-generic case as ``miraculous''.  In this
case, our polynomial solutions can have leading behavior
$(z-1)^{-s-N_+/2}$.  However, unlike the generic case, this does not
guarantee that our solution is a QNM.  Because a term like
$(z-1)^{N_+/2}$ can be part of the same solution, the polynomial
solution may be a linear combination of ingoing and outgoing modes at
the horizon.  If so, then our solution is neither a QNM nor a
TTM${}_L$.  In Fig.~\ref{fig:l234omegapN4-133}, all ``miraculous''
solutions are displayed with a gray dot.  These behaviors are, to say
the least, counter-intuitive.  To understand the implications of these
two possibilities, it is useful to view them from the perspective of
scattering theory.

Begin with a free mode propagating into the horizon at $z=1$ and then
construct an $n^{\rm th}$-order Born approximation keeping
$\bar\omega$ and $\bar{a}$ unspecified.  Let us denote the Born series
solution for this {\em outgoing} mode as
$R_\minus(z,\bar\omega,\bar{a})$.  Since we are only concerned with
the behavior near the horizon, we need only construct the Born
approximation locally.  This can be simply constructed using the
recurrence relation, Eq.~(\ref{eq:local_a_3term}), for the local
solution (\ref{eq:local_sol_z1a}) with the parameter choice
$\{\bar\zeta_+,\xi_\minus,\eta_+\}$.  For example, the $3^{\rm
  rd}$-order local Born approximation would be
\begin{align}\label{eqn:3rd-born}
R_\minus^{(3)}(z,\bar\omega,\bar{a}) \sim (z-&1)^{-s-i\sigma_+}\Big[1 
- \frac{g_0}{f_0}(1-z) \\
&\mbox{}+ \frac{g_0g_1-f_0h_1}{f_0f_1}(1-z)^2 \nonumber \\
&\mbox{}- \frac{g_0g_1g_2-f_0h_1g_2-g_0f_1h_2}{f_0f_1f_2}(1-z)^3\Big]
 \nonumber
\end{align}
The denominator of the $n^{\rm th}$-order Born approximation is
$\prod_{i=0}^{n-1}f_i$, and we must consider when this can vanish.
From Eq.~(\ref{eq:local_a_f}), we see that $f^{(a)}_k$ will vanish if
$k=-\gamma$.  In Eq.~(\ref{eq:local_a_sol_series}), $\gamma$ is the
third parameter, but in Eq.~(\ref{eq:local_sol_z1a}) the third
parameter is $\delta$.  For our parameter choice,
$\delta=1-s-2i\sigma_+$, with $\sigma_+$ is a function of $\bar\omega$
and $\bar{a}$.  Recall that any time $\bar\omega=\bar\omega_+$, we
find that $2i\sigma_+=N_+$ and $\delta$ is an integer.  In summary,
whenever $\bar\omega=\bar\omega_+$ with $N_+>-s$, $\delta$ is a
non-positive integer and $f^{(a)}_{\minus\delta}=0$ causing the
$(1-\delta)^{\rm th}$-order and higher terms in the Born approximation
to have a vanishing denominator.

As we saw above, Frobenius theory tells us that our solution is
non-generic at $z=1$ if $\delta$ is an integer.  Moreover,
$(z-1)^{\xi_\minus}$ corresponds to the smaller root of the indicial
equation when $\delta$ is a negative integer.  So, our non-generic
solutions correspond to the case where the denominator vanishes in the
$(1-\delta)^{\rm th}$-order and higher terms in the Born series.

Of course, there are two types of non-generic behavior.  We first
consider the ``anomalous'' case.  As Maassen van den Brink has
shown\cite{van_den_brink-2000}, the vanishing denominator means the
mode propagating into the black hole scatters so strongly off of the
potential tail as it approaches the event horizon that the normally
dominant behavior of the outgoing mode is overwhelmed and the outgoing
mode has exactly the same local behavior as the incoming mode.  More
precisely, in the {\em anomalous case}, the mode propagating into the
black holes behaves like
\begin{equation}\label{eq:anom-rescale}
  {\bf\tt R}^{(A)}_\minus(z,\bar{a}) \sim \lim_{\bar\omega\rightarrow\bar\omega_+}
  f^{(a)}_{-\delta}R_\minus(z,\bar\omega,\bar{a}),
\end{equation}

The Born series for the mode traveling out of the black hole can be
constructed in a similar way using the parameter choice
$\{\bar\zeta_+,\xi_+,\eta_+\}$.  Let us denote the Born series
solution for this {\em incoming} mode as $R_+(z,\bar\omega,\bar{a})$.
For the anomalous solutions, we find that ${\bf\tt
  R}^{(A)}_\minus(z,\bar{a})\propto R_+(z,\bar\omega_+,\bar{a})$.
Hence, the {\em anomalous solutions are simultaneously QNMs and
  TTM${}_L$s}.

For the ``miraculous'' case, in addition to the denominator of the
$n^{\rm th}$-order Born approximation vanishing, we find that the
numerator also vanishes.  The value of the coefficient of the $n^{\rm
  th}$-order term in the Born series is obtained by taking the limit
as $\bar\omega\rightarrow\bar\omega_+$ and
$\bar{a}\rightarrow\bar{a}_p$, where $\bar{a}_p$ is the angular
momentum parameter of the polynomial mode.  The result for the $n^{\rm
  th}$ coefficient depends on the order in which the limits are taken,
however the conclusion is the same regardless of the ordering.  By
definition\cite{van_den_brink-2000}, in the {\em miraculous case}, the
mode propagating into the black hole behaves like
\begin{equation}\label{eq:mirac-limit}
  {\bf\tt R}^{(M)}_\minus(z) \sim \lim_{\bar\omega\rightarrow\bar\omega_+}
  \left(\lim_{\bar{a}\rightarrow\bar{a}_p}
  R_\minus(z,\bar\omega,\bar{a})\right).
\end{equation}
Interestingly, we find that the Born series ${\bf\tt
  R}^{(M)}_\minus(z)$ {\em does not terminate} when $\bar\omega_+$ and
$\bar{a}$ correspond to our confluent Heun polynomial solutions.
Instead, it is a linear combination, ${\bf\tt
  R}^{(M)}_\minus(z)+C_MR_+(z,\bar\omega_+,\bar{a}_{{}_P})$, that
terminates.  In fact, the same is true if the limits are taken in the
reverse order, except that $C_M$ changes.  So the commutation of
the limits gives
\begin{align}
  \lim_{\bar\omega\rightarrow\bar\omega_+}
  \lim_{\bar{a}\rightarrow\bar{a}_p}
  R_\minus(z,\bar\omega,\bar{a}) -
  \lim_{\bar{a}\rightarrow\bar{a}_p}
  \lim_{\bar\omega\rightarrow\bar\omega_+}&
  R_\minus(z,\bar\omega,\bar{a}) \\
  &\propto R_+(z,\bar\omega_+,\bar{a}_{{}_P}). \nonumber
\end{align}

We should point out that the non-terminating Born series ${\bf\tt
  R}^{(M)}_\minus(z)$ and $R_+(z,\bar\omega_+,\bar{a}_{{}_P})$ are
both local series solutions.  We have seen for QNM solutions on the
NIA, a confluent Heun function cannot exist (ie the series will not
converge) unless the series terminates.  A similar argument holds for
TTM${}_L$ solutions\footnote{To consider TTM${}_L$s, we follow the
  method of Sec.~\ref{sec:radial-teukolsky-Leaver}, but let
  $\xi=\xi_+$. $p$ remains unchanged and only $u_2$ in
  Eq.~(\ref{eq:asymptotic_a}) is different.  This term has no effect
  on the convergence argument.}, and so neither of the Born series are
convergent for large $z$.

As a concrete example of this, consider the case with $N_+=5$.
A polynomial solution is found when $\bar{a}=0.931905$ with
${}_{-2}A_{20}=4.199325$.
The radial solution is
\begin{align}
  R_{51}(z)=& z^{\eta_+}(z-1)^{-1/2}e^{(\bar{r}_+-\bar{r}_\minus)\bar\zeta_+z}\big(1 \\
    &+ 1.06819(1-z) + 0.580484(1-z)^2 \nonumber\\
    &+ 0.191606(1-z)^3 + 0.117043(1-z)^4 \nonumber\\
    &+ 0.0337923(1-z)^5 + 0.0200664(1-z)^6\big).\nonumber
\end{align}
The first few terms of the local Born series for the outgoing mode are
\begin{align}
  {\bf\tt R}^{(M)}_\minus(z)=& (z-1)^{-1/2}\big(1 + 1.06819(1-z) \\
    &+ 0.580484(1-z)^2 - 20.425330(1-z)^3  \nonumber\\
    &+ 47.821179(1-z)^4 - 42.3949518(1-z)^5 \nonumber\\
    &+ 14.5536633(1-z)^6 - 0.6166794(1-z)^7+\cdots\big),\nonumber
\end{align}
while the first few terms of the local Born series for the incoming
mode are
\begin{align}
  R_+(z)=& (z-1)^{5/2}\big(1 - 2.313832(1-z) \\
    &+ 2.0546777(1-z)^2 - 0.7049349(1-z)^3 \nonumber\\
    &+ 0.0299113(1-z)^4+\cdots\big).\nonumber
\end{align}
We find that $R_{51}(z)\sim {\bf\tt R}^{(M)}_\minus(z)+20.616936R_+(z)$
with all terms of order $(z-1)^{13/2}$ and higher canceling.

At $z=1$, $R_{51}(z)$ behaves as a linear combination of an incoming
and an outgoing mode, and by itself cannot be a QNM.  However, if a
TTM${}_L$ mode exists at the same values of $\bar\omega$ and
$\bar{a}$, we could take a linear combination of $R_{51}(z)$ with this
mode to remove the incoming mode behavior at $z=1$ without introducing
an incoming mode a infinity.  Given the explicit, closed form solution
$R_{51}(z)$ for the radial function, we can construct a second,
linearly independent solution $\tilde{R}(z)=v(z)R_{51}(z)$ to the
radial Teukolsky equation.  Using standard methods, we can obtain
$v(z)$ as an integral and determine the local behavior of
$\tilde{R}(z)$ at both $z=1$ and $z=\infty$.  We find that
$\lim_{z\rightarrow1}\tilde{R}(z)\sim R_+(z)$, representing an
incoming mode at the event horizon.  At infinity, $\tilde{R}(z)$ also
corresponds, to an {\em incoming} mode, although the solution seems
ill-behaved here.  Thus, $\tilde{R}(z)$ is {\em not} a TTM${}_L$ mode
and we must conclude that no QNM or TTM${}_L$ modes exists at the
values of $\bar\omega$ and $\bar{a}$ associated with $R_{51}(z)$.

We have performed the same analysis for numerous cases of
``miraculous'' polynomial solutions (gray dots in
Fig.~\ref{fig:l234omegapN4-133}) and have found the same result in
each case.  This does not prove that the ``miraculous'' cases are not
QNMs (or TTM${}_L$s) in all cases.  It is possible that under certain
circumstances ${\bf\tt R}^{(M)}_\minus(z)$ itself terminates and
contains no contribution from $R_+(z)$.  However, we conjecture that
whenever ${\bf\tt R}^{(M)}_\minus(z)$ does not terminate, {\em the
  ``miraculous'' solutions are neither QNMs nor TTM${}_L$s}.

While examining the Born series is helpful in understanding the
behavior of the anomalous and miraculous solutions, it is somewhat
simpler to determine their behavior by examining the properties of the
coefficient matrix, (\ref{eq:tridiag_cond}), used to locate the
Heun polynomial solutions.  For clarity, we will illustrate this
for the specific cases discussed above. Assume that we choose the
parameter set $\{\bar\zeta_+,\xi_\minus,\eta_+\}$ and use the local
solution of Eq.~(\ref{eq:local_sol_z1a}) to construct the coefficient
matrix.  For the case of $\bar\omega=\bar\omega_+$, the upper-left
block of (\ref{eq:tridiag_cond}) will be a
$(q+1)\times(q+1)$-dimensional tridiagonal matrix.  As outlined above,
we can easily see that the coefficient $f^{(a)}_{-\delta}=0$, and so long as
$s<0$ this vanishing element will be part of the upper-left block of
(\ref{eq:tridiag_cond}).  This submatrix takes the block form
\begin{equation}\label{eq:tridiag_submatrix}
  \left[\begin{array}{cccc|ccccc}
  g_0 & f_0 & 0 & \cdots & 0 & 0 & 0 & \cdots & 0 \\
  h_1 & g_1 & f_1 & \ddots & 0 & 0 & 0 & \cdots & 0  \\
  0  & \ddots & \ddots & \ddots & \ddots & \ddots & \ddots & \ddots & \vdots \\
  0  & \ddots & h_{-\delta} & g_{-\delta} & 0 & 0 & 0 & \cdots & 0  \\
\hline
  0  & \cdots & 0 & h_{1-\delta} & g_{1-\delta} & f_{1-\delta} & 0 & \cdots & 0 \\
  0  & \cdots & 0 & 0 & h_{2-\delta} & g_{2-\delta} & f_{2-\delta} & \ddots & \vdots \\
  \vdots & \ddots & \vdots & \vdots & \ddots & \ddots & \ddots & \ddots & 0 \\
  0  & \cdots & 0 & 0 & 0 & 0 & h_{q-1} & g_{q-1} & f_{q-1} \\
  0  & \cdots & 0 & 0 & 0 & 0 & 0 & h_q & g_q \\
  \end{array}\right],
\end{equation}
with $q=N_+-s-1$ and $-\delta=q+2s$.  Because of its tridiagonal form
and the vanishing of the off-diagonal element $f^{(a)}_{-\delta}$, it is
easy to see that the determinant of the matrix
(\ref{eq:tridiag_submatrix}) can be written as the product of the
determinants of the two diagonal block elements\cite{ElMikkawy-2004}.
To simplify the discussion, let us denote the determinant of the
upper-left diagonal block as $\Delta_u$ and the determinant of the
lower-right diagonal block as $\Delta_d$.  So, the necessary and
sufficient condition for a polynomial solution can be written as
$\Delta_{q+1}=\Delta_u\Delta_d=0$.

First, consider the case when $\Delta_d=0$ while $\Delta_u\ne0$.
Because $\Delta_u\ne0$, the first $1-\delta=N_++s$ coefficients,
$c^{(a)}_k$, of Eq.~(\ref{eq:local_a_sol_series}) must vanish.  The
next $-2s$ coefficients, $c^{(a)}_k$ with $1-\delta\le k\le q$,
associated with the lower-right block need not vanish.  The vector of
values for $c_k^{(a)}$ can be determined using standard matrix
methods.  We are free to normalize this solution vector so that
$c^{(a)}_{1-\delta}=1$.  If we were to simply evaluate the coefficients
directly via the recurrence relation, we would find that
\begin{equation}\label{eq:non-gen-coef}
  c^{(a)}_{1-\delta} = \frac{\Delta_u}{\prod_{i=0}^{-\delta}f_i}.
\end{equation}
Since $\Delta_u\ne0$ and $f^{(a)}_{-\delta}=0$, $c^{(a)}_{1-\delta}$
diverges.  This is the ``anomalous'' case, but computing the
coefficients via matrix methods automatically rescales the solution as
in Eq.~(\ref{eq:anom-rescale}).

We have seen that an anomalous solution is simultaneously a QNM and a
TTM${}_L$, so it is not surprising that the lower-right block of
(\ref{eq:tridiag_submatrix}) is a $(-2s)\times(-2s)$ matrix with
coefficients that are identical to the coefficients of one of the
matrices that can be constructed to consider TTM${}_L$ polynomial
solutions.  To construct this matrix, we continue to use
Eq.~(\ref{eq:local_sol_z1a}), but change the choice of $\xi$ to
$\xi_+$.  The local solution is now associated with modes propagating
out from the event horizon.  The necessary condition for a polynomial
solution in this case is $q=-1-2s$ (see Ref.\cite{cook-zalutskiy-2014}
for details).  While this condition does not constrain $\bar\omega$,
we may choose to look for solutions when $\bar\omega=\bar\omega_+$.
Under these conditions, the matrix is identical to the lower-right
block of (\ref{eq:tridiag_submatrix}).

Now consider the reverse situation in which $\Delta_d\ne0$ while
$\Delta_u=0$.  In this case, we can determine the first
$1-\delta=N_++s$ coefficients, $c^{(a)}_k$, of
Eq.~(\ref{eq:local_a_sol_series}) using standard matrix methods.  The
matrix element $h^{(a)}_{1-\delta}$ in the lower-left block cannot
vanish because we have chosen $N_+$ so that $h^{(a)}_{q+1}=0$, and
from Eq.~(\ref{eq:local_a_h}) it is clear that only one coefficient
$h^{(a)}_k$ can vanish unless they all vanish with $p=0$.  Now, so
long as $\Delta_d\ne0$, there exists a unique solution for the
remaining $-2s$ coefficients, $c^{(a)}_k$.  If we again simply
evaluate the coefficients directly via the recurrence relation,
$\Delta_u=f^{(a)}_{-\delta}=0$ means that we must evaluate
Eq.~(\ref{eq:non-gen-coef}) in the limit as in
Eq.~(\ref{eq:mirac-limit}).  This is the ``miraculous'' case.
However, the value of $c^{(a)}_{1-\delta}$ computed via matrix methods
will not, in general, equal the limit value of
Eq.~(\ref{eq:non-gen-coef}).  When the values differ, the behavior
local to $z=1$ is a linear combination of incoming and outgoing modes.

What if the determinants of both diagonal blocks vanish?  First, it is
not clear that the ``miraculous'' solution will persist in this case.
If it could, then the vanishing of the determinant of the lower-right
block implies that a polynomial TTM${}_L$ mode also exists.  As
discusses earlier, the existence of a TTM${}_L$ solution in addition
to a ``miraculous'' solution would guarantee that we could construct a
QNM through a linear combination of these solutions.  This would give
us two linearly independent polynomial solutions, one a QNM and one a
TTM${}_L$.  If the ``miraculous'' solution does not persist, then
the solution reverts to being of ``anomalous'' type.

\section{Summary and Discussion}
\label{sec:summary}

The main goal of this paper has been to understand when QNMs can exist
with frequencies precisely on the NIA.  We have shown that there exist
countably infinite sets of QNMs with frequencies on the NIA.  These
exist in two distinct families.  The first family obeys the constraint
in Eq.~(\ref{eq:omega_minus}) that $\bar\omega=\bar\omega_\minus$.  It
seems that all of these polynomial solutions are ``generic''\footnote{
  Our methods do not rule out the possibility that a particular
  solution could be non-generic.  However, we have not encountered
  any.} and correspond to QNMs with purely imaginary frequencies.
Examples of this family of modes, for $\ell=2$, $3$, and $4$, are
displayed in Fig.~\ref{fig:l234omegamN9-133}.  The second family obeys
the constraint in Eq.~(\ref{eq:omega_plus}) that
$\bar\omega=\bar\omega_+$.  All of these modes are ``non-generic'' and
split into two types.  One type is ``anomalous'', in which case
each polynomial solution corresponds simultaneously to a QNM and a
TTM${}_L$.  The second type is ``miraculous'', in which case each
polynomial solution is neither a QNM nor a TTM.  Examples of both types
of this family of modes, for $\ell=2$, $3$, and $4$, are displayed in
Fig.~\ref{fig:l234omegapN4-133}.  There are additional interesting
points to consider for each case.

\subsection{The anomalous cases}
\label{sec:summary_anom}
It has been know for some time that the algebraically special modes of
Schwarzschild (see Eq.~(\ref{eq:alg-spec-sch})) are simultaneously
QNMs and TTM${}_L$s.  This was proven by Maassen van den
Brink\cite{van_den_brink-2000} by exploiting the supersymmetric
relationship between the Regge-Wheeler\cite{regge-wheeler-1957} and
Zerilli\cite{zerilli-1970} equations for the odd- and even-parity
gravitational perturbations of Schwarzschild.  More precisely, he
found that the frequencies $\bar\Omega_\ell$ yielded solutions that
were simultaneously QNMs and TTM${}_L$s when considering the Zerilli
equation, but that the Regge-Wheeler equation had no QNM or TTM
solutions at this frequency.  The latter proving that there are no
TTM${}_R$ modes for the algebraically special frequencies of
Schwarzschild.  In terms of the Teukolsky equation, he was able to
show the analogous result in the limit that $\bar{a}=0$.
Specifically, for $s=-2$ the algebraically special modes of
Schwarzschild are simultaneously QNMs and TTM${}_L$s, while for $s=+2$
they do {\em not} correspond to TTM${}_R$s.  However, in considering
the extension of these arguments to $0<\bar{a}<1$, his approach lead
him to conclude that the algebraically special modes were all of
generic type for $\bar{a}>0$.

Using the theory of confluent Heun polynomials, we have shown
conclusively that this is not true.  The {\em anomalous} cases of
$\bar\omega=\bar\omega_+$ (black dots in
Fig.~\ref{fig:l234omegapN4-133}) are all discrete algebraically
special modes with $\bar{a}>0$.  As we discussed in
Sec.~\ref{lab:nongeneric}, for the anomalous case, the determinant
$\Delta_d$ vanishes and this is identical to the $\Delta_{q+1}=0$
condition when considering TTM${}_L$s.  We have shown in
Ref.\cite{cook-zalutskiy-2014}, that this is identical to the
vanishing of the square of the Starobinski constant which is the
equation satisfied by the algebraically special modes for all values
of $\bar{a}$\cite{chandra-1984}.

We find, then, that the $m=0$ algebraically special modes have a
particularly interesting behavior.  Chandrasekhar\cite{chandra-1984}
provided the first table of the mode frequencies for $\bar{a}>0$,
showing that the $\ell=2$, $m=0$ mode frequencies moved along the NIA
(at least initially) as $\bar{a}$ increased.
Onozawa\cite{onozawa-1997} plotted in his Fig.~7 the $\ell=2$, $m\ge0$
algebraically special mode frequencies.  In this plot it is clear that
when $\bar{a}\sim0.494446$, the $m=0$ sequence turns off of the NIA
and the mode frequencies are complex.  This plot was extended to show
similar behavior for the $\ell=3$ modes as well in Fig.~24 of
Ref.\cite{cook-zalutskiy-2014}.  Comparing this to the behavior of the
anomalous points (black dots) in Fig.~\ref{fig:l234omegapN4-133}, we
see that the first few points, starting at $\bar\Omega_\ell$ and
moving to the right (increasing $\bar{a}$) correspond to specific
points along the algebraically special sequences as they move along
the NIA.  While each mode frequency along the sequence represents, in
general, a TTM${}_L$, at these particular points, the mode is
simultaneously a QNM and a TTM${}_L$.\footnote{Along the corresponding
  $s=+2$ sequence of algebraically special modes, in general each mode
  is TTM${}_R$.  However, as with the $\bar{a}=0$ case, modes at the
  same set of frequencies are ``miraculous'' and are {\em not}
  TTM${}_R$s.}

The anomalous polynomial mode frequencies with the largest values of
$\bar{a}$ correlate with the point where the algebraically special
sequences turn off of the NIA as $\bar{a}$ increases.  But,
Fig.~\ref{fig:l234omegapN4-133} shows something else that is new.
{\em There is a second branch of the algebraically special modes
  which, to our knowledge, has never before been noticed}.  To be
clear, let us consider the case of the $\ell=2$, $m=0$ algebraically
special modes.  The sequence begins at $\bar\omega=\bar\Omega_2$ with
$\bar{a}=0$.  For $0<\bar{a}\alt0.494446$, $-{\rm Im}(\bar\omega)$
increases toward $\bar\omega\sim-3.3308i$.  At this point, if
$\bar{a}$ increases, the sequence moves off of the NIA.  But, {\em
  there is a new branch of the sequence moving along the NIA with
  $-{\rm Im}(\bar\omega)$ increasing beyond this point as $\bar{a}$
  decreases back to zero}.  Numerical evidence shows that this new
branch of the algebraically special modes behaves as an inverse-power
law in $\bar{a}$.  So, we find that for $0<\bar{a}\alt0.494446$, there
are two distinct sequences of algebraically special modes which
together have frequencies that cover the entire NIA below $-2i$.  And,
at a countably infinite number of frequencies, the $s=-2$ modes become
anomalous.  These anomalous solutions correspond with points where the
$m=0$ QNM frequency sequences terminate at (or emerge from) the NIA.
At all of these points, the $s=-2$ mode is simultaneously a QNM and a
TTM${}_L$, but the $s=+2$ mode is not a TTM${}_R$.

Similar behavior is seen for the $\ell=3$, $m=0$ algebraically special
modes.  However, Fig.~\ref{fig:l234omegapN4-133} shows that the
$\ell=4$ case is a bit more complicated.  It suggests that the
algebraically special mode frequencies will first move to smaller
values of $-{\rm Im}(\bar\omega)$, then reverse course and increase,
as $\bar{a}$ is increased.  Following the point where continuing to
increase $\bar{a}$ causes the mode frequencies to move off of the NIA,
we see the same behavior of the mode frequencies where they continue
to moving along the NIA as $\bar{a}$ decreases again toward zero.

\subsection{The miraculous cases}
\label{sec:summary_mirac}
The miraculous subset of the $\bar\omega=\bar\omega_+$ polynomial
solutions seem to represent a new and unusual set of modes.  They are
neither QNM nor TTM unless some third level of non-generic behavior
intervenes to make them anomalous, and this has not been seen so far.
These solutions are countably infinite, with a possibly finite number
of them (see Fig.~\ref{fig:l2omegap-cmp}) for each $\ell$
corresponding to points along certain QNM sequences that become
tangent to the NIA.  For every $m=0$ looping QNM sequence we have seen
(see Figs.~\ref{fig:m0l2n09-26}, \ref{fig:m0l3n18-31}, and
\ref{fig:m0l4n25-31}), {\em each point of tangency with the NIA represents
a point that must be missing from that QNM sequence}.

Two recent papers\cite{Yang-et-al-2013a,Hod-2013} have claimed to find
gravitational QNMs with frequencies on the NIA.  We have shown that no
actual QNMs exist with frequencies on the NIA at frequencies
corresponding to their solutions.  However, in both cases there is
some correlation with our ``miraculous'' solutions.  In the first
paper\cite{Yang-et-al-2013a}, using both WKB and
matched-asymptotic-expansion methods, the authors find an
approximation for $\bar\omega$ for what they refer to as zero damped
modes (ZDMs) and damped modes (DMs).  With $\epsilon\equiv1-\bar{a}$,
and with $m=0$, they find
$\bar\omega\approx-i(n+1/2)\sqrt{\epsilon/2}$.  If we consider
$\bar\omega_+$ in the limit that $\bar{a}\rightarrow1$, then we find
$\bar\omega_+\approx-iN_+\sqrt{\epsilon/2}$.  As
Fig.~\ref{fig:l2omegapN4-16} makes clear, for $N_+\gg1$ there are a
large number of curves with small $-{\rm Im}(\bar\omega)$ as we
approach $\bar{a}\sim1$.  Clearly, the difference between $n+1/2$ and
$N_+$ is small for large $n=N_+$, so their approximation can be a
reasonably good approximation for our necessary (but not sufficient)
condition for having a polynomial QNM solution with a purely imaginary
frequency.  In the second paper\cite{Hod-2013}, the author uses a more
tailored matched asymptotic expansion to find $m=0$ resonances in the
limits that $\bar{a}\approx1$ and $-{\rm Im}(\bar\omega)$ is small.
His result is nearly identical with our expression for $\bar\omega_+$,
except that $N_+$ is replaced by $\ell+1+n$.  Here, $n$ is an integer
and if $\ell$ were simply our harmonic index, the result would be
essentially identical to ours.  However, in this case $\ell$ is the
integer harmonic index {\em plus} a non-integer correction.
Never-the-less, the correction is small, and we again find that his
resonance is a good approximation for our necessary (but not
sufficient) condition for having a polynomial QNM solution with a
purely imaginary frequency.

Clearly, only the ``miraculous'' modes are reasonably close to either
of the claimed solutions with frequencies on the NIA.  But we have
seen that none of these modes are QNMs.  The authors of
Ref.~\cite{Yang-et-al-2013a} offered additional evidence that they had
found QNMs with a purely imaginary frequency by finding numerical
solutions which seemed consistent with their approximate expressions
for those frequencies.  However, in finding numerical solutions with
frequencies on the NIA, the authors used Leaver's continued fraction
method on the NIA.  As we have shown in
Sec.~\ref{sec:radial-teukolsky-Leaver}, the continued fraction does
not converge when evaluated with a frequency on the NIA and cannot be
used to locate QNMs.

In Ref.~\cite{Hod-2013}, the author's finding of a continuum of QNMs
for $\bar{a}\sim1$ is accompanied by a continuum of ``total reflection
modes'' that correspond to our TTM${}_L$s.  These nearly coinciding
QNM/TTM${}_L$ pairs seem like good approximations for the anomalous
QNM/TTM${}_L$ solutions we have found, except that we find no such
solutions in the small $\bar\omega$, large $\bar{a}$ limit in which
the author's approximations are valid.  It could be very informative
to understand why these analytic approximation methods both seem to
find QNM solutions when none exist.  At present, we do not have a
clear understanding of this.  It seems clear that both methods are
finding a reasonable approximation of the necessary condition that
$\bar\omega=\bar\omega_+$.  However, neither seems to incorporate
anything analogous to the $\Delta_{q+1}=0$ condition to restrict the
solutions.  Perhaps more to the point, none of these approximate
methods deal with the very subtle aspects of determining the nature of
the solution at the horizon boundary when the solutions are
non-generic.

\acknowledgments
We would like to thank Emanuele Berti, Aaron Zimmerman, and Shahar Hod
for helpful discussions.

\end{document}